\documentclass[pdflatex,sn-mathphys-num]{sn-jnl}

\usepackage{graphicx}%
\usepackage{multirow}%
\usepackage{amsmath,amssymb,amsfonts}%
\usepackage{amsthm}%
\usepackage{mathrsfs}%
\usepackage[title]{appendix}%
\usepackage{xcolor}%
\usepackage{textcomp}%
\usepackage{manyfoot}%
\usepackage{booktabs}%
\usepackage{algorithm}%
\usepackage{algorithmicx}%
\usepackage{algpseudocode}%
\usepackage{listings}%

\theoremstyle{thmstyleone}%
\newtheorem{theorem}{Theorem}
\newtheorem{proposition}[theorem]{Proposition}%

\theoremstyle{thmstyletwo}%
\newtheorem{remark}{Remark}%

\theoremstyle{thmstylethree}%
\newtheorem{lemma}{Lemma}%

\begin{document}

\title[Intrinsic FEM for Fluids on Manifolds]{Intrinsic Finite Element Methods for Fluids on Riemannian Manifolds Compared with Surface FEM}


\author*[1]{\fnm{Yongxing} \sur{Wang}}\email{scsywan@leeds.ac.uk}



\affil*[1]{\orgdiv{School of Computer Science}, \orgname{University of Leeds}, \orgaddress{ \city{Leeds}, \country{UK}}}




\abstract{We present an intrinsic finite element formulation for the incompressible Navier--Stokes equations on Riemannian manifolds. We derive the corresponding weak formulation and prove that the backward Euler discretisation is energy stable. The proposed framework is validated on several representative manifolds, with particular attention paid to the long-time behaviour of the flow and its convergence to steady-state solutions represented by Killing vector fields. Comprehensive comparisons are performed with the surface finite element method and a corresponding eigenvalue formulation for Killing vector fields. The numerical results demonstrate that the intrinsic formulation provides an accurate, computationally efficient, and geometrically transparent alternative to embedded surface finite element formulations, while naturally extending to higher-dimensional Riemannian manifolds.}

\keywords{Navier--Stokes equations, Riemannian manifolds, surface finite element method, intrinsic finite element method, Killing vector fields}



\maketitle

\section{Introduction}
\label{sec_introduction}

Fluid flows constrained to curved manifolds arise in a wide range of physical and computational settings, including thin liquid films \cite{craster2009dynamics,ledda2022gravity,olshanskii2023equilibrium}, viscous membranes and biological surface flows \cite{bothe2010surface,tasso2013viscous,barrett2016stable,wittwer2023shape,olshanskii2023equilibrium}, as well as geophysical flows on planetary surfaces \cite{fries2018surface,reuther2018solving}. These problems naturally lead to Navier--Stokes (NS) equations in which the usual Euclidean differential operators are replaced by their Riemannian counterparts. The velocity is a tangent vector field, incompressibility is expressed through the covariant divergence, and viscous stresses are defined using the intrinsic strain-rate tensor associated with the Levi--Civita connection. The resulting equations have attracted increasing attention in both analysis and numerical computation \cite{cao1999navier,reuther2018solving,olshanskii2018finite,samavaki2020navier,pruss2021navier,olshanskii2022tangential,olshanskii2023equilibrium,shao2025navier,shao2025coriolis,hardering2025parametric}.

A key geometric feature of incompressible viscous flows on compact manifolds is the role of Killing vector fields \cite{petersen2016riemannian,shimizu2024hydrodynamic}. For a Killing field the symmetric part of the covariant derivative vanishes, and hence the viscous dissipation associated with the strain rate tensor is zero. Analytical results show that, under suitable assumptions, equilibria of the surface NS equations are precisely Killing vector fields and that any initial velocity field converges exponentially towards such equilibria \cite{pruss2021navier,shao2025navier,shao2025coriolis}. The geometry of manifolds admitting hydrodynamic Killing vector fields has also been investigated in \cite{shimizu2024hydrodynamic}. From a computational viewpoint, Killing fields provide a useful benchmark because they test whether a discretisation can preserve the nullspace of the viscous operator and the long-time structure of the flow. However, numerical approaches to Killing vector fields remain limited, with an eigenvalue formulation being investigated on Enneper's surface, the torus, and the Klein bottle in \cite{brunet2019numerical}.

Most recent numerical work on the surface NS equations follows an embedded surface finite element philosophy. The governing equations are usually formulated in intrinsic differential-geometric notation using operators such as the surface divergence, covariant derivative, Bochner Laplacian and Hodge Laplacian, whereas the numerical discretisation is subsequently performed on a triangulated surface embedded in $\mathbb{R}^3$. Representative approaches include the classical surface finite element method \cite{olshanskii2010finite,dziuk2007finite,dziuk2012fully,dziuk2013finite,elliott2015evolving,reuther2018solving}, together with TraceFEM and CutFEM based on background volumetric meshes and level-set representations of the surface \cite{olshanskii2009trace,demlow2012adaptive,olshanskii2014evolving,olshanskii2015analysis,olshanskii2017trace,olshanskii2018finite,olshanskii2022tangential}, as well as a recently proposed method based on symbolic computation that performs all differential operations entirely within a Cartesian framework \cite{herrada2025goodbye}. These embedded formulations have proved highly successful for complex geometries and surfaces without global parameterisations. However, they inevitably rely on an explicit surface triangulation in the ambient space and do not exploit the intrinsic parameterisation when one is available. Consequently, quantities such as the metric tensor, covariant derivative and Christoffel symbols, which naturally describe the geometry of the manifold, play little role in the numerical discretisation. Furthermore, extending embedded formulations to higher-dimensional manifolds is considerably less straightforward. These observations motivate the development of numerical methods formulated directly in intrinsic parameter space.

Notably, a fully intrinsic discretisation in local parameter coordinates, although seemingly the most natural approach, has received relatively little attention and remains largely unexplored for incompressible flows on manifolds. In such an approach, the computational domain is the parameter space, and all geometric information is incorporated through the metric tensor, its inverse, the Riemannian volume element, and the Christoffel symbols. This formulation is closely aligned with the coordinate-based framework commonly used in Riemannian geometry, geophysical fluid dynamics, and relativistic fluid mechanics. Moreover, it extends naturally beyond two-dimensional embedded surfaces to genuinely higher-dimensional manifolds, including relativistic hydrodynamics in four-dimensional spacetime \cite{gourgoulhon2006introduction,hoult2020stable,denicol2022relativistic}. A very recent intrinsic finite element study on manifolds focused primarily on geometric analysis and finite element exterior calculus rather than incompressible fluid dynamics \cite{Gawlik2026}.

The objective of this work is to develop an intrinsic finite element framework for the numerical simulation of incompressible NS equations on Riemannian manifolds and to demonstrate its capability through stability analysis and numerical experiments. The main contributions are summarised as follows:
\begin{itemize}
    \item We derive a fully intrinsic weak formulation of the incompressible NS equations on Riemannian manifolds and prove that the backward Euler time discretisation is energy stable.
    
    \item We perform a systematic comparison between the intrinsic formulation, surface FEM, and a corresponding eigenvalue formulation, demonstrating the respective advantages and limitations of these computational frameworks.

    \item We employ Killing vector fields as exact steady-state solutions of the NS equations to validate the long-time behaviour of the proposed method.
    
    \item We demonstrate that, unlike surface FEM, the proposed formulation extends naturally to higher-dimensional Riemannian manifolds without requiring an ambient Euclidean embedding, as illustrated by numerical simulations on a three-dimensional manifold equipped with the Schwarzschild spatial metric.
\end{itemize}

The remainder of the paper is organised as follows. Section~\ref{sec_surface_ns} introduces the covariant form of the incompressible Navier--Stokes equations on Riemannian manifolds. Section~\ref{sec_weak form} derives the finite element weak formulationulation, and Section~\ref{sect_stability} presents the time discretisation and energy stability result. Section~\ref{sec_sphere} studies flows on spherical geometries and compares the intrinsic formulation with surface FEM and an eigenvalue approach. Section~\ref{sec_torus} considers the torus and helicoid, where no coordinate singularity is present. Section~\ref{sec_schwarzschild} illustrates the extension to a three-dimensional manifold with Schwarzschild spatial metric. Conclusions and perspectives are given in Section~\ref{sec_conclusion}.

\section{Navier--Stokes equations on Riemannian manifolds}\label{sec_surface_ns}

We first recall the standard incompressible Navier--Stokes equations in the Euclidean space $\mathbb{R}^d$, where $d$ denotes the spatial dimension:
\begin{align}
\label{ns_flat1}
\rho\left(\partial_t u_i + \sum_{j=1}^d u_j\partial_j u_i\right)
-
\sum_{j=1}^d \partial_j\sigma_{ij}
&=0,\\
\label{ns_flat2}
\sum_{j=1}^d\partial_j u_j&=0,
\end{align}
where
\begin{equation}
\label{stress_tensor_euclidean}
\sigma_{ij}
=
\mu(\partial_j u_i+\partial_i u_j)-\pi\delta_{ij}
\end{equation}
is the Cauchy stress tensor. Here, $\rho$ denotes the density, $\mu$ the dynamic viscosity, $u_i$ the velocity components, $\pi$ the pressure, and $\delta_{ij}$ the Kronecker delta. The purpose of this section is to replace these Euclidean differential operators by their covariant counterparts on a Riemannian manifold $(M,g)$ with $g$ being the metric tensor.

The very first step of extension of the equations (\ref{ns_flat1}) and (\ref{ns_flat2}) to a Riemannian manifold $M$ is to distinguish variables or tensors with upper indices and lower indices since they represent different geometric objects. Please refer to \ref{sec_notations} for details. The velocity vector $u=u^i\partial_i$ (Einstein summation is adopted) is defined in the tangent space $T_pM$ of manifold $M$ at point $p$, where $\partial_i$ is the coordinate basis of $T_pM$. By introducing the covariant derivative, the convection term and divergence free equation are extended straightforward: 
\begin{equation}\label{convection}
\sum\nolimits_{j=1}^d u_j\partial_ju_i \rightarrow   u^j\nabla_j u^i,
\end{equation}
\begin{equation}\label{divergence}
\sum\nolimits_{j=1}^d\partial_j u_j 
\rightarrow
\nabla_j u^j ,
\end{equation}
where $\nabla_j$ represent the covariant derivative with respect to the $j^{th}$ coordinate.

The extension of the divergence of stress is subtle. From the expression of the convection term in (\ref{convection}), we seek a final system written in terms of the contravariant velocity components $u^i$. Therefore, the divergence of stress must also have a free upper index $i$. It's not difficult to see the stress tensor need to be defined as a $(2, 0)$ tensor $\mathcal{T}^{ij}$, so that its divergence $\nabla_j \mathcal{T}^{ij}$ would have a free upper index $i$. This is exactly the so-called Boussinesq–Scriven surface stress \cite{scriven1960dynamics,koba2018derivation,pruss2021navier}:
\begin{equation}
\mathcal{T}^{ij}(u, \pi)=\mu \left(\nabla^j u^i+ \nabla^i u^j \right)  - \pi g^{ij},
\end{equation}
where $\nabla^i = g^{ik}\nabla_k$ and $g^{ij}$ is the components of the inverse of the metric tensor. Alternative the stress tensor may be written as following using the short-hand notation ``;" for covariant derivative (while ``," is used for partial derivative) which has been widely used in literature:

\begin{equation}\label{strees_tensor_curved}
{\mathcal{T}^{ij}(u, \pi)=\mu \left(g^{jk}u_{;k}^i + g^{ik}u_{;k}^j \right)  - \pi g^{ij},}
\end{equation}

The computation of the divergence of the pressure term is straightforward:
\[
\nabla_j\left(\pi g^{ij}\right) =  g^{ij} \partial_j\pi,
\]
based on the metric compatibility condition $\nabla_k g^{ij} =0$.

While the computation of the divergence of the viscous strain term is complicated, and please refer to \ref{sec_hodge_laplacian} for the details of this derivation. The results are very tidy and neat: the divergence of the viscous strain yields the so-called Hodge Laplacian:
\begin{equation}\label{hodge_laplacian_paper}
\Delta_H (u)^i = \Delta_B (u)^i + R^i_j u^j,
\end{equation}
where $R^i_j$ is the Ricci curvature tensor. The first term in (\ref{hodge_laplacian_paper}) is called the Bochner Laplacian or rough Laplacian
\begin{equation}\label{bochner_paper}
     \Delta_B (u)^i = \nabla_j\left(g^{jk}u_{;k}^i\right)
 = g^{jk} u^i_{; jk}.
\end{equation}
which is from the divergence of the first term in (\ref{strees_tensor_curved}). The Ricci term in (\ref{hodge_laplacian_paper}) is from the divergence of the second term in (\ref{strees_tensor_curved}). 

Finally, we can write down the Navier-Stokes equation on a manifold as follows:

\begin{align}\label{ns_nonflat1}
\rho(\partial_t u^i + u^j\nabla_ju^i) - \nabla_j\mathcal{T}^{ij}  &= 0, \\
\label{ns_nonflat2}
\nabla_j u^j &= 0,
\end{align}
which shares the same structure as NS equations (\ref{ns_flat1}) and (\ref{ns_flat2}) in the Euclidean space, except that the partial derivative becomes covariant derivative. In particular, there are more terms involving the metric tensor and Christoffel symbols (\ref{christoffel_symbol}) appearing in the equations (\ref{ns_nonflat1}) and (\ref{ns_nonflat2}). For example, expansion of the convection term yields:
\[
u^j\nabla_j u^i
=  u^j\left(\partial_j u^i + u^k \Gamma_{kj}^i\right).
\]
The second-order covariant derivative in the Bochner Laplacian (\ref{bochner_paper}) also ends up with many tems involving metric tensor and Christoffel symbols as shown in (\ref{second_order_covariant_derivative}).
\section{Finite element weak formulation}\label{sec_weak form}
Fortunately, the weak form can be derived directly using the divergence theorem on manifolds \cite{lee2018introduction,petersen2006riemannian}, avoiding working with metric tensor and Christoffel symbols in (\ref{ns_nonflat1}) and (\ref{ns_nonflat2}).

Using the test functions $v_i=g_{ik}v^k$ for the velocity and $q$ for the pressure, the lower index $i$ in $v_i$ naturally contracts with the upper index $i$ in (\ref{ns_nonflat1}). Integrating the stress term by parts, the weak formulation of (\ref{ns_nonflat1}) and (\ref{ns_nonflat2}) reads:

\begin{equation}\label{ns_weak_form_manifold}
\begin{split}
&
\rho
\int_M
\left(
\partial_t u^i
+
u^j\nabla_j u^i
\right)
v^k g_{ik}
\, dM
\\
&
+
\frac{\mu}{2}
\int_M
\left(
g^{jk}u_{;k}^i
+
g^{ik}u_{;k}^j
\right)
\left(
g_{ik} v_{;j}^k
+
g_{jk} v_{;i}^k
\right)
\, dM
\\
&
-
\int_M q\, \nabla_ju^j\, dM
-
\int_M \pi\, \nabla_jv^j\, dM
\\
&
=
\int_{\partial M}
\mathcal{T}^{ij} \tau_j v^k g_{ki}
\, ds,
\end{split}
\end{equation}
where $dM$ is the Riemannian surface element and $ds$ is the induced line element on the boundary $\partial M$. Here, $\tau_j$ denotes the components of the outward unit normal to $\partial M$. Since it carries a lower index, $\tau_j$ is a covector.

\begin{figure}[h!]
\centering
\includegraphics[width=0.9\linewidth]{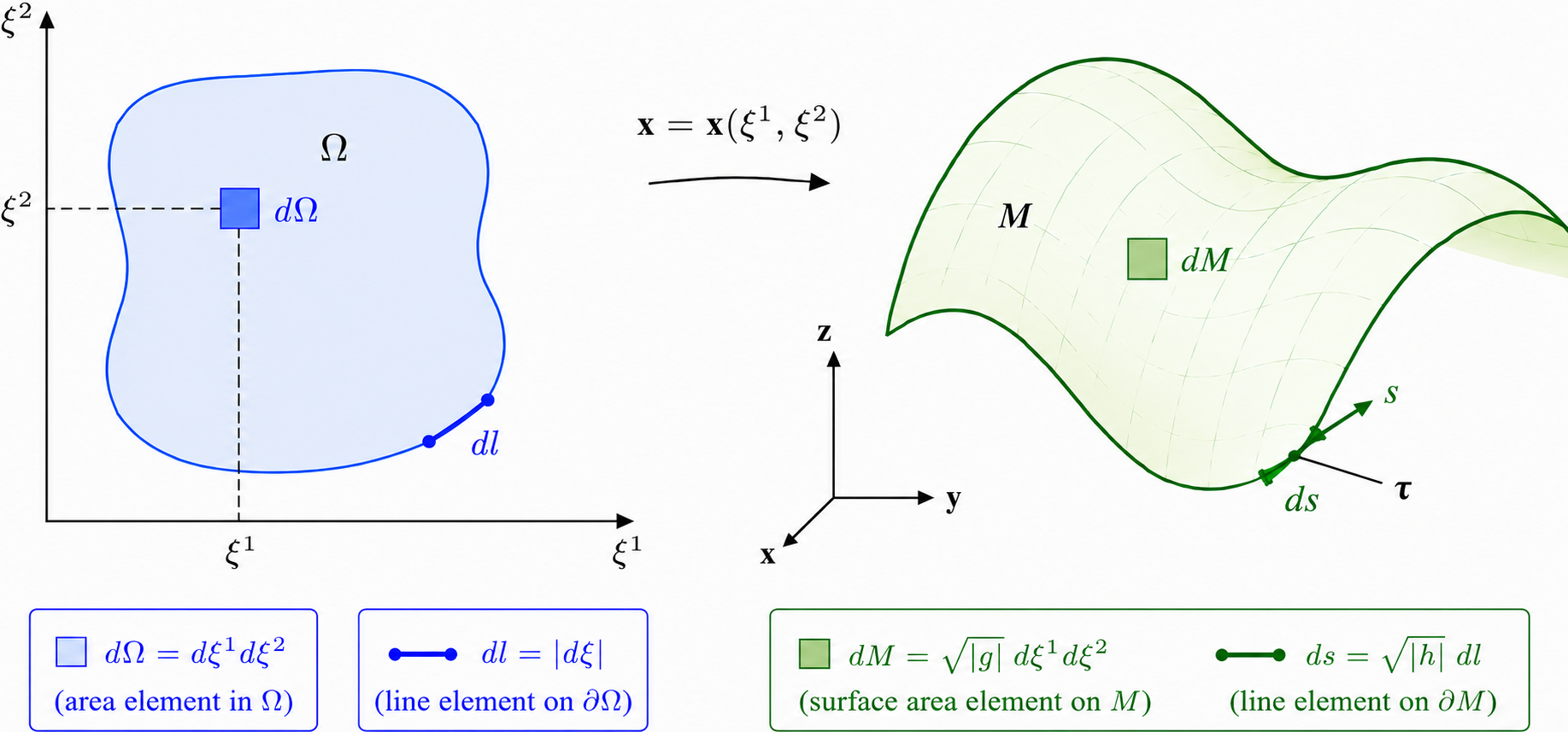}
\caption{Mapping from the parameter space $\Omega$ to a manifold $M$, where $h=g_{ij}s^is^j$ is the induced metric on $\partial M$.}
\label{fig_mapping_m_omega}
\end{figure}

For a two-dimensional surface, one may embed it into $\mathbb{R}^3$, directly triangulate the surface to approximate $M$, and use the surface finite element method (surface FEM) to discretise equation (\ref{ns_weak_form_manifold}). Alternatively, one may map the surface to a parameter domain $\Omega$, as shown in Figure \ref{fig_mapping_m_omega}, and compute the integrals in (\ref{ns_weak_form_manifold}) in the parameter space. In this case, the Jacobian factor $\sqrt{|g|}$ (or $\sqrt{|h|}$) appears in the surface (or line) element. The second approach is an intrinsic formulation, which is the focus of this work and can also be applied to higher-dimensional manifolds.

\section{Discretisation and energy stability}\label{sect_stability}
We adopt the backward Euler scheme for time discretisation of equation (\ref{ns_weak_form_manifold}), i.e.:

\begin{equation}\label{ns_weak_form_time_discretisation}
\begin{split}
&
\rho
\int_M
\left(
\frac{u^i_{n+1} - u^i_n}{\delta t}
+
u^j_{n+1}\nabla_j u^i_{n+1}
\right)
v^k g_{ik}
\, dM
\\
&
+
\frac{\mu}{2}
\int_M
\left(
g^{jk}u_{{n+1};k}^i
+
g^{ik}u_{{n+1};k}^j
\right)
\left(
g_{ik} v_{;j}^k
+
g_{jk} v_{;i}^k
\right)
\, dM
\\
&
-
\int_M q\, \nabla_ju^j_{n+1}\, dM
-
\int_M \pi_{n+1}\, \nabla_jv^j\, dM
\\
&
=
\int_{\partial M}
\mathcal{T}^{ij}_{n+1} \tau_j v^k g_{ki}
\, ds,
\end{split}
\end{equation}
for $t=t_0=0, t_1, t_2, \ldots$ with time step $\delta t=t_{n+1}- t_n$ ($n = 0, 1, 2, \ldots $). We shall prove that the above temporal discretisation (\ref{ns_weak_form_time_discretisation}) is energy-stable, for an enclosed flow: $\tau_iu^i =0$.

\begin{lemma}
If $(u_{n+1}, \pi_{n+1})$ is a solution of (\ref{ns_weak_form_time_discretisation}) at $t=t_{n+1}$, then
\begin{equation}\label{lemma_1}
 \int_M
\left(u^j_{n+1}\nabla_j u^i_{n+1}\right)
u^k_{n+1} g_{ik}
\, dM   =0.
\end{equation}
\end{lemma}
\begin{proof}
Using the condition of enclosed flow, the convection term can be integrated by parts and rewritten as:
\[
 \int_M
\left(u^j_{n+1}\nabla_j u^i_{n+1}\right)
v^k g_{ik}
\, dM   =  
-
\int_M u^i_{n+1}
\nabla_j\left(u^j_{n+1}
v^k g_{ik}\right)
\, dM.
\]
Using the metric compatibility condition $\nabla_j g_{ik}=0$, we further have:
\[
 \int_M
\left(u^j_{n+1}\nabla_j u^i_{n+1}\right)
v^k g_{ik}
\, dM   =  
-
\int_M u^i_{n+1}u^j_{n+1}
\left(\nabla_jv^k 
\right) g_{ik}
\, dM
-
\int_M u^i_{n+1}
\left(\nabla_ju^j_{n+1}
\right)v^k g_{ik}
\, dM.
\]
Let $v^k = u^k_{n+1}$ in the above equation, we have
\begin{equation}\label{proof_convection3}
 \int_M
\left(u^j_{n+1}\nabla_j u^i_{n+1}\right)
u^k_{n+1} g_{ik}
\, dM   =  
-\frac{1}{2}
\int_M u^i_{n+1}
\left(\nabla_ju^j_{n+1}
\right)u^k_{n+1} g_{ik}
\, dM.
\end{equation}
Finally, we let $v^k=0$ and $q=\left(u_{n+1},u_{n+1}\right)_g = u^i_{n+1}u^k_{n+1} g_{ik}$ in (\ref{ns_weak_form_time_discretisation}) to obtain:
\[
\int_M \left(u^i_{n+1}u^k_{n+1} g_{ik}\right)
\left(\nabla_ju^j_{n+1}
\right)
\, dM =0,
\]
which, together with (\ref{proof_convection3}), proves the lemma (\ref{lemma_1}).
\end{proof}

\begin{proposition}
If $(u_{n+1}, \pi_{n+1})$ is a solution of (\ref{ns_weak_form_time_discretisation}) at $t=t_{n+1}$, then the following energy estimate holds:
\begin{equation}\label{proposition_eq}
\frac{\rho}{2}\int_M u^i_{n+1}u^k_{n+1}g_{ik}dM 
+  2\delta t\mu\int_M\left(\epsilon_{n+1}, \epsilon_{n+1}\right)_g dM
\leq     
\frac{\rho}{2}\int_M u^i_{n}u^k_{n}g_{ik}dM,
\end{equation}
where 
\begin{equation}\label{viscous_strain_rate}
\epsilon(u)^{ij} = \frac{1}{2}\left(
g^{jk}u_{;k}^i
+
g^{ik}u_{;k}^j
\right)
\end{equation}
is the viscous strain rate tensor, $\epsilon_n = \epsilon(u_n)$, and $(\epsilon, \epsilon)_g = \epsilon^{ij}\epsilon_{ij} = \epsilon^{ij}\epsilon^{mn}g_{mi}g_{nj}$.
\end{proposition}
\begin{proof}
First notice that
\[
\left(
g^{nk}u_{;k}^m
+
g^{mk}u_{;k}^n
\right)g_{mi}g_{nj}
=g_{mi}\delta^k_j u_{;k}^m
+
g_{nj}\delta^k_iu_{;k}^n
=g_{mi} u_{;j}^m
+
g_{nj}u_{;i}^n.
\]
Then, setting $v^k = u^k_{n+1}$ and $q=-\pi_{n+1}$ in (\ref{ns_weak_form_time_discretisation}) yields:
\begin{equation}\label{proof_proposition1}
\begin{split}
&
\rho
\int_M
\left(
\frac{u^i_{n+1} - u^i_n}{\delta t}
+
u^j_{n+1}\nabla_j u^i_{n+1}
\right)
u^k_{n+1} g_{ik}
\, dM
\\
&
+
2\mu
\int_M
\left(\epsilon_{n+1}, \epsilon_{n+1}\right)_g\,dM
=
\int_{\partial M}
\mathcal{T}^{ij}_{n+1} \tau_j u^k_{n+1} g_{ki}
\, ds.
\end{split}
\end{equation}
Considering an enclosed flow without any boundary force prescribed, we drop the boundary term in (\ref{proof_proposition1}). Finally, we use the Cauchy–Schwarz inequality:
\begin{equation}\label{cauchy-schwarz}
\begin{split}
\int_M u^i_n u^k_{n+1} g_{ik} dM
&\leq
\left(\int_M u^i_n u^k_n g_{ik} dM\right)^{1/2}
\left(\int_M u^i_{n+1} u^k_{n+1} g_{ik} dM\right)^{1/2} \\
&\leq\frac{1}{2}\left(\int_M u^i_n u^k_n g_{ik} dM + \int_M u^i_{n+1} u^k_{n+1} g_{ik} dM\right),
\end{split}
\end{equation}
and lemma (\ref{lemma_1}), (\ref{proof_proposition1}) becomes
\begin{equation}\label{propostion_proof2}
\begin{split}
&\rho
\int_M
u^i_{n+1}
u^k_{n+1} g_{ik}
\, dM
+
2\delta t\mu
\int_M
\left(\epsilon_{n+1}, \epsilon_{n+1}\right)_g\,dM \\
&=\rho
\int_M
u^i_n
u^k_{n+1} g_{ik}
\, dM
\leq\frac{\rho}{2}\left(\int_M u^i_n u^k_n g_{ik} dM + \int_M u^i_{n+1} u^k_{n+1} g_{ik} dM\right).
\end{split}
\end{equation}
The energy estimate (\ref{proposition_eq}) is obtained after rearranging the terms in (\ref{propostion_proof2}).
\end{proof}

\begin{remark}
 For a steady-state solution, the viscous energy would disappear, so there $\exists N$, so that $\forall n>N$ the energy estimate (\ref{proposition_eq}) becomes:
 \begin{equation}\label{steady_state_estimate1}
\frac{\rho}{2}\int_M u^i_{n+1}u^k_{n+1}g_{ik}dM 
\leq     
\frac{\rho}{2}\int_M u^i_{n}u^k_{n}g_{ik}dM.
\end{equation}
On the other hand, steady-state also indicates $u_{n+1} \approx u_n$ for $\forall n>N$. This is also equal condition in the Cauchy-Schwarz inequality (\ref{cauchy-schwarz}), which leads to 
 \begin{equation}\label{steady_state_estimate2}
\frac{\rho}{2}\int_M u^i_{n+1}u^k_{n+1}g_{ik}dM 
\approx     
\frac{\rho}{2}\int_M u^i_{n}u^k_{n}g_{ik}dM,
\end{equation}
for $n>N$. While (\ref{steady_state_estimate1}) guarantees the numerical stability of (\ref{steady_state_estimate2}).
\end{remark}

\section{Numerical experiments}
In this section, we perform several numerical tests. Including 2 and 3 dimensional geometries, and geometries with and without boundaries or singularities. We will study the Killing vector fields corresponding the steady-state of the fluids. We will compare the intrinsic FEM with surface FEM, where The Taylor-Hood $P2/P1$ element will be use for the space discretisation. We will also solve the corresponding eigenvalue problems for comparison, where both the $P2$ and $P1$ elements will be tested. All numerical tests are implemented in \texttt{FreeFEM++}~\cite{MR3043640}, and the corresponding code and animations are available in the public GitHub repository: \href{https://yongxingwang.github.io/surfacefluids/}{https://yongxingwang.github.io/surfacefluids/}.

\subsection{Fluids on a 2-sphere $S_r^2$}\label{sec_sphere}
Consider $M = S_r^2$, the sphere in $\mathbb{R}^3$ of radius $r$ centred at the origin. Let
\begin{equation}\label{sphere_parameterised}
x = r \sin\phi \cos\theta, \quad
y = r \sin\phi \sin\theta, \quad
z = r \cos\phi,
\qquad
\theta \in [0, 2\pi), \ \phi \in (0, \pi),
\end{equation}
be the standard spherical coordinates on $S_r^2$, where $\theta$ is the longitude and $\phi$ is the colatitude.  

Assign the index correspondence $\theta \leftrightarrow 1$ and $\phi \leftrightarrow 2$. Then the local coordinate basis vectors are
\[
\mathbf{e}_1 = \partial_1 = \partial_\theta
= r\left(-\sin\phi \sin\theta, \, \sin\phi \cos\theta, \, 0\right),
\]
\[
\mathbf{e}_2 = \partial_2 = \partial_\phi
= r\left(\cos\phi \cos\theta, \, \cos\phi \sin\theta, \, -\sin\phi\right),
\]
and the components of the metric tensor are given by
\[
g_{11} = r^2 \sin^2, \quad g_{22}=r^2,\quad g_{12}=g_{21}=0.
\]
The components of the inverse metric tensor are:
\begin{equation}\label{sphere_inverse_g}
g^{11}=\frac{1}{r^2 \sin^2\phi},\quad g^{22} = \frac{1}{r^2}, \quad g^{12}=g^{21}=0.
\end{equation}

The Christoffel symbol has $8$ components, and only 3 of them are non-zero:
\begin{equation*}
\begin{split}
&\Gamma_{12}^1=\Gamma_{21}^1=\frac{1}{2} g^{1j}
\left(
\partial_1 g_{2j}
+
\partial_2 g_{1j}
-
\partial_j g_{21}
\right)=\frac{1}{2} g^{11}\partial_2 g_{11}=\frac{1}{2a^2\sin^2\phi}\left(2a^2\sin\phi\cos\phi\right)= \cot{\phi},\\
&\Gamma_{11}^2=\frac{1}{2} g^{2j}
\left(
\partial_1 g_{1j}
+
\partial_1 g_{1j}
-
\partial_j g_{11}
\right) = -\frac{1}{2} g^{22} \partial_2 g_{11}=-\frac{1}{2}\sin(2\phi).
\end{split}
\end{equation*}

For numerical test in this section, we simply use the following parameters: radius $r=1$, fluid density $\rho=1$ and viscosity $\mu=1$. 

\subsubsection{Simulations on half of a sphere without singularity}

It can be seen from the expression of the inverse metric tensor~(\ref{sphere_inverse_g}) that there are two singular points corresponding to $\phi=0$ (the North Pole) and $\phi=\pi$ (the South Pole). To avoid these singularities, we begin our numerical experiments with a hemisphere in which the North Pole is excluded, i.e.: $(\theta, \phi) \in [0,2\pi] \times [\pi/8, \pi/2]$. Periodic boundary conditions are applied at $\theta=0, 2\pi$, and slip boundary conditions are applied at $\phi=\pi/8, \pi/2$.

We start from an arbitrary initial velocity profile to run the simulation sufficiently long to investigate the steady-state solution. According a recent analysis \cite{shao2025navier}, an initial fluid field is expected to converge to a rotational Killing vector field. 

\begin{figure}[h!]
\centering   \includegraphics[width=1\linewidth]{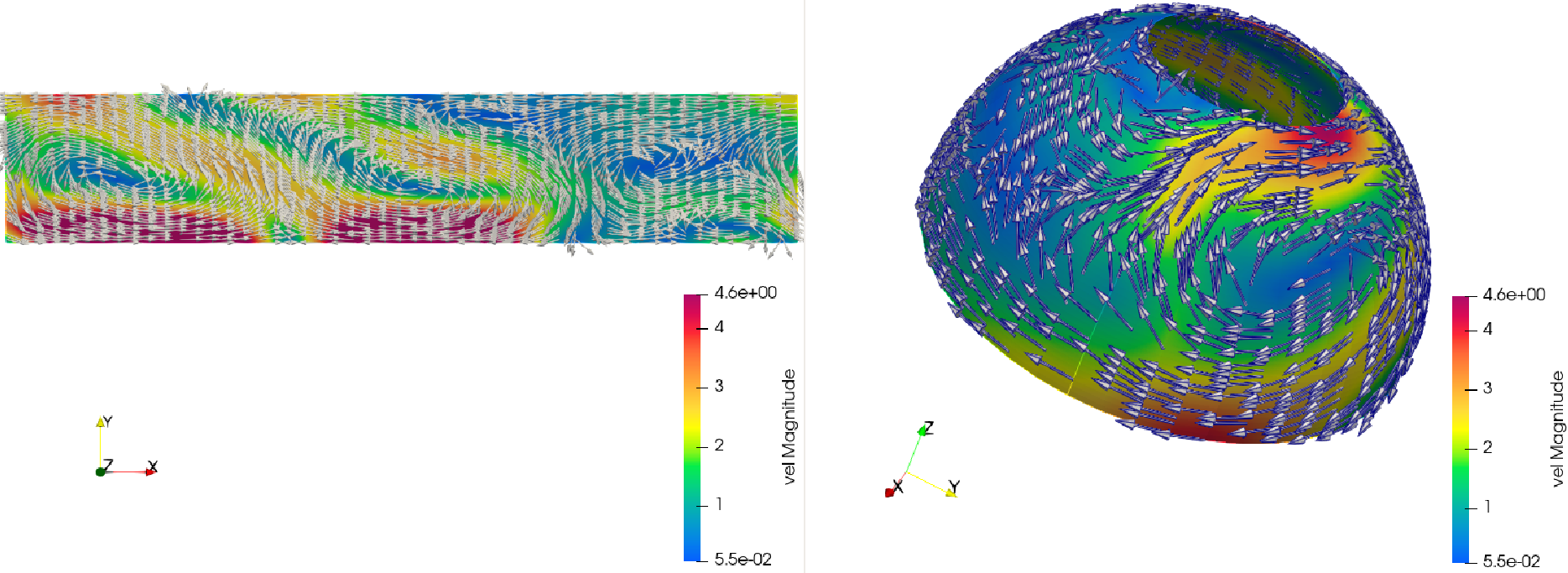}
\includegraphics[width=1\linewidth]{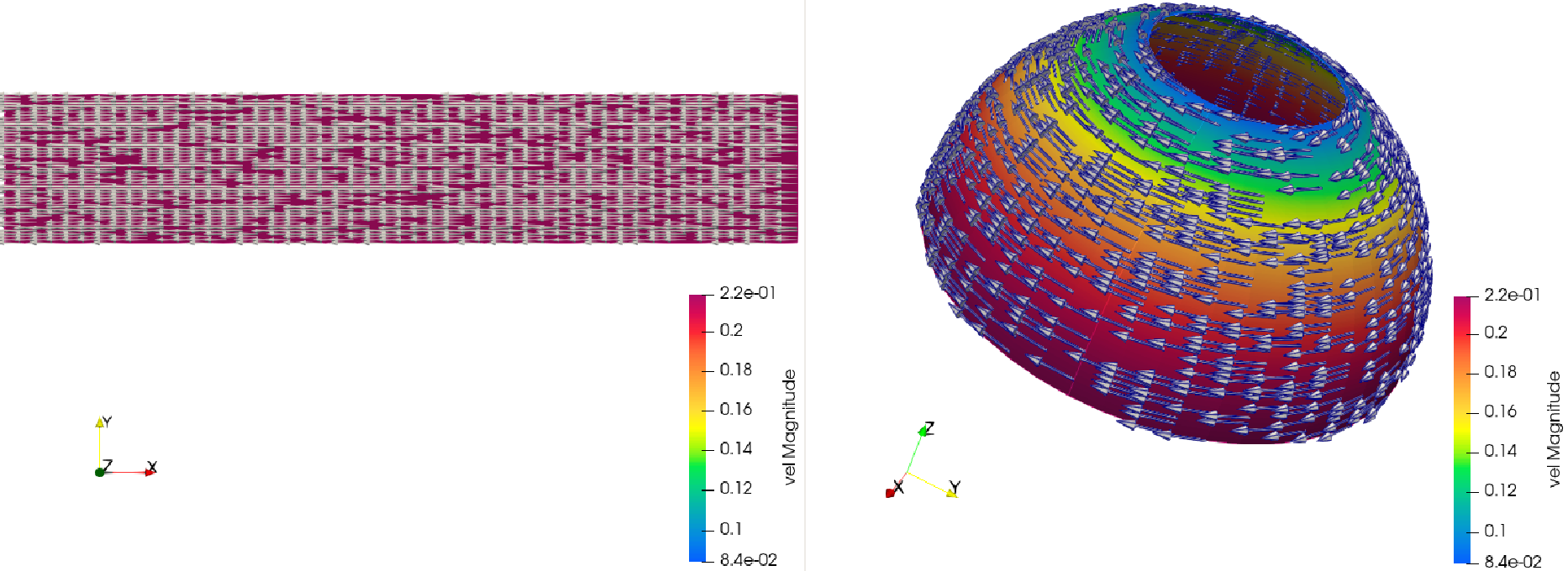}
\caption{The initial flow profile $u = 10\left(\cos(2\theta)\sin(2\phi),\, \sin(\theta\phi)\right)$ converges to a steady state given by a Killing field (rotational flow). The top and bottom figures correspond to the initial flow profile at $t = 0$ and the converged velocity profile at $t = 1$, respectively. In each case, the left figure shows the flow in parameter space, while the right figure shows the flow on the surface.}
\label{fig_ns_halfsphere_killing}
\end{figure}

Figure \ref{fig_ns_halfsphere_killing} shows an initial flow profile $u = 10\left(\cos(2\theta)\sin(2\phi),\, \sin(\theta\phi)\right)$ converges to a steady state given by a Killing field (rotational flow), using the intrinsic FEM. Numerically, we found that the initial flow is not required to be divergence-free, although a divergence-free condition is required in the \cite{shao2025navier}, and a designed Navier boundary condition is used for the proof.

\begin{figure}[h!]
\centering   \includegraphics[width=0.32\linewidth]{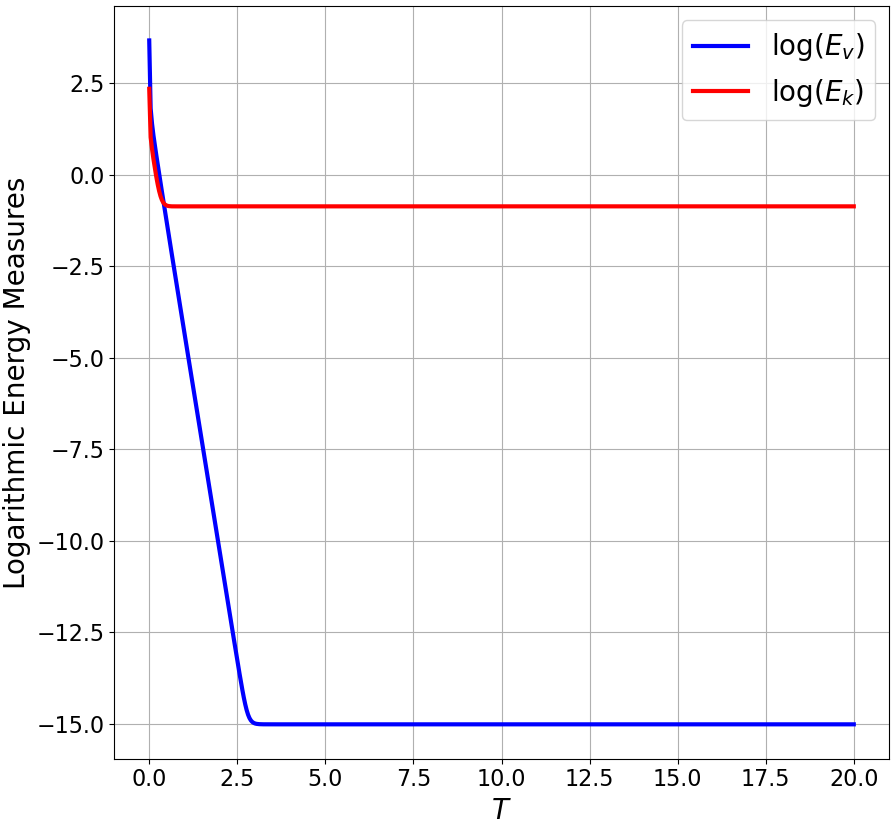}
\includegraphics[width=0.32\linewidth]{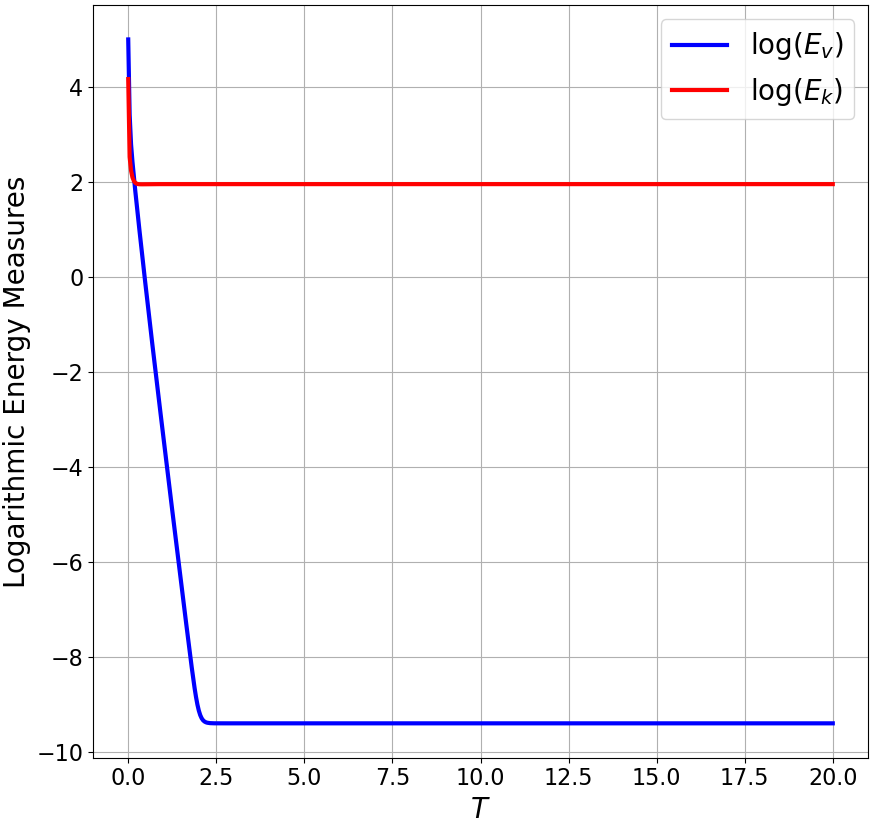}
\includegraphics[width=0.32\linewidth]{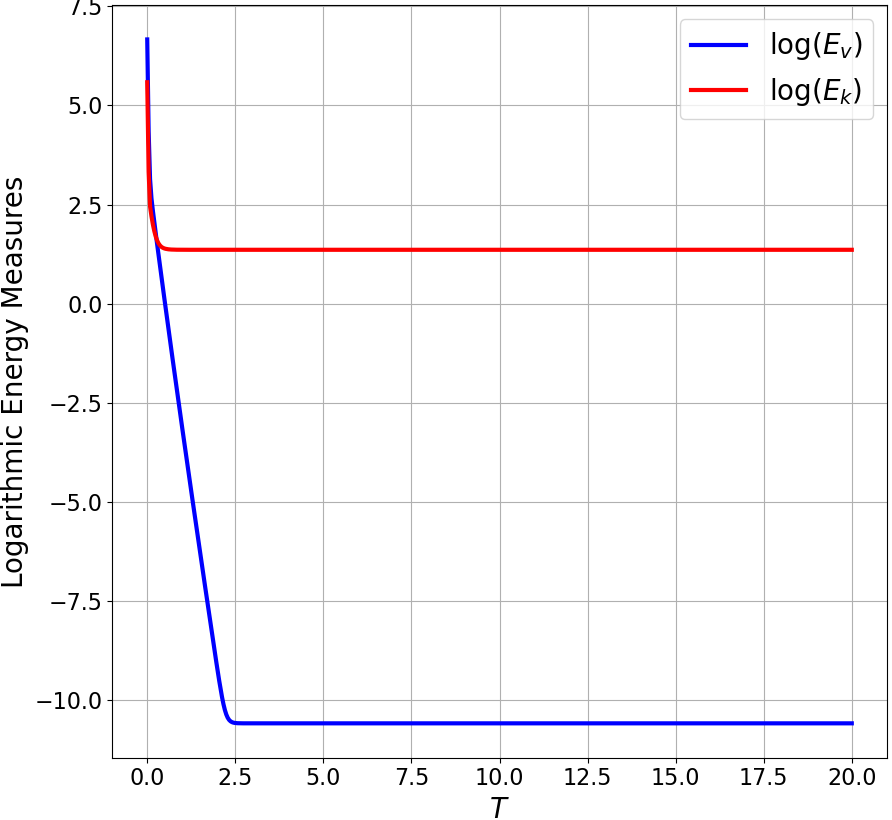}
\caption{The energy evolution of the flow on a hemisphere (with the North Pole excluded) for three different initial conditions. From left to right: $10\left(\cos(2\theta)\sin(2\phi),\, \sin(\theta\phi)\right)$, $100\left(\cos(2\theta)\sin(2\phi),\, -\sin(2\theta)\cos(2\phi)\right)$, and $\left(0,\,100\theta\right)$.}
\label{fig_halfsphere_longrun}
\end{figure}

To test the stability, we run the simulation for a long time up to $t=20$, and plot the viscous energy
\begin{equation}\label{viscous_energy}
E_\text{v}(t)=2\mu\int_\Omega\sqrt{|g|}\left(\epsilon, \epsilon\right)_g d\Omega,
\end{equation}
and the Kinetic energy
\begin{equation}
E_\text{k}(t) = \frac{\rho}{2} \int_\Omega \sqrt{|g|}(u, u)_g\, d\Omega,
\end{equation}
for three different initial velocity profiles. We found all of them converge a steady-state solution corresponding to a Killing vector field, as shown in Figure \ref{fig_halfsphere_longrun}. It can be seen that the numerical steady state is very stable, and there is no numerical damping or instability,  which is consistent with the analysis in (\ref{proposition_eq}) and (\ref{steady_state_estimate2}).

\begin{figure}[h!]
\centering   \includegraphics[width=0.35\linewidth]{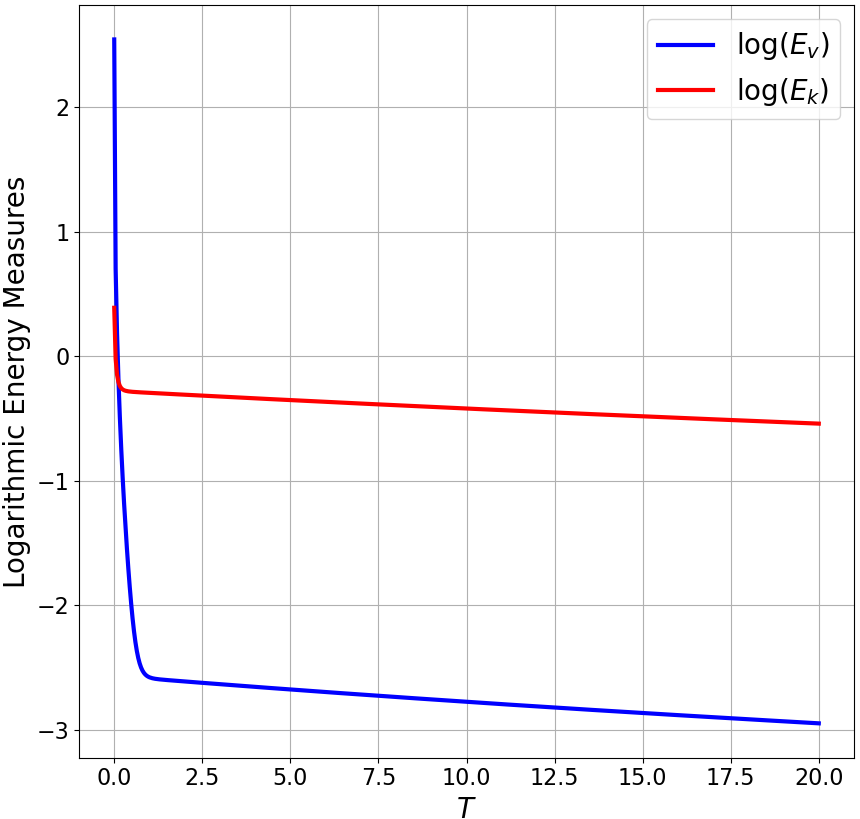}
\quad
\includegraphics[width=0.35\linewidth]{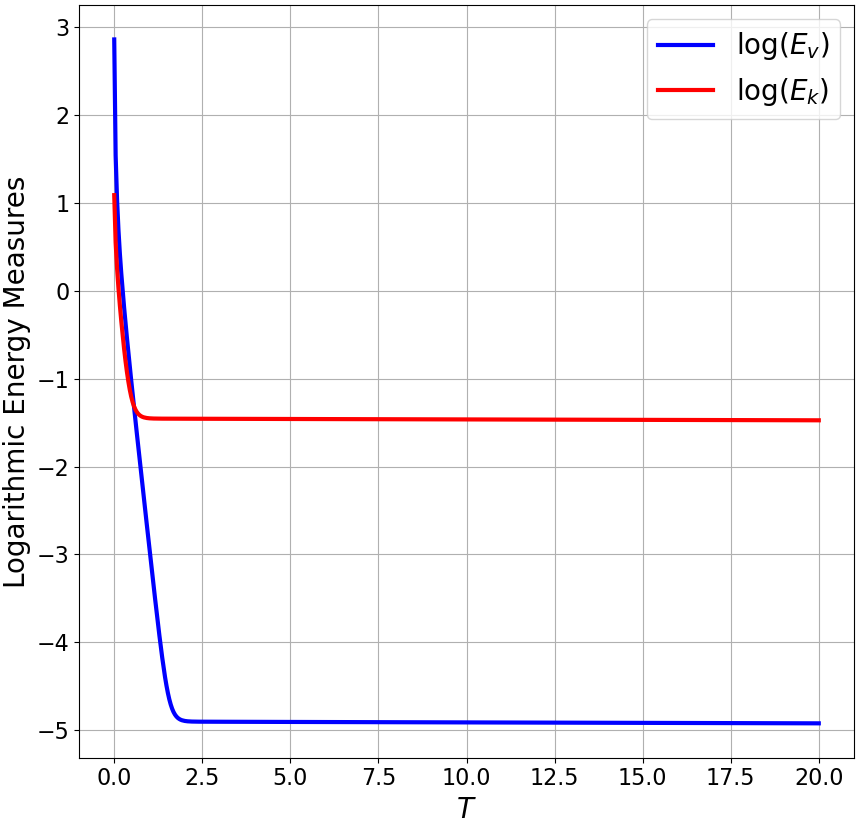}
\caption{The energy evolution of flow on half of a sphere using surface FEM on a coarse (left) mesh and a fine (right) mesh. The sides of domain $[0,2\pi]\times[\pi/8, \pi/2]$ are divided by $60\times40$ (coarse) $80\times60$ (fine) segments.}
\label{fig_halfsphere_longrun_surfacefem}
\end{figure}

To compare with surface FEM, we map the initial velocity $u(\theta, \phi) = 10\left(\cos(2\theta)\sin(2\phi),\, \sin(\theta\phi)\right)$ to the hemisphere based on (\ref{sphere_parameterised}). The mesh is also mapped to the sphere from the parameter space, so that the same mesh size is adopted for comparison. An immediate challenge is to implement the slip boundary condition: $u^i\tau_i =0$. At $\phi=\pi/2$, $u^i\tau_i = u_z = 0$ which is easy. However, $u^i\tau_i =0$ at $\phi=\pi/8$ is not trivial, we introduce a Lagrange multiplier to enforce this condition. First, the surface FEM is much slower as expected since it solve for a three dimensional velocity vector. In addition, the viscous energy has not decreased to zero after a long run, and the ``steady-state" solution is not stable, as shown in the left figure in Figure \ref{fig_halfsphere_longrun_surfacefem}. There are numerical damping, which keeps reducing both the kinetic and viscous energy. 

The reason of this observed problems should not due the boundary conditions, because the same problems are observed in the case of a whole sphere as seen in Figure \ref{fig_viscousenergy_longrun}. We think the numerical damping and the non-zero viscous could be due to the surface mesh itself - an isoparametric mesh is expected to be better which is unfortunately not available in FreeFEM++ for a comparison. Therefore, we expect the results can be improved by a mesh refinement, which is shown in the right figure in Figure \ref{fig_halfsphere_longrun_surfacefem} after the mesh size halved, from which we can see the a steady state has been achieved. However, the viscous energy is still far from zero. In addition, the numerical damping is higher compared with the first figure in Figure \ref{fig_halfsphere_longrun} - one magnitude difference in the kinetic energy.

\subsubsection{Simulations on the whole sphere}

\begin{figure}[h!]
\centering   
\includegraphics[width=1\linewidth]{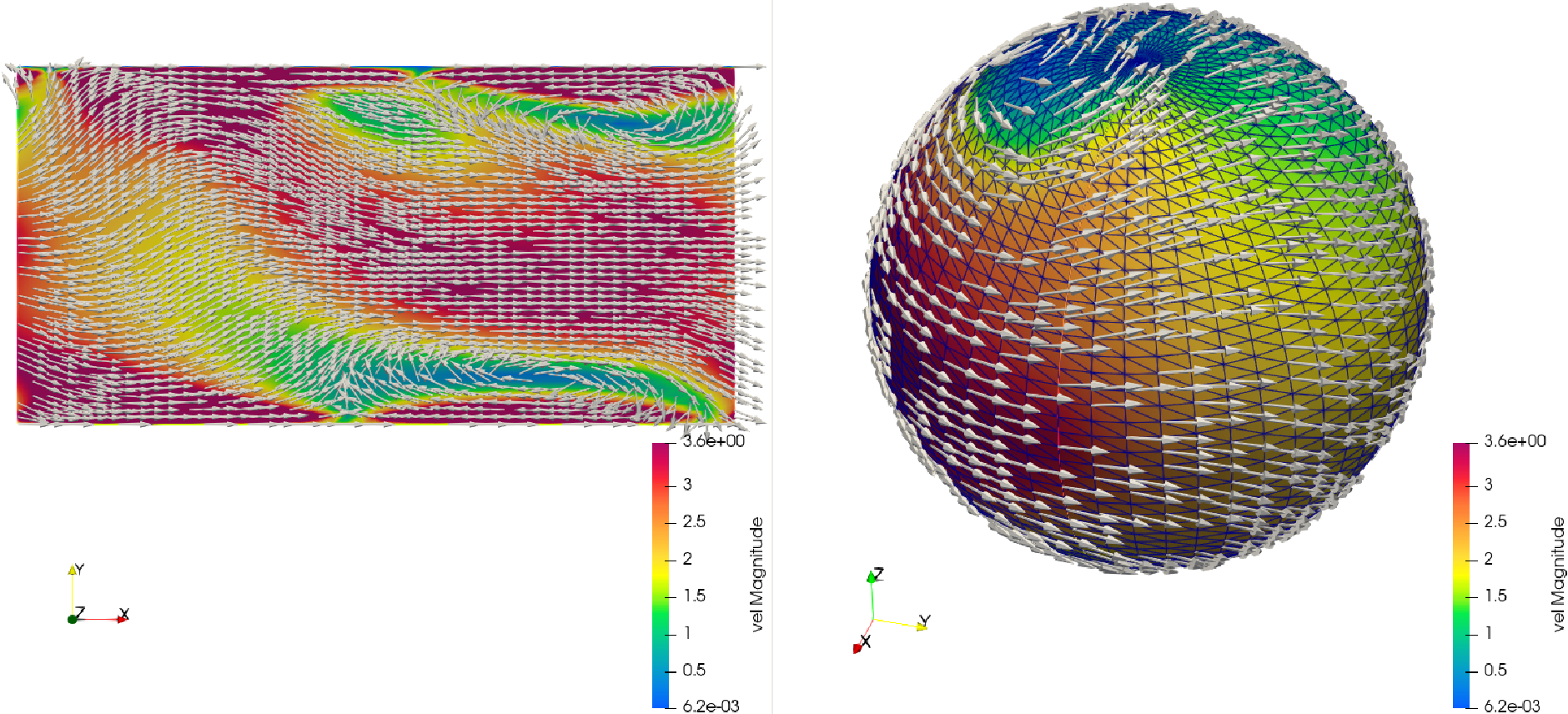}
\includegraphics[width=1\linewidth]{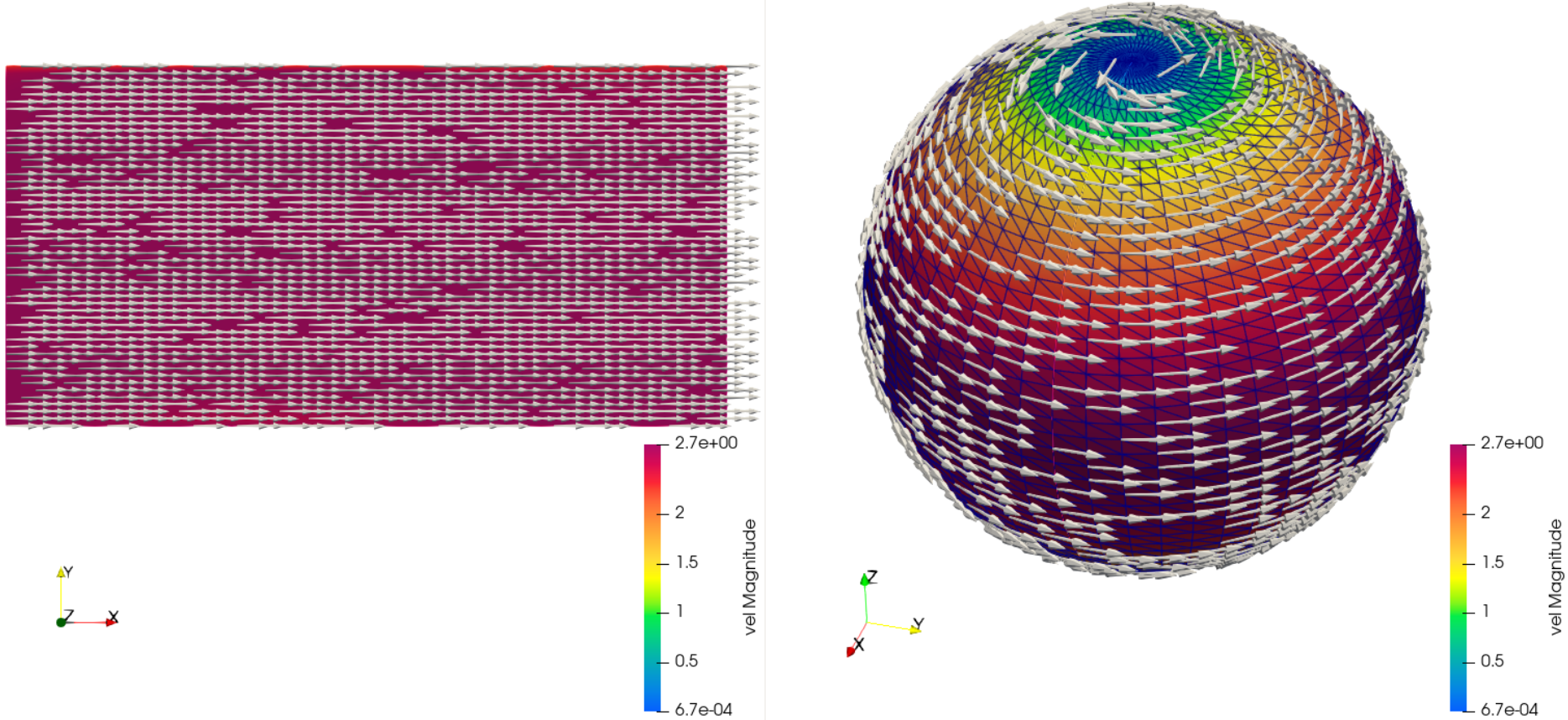}
\caption{Initial flow profile $u = (\theta\sin\phi , 2\sin (\theta\phi))$ converges to the steady state given by a Killing field rotational around the $z$ axis. The top and bottom figures correspond to $t = 0$ and $t = 1$.}
\label{fig_ns_sphere_killing1}
\end{figure}

\begin{figure}[h!]
\centering   \includegraphics[width=0.9\linewidth]{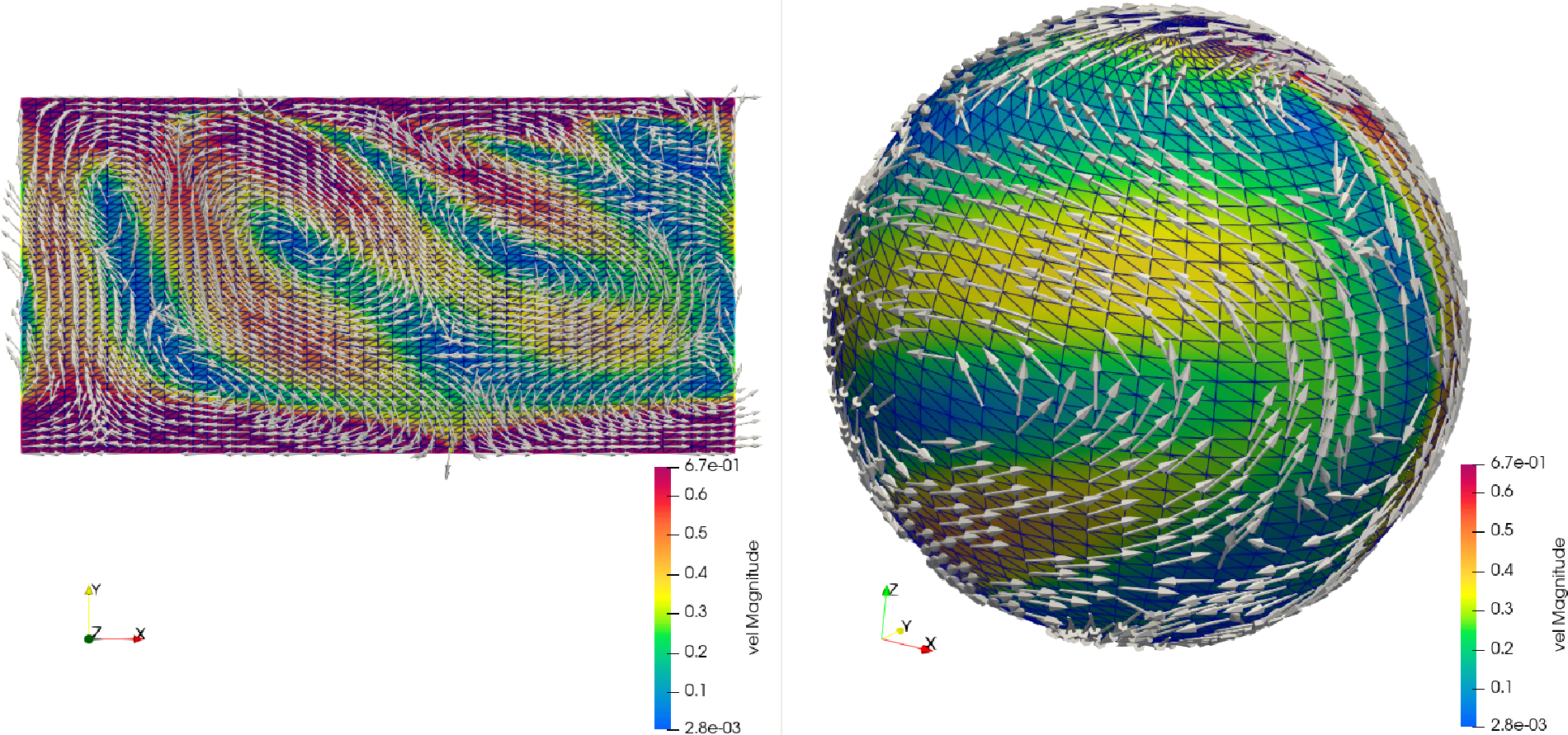}
\includegraphics[width=0.9\linewidth]{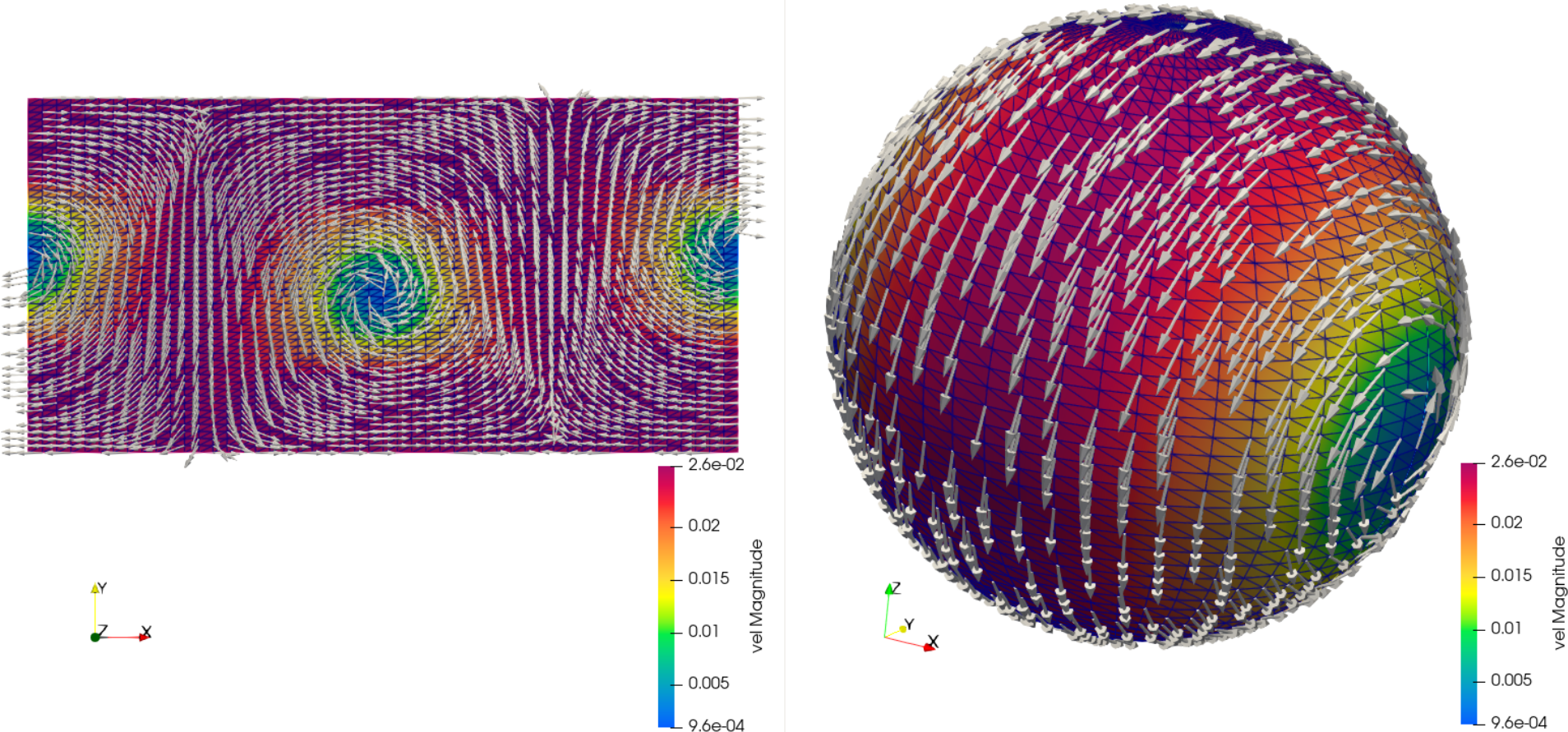}
\caption{Initial flow profile $u = (\sin\theta\cos\phi,\, \cos(\theta\phi))$ converges to the steady state given by a Killing field rotational around the $x$ axis. The top and bottom figures correspond to $t = 0$ and $t = 1$.}
\label{fig_ns_sphere_killing2}
\end{figure}

\begin{figure}[h!]
\centering   \includegraphics[width=1\linewidth]{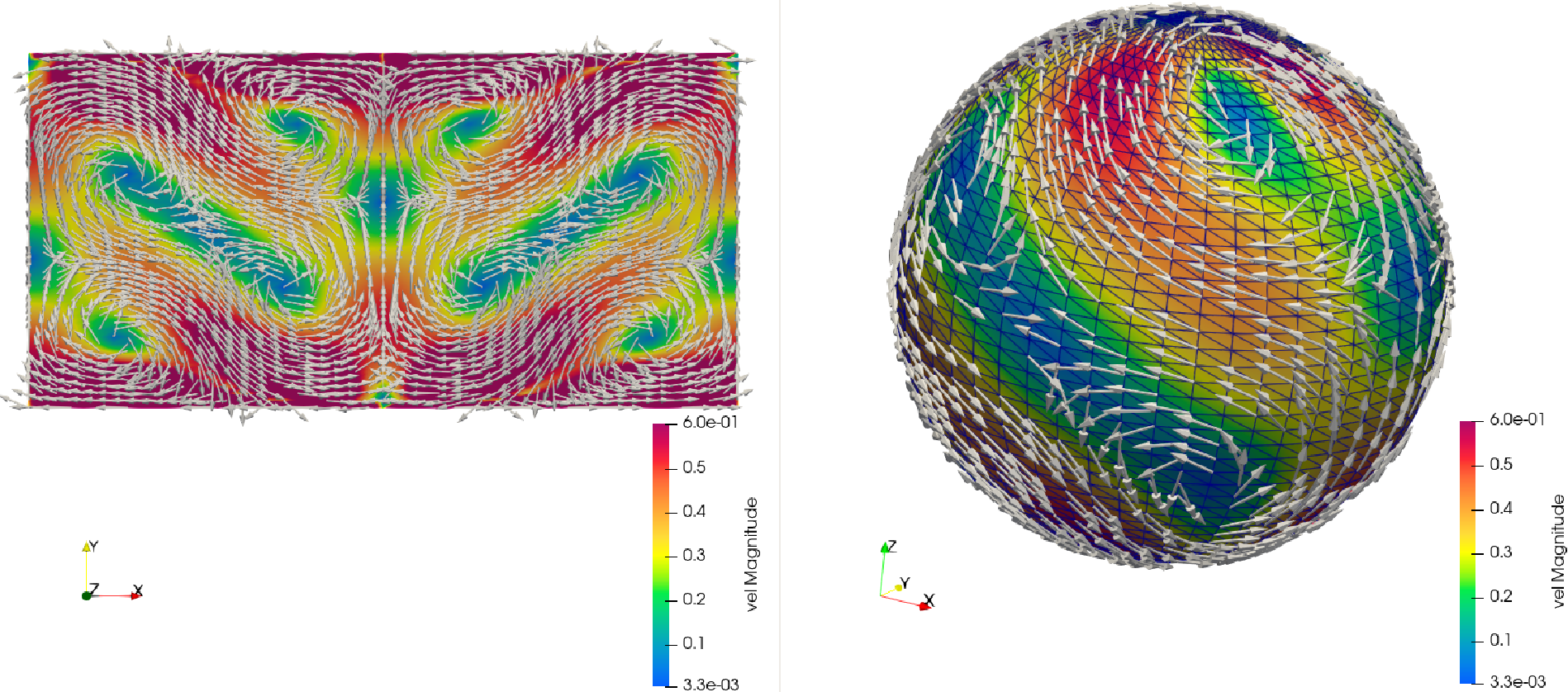}
\includegraphics[width=1\linewidth]{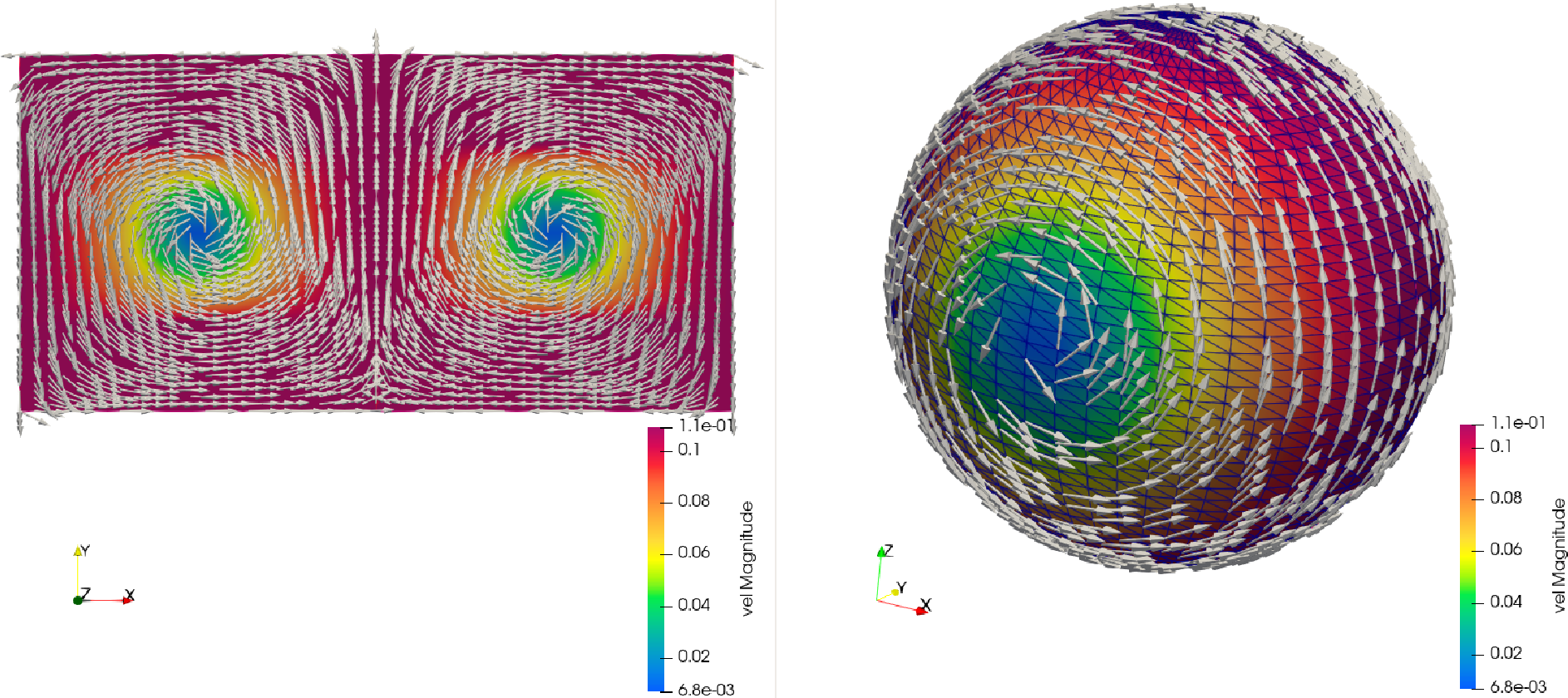}
\caption{Initial flow profile $u = (\sin\theta\cos\phi,\, \cos(2\theta)\sin(2\phi))$ converges to the steady state given by a Killing field rotational around the $y$ axis. The top and bottom figures correspond to $t = 0$ and $t = 1$.}
\label{fig_ns_sphere_killing3}
\end{figure}

Although $\phi = 0$ and $\phi = \pi$ correspond to singular values in the inverse metric tensor, the Gauss quadrature points always lie strictly inside the triangles, thereby conveniently avoiding the singularities. Three different initial velocity profiles are selected, and they converge to different steady-state solutions corresponding to three different rotational Killing vector fields, as shown from Figure \ref{fig_ns_sphere_killing1} to \ref{fig_ns_sphere_killing3}. 

Periodic boundary conditions are imposed at $\theta = 0$ and $\theta = 2\pi$. However, since the north pole (a single point on the sphere) corresponds to the line $\phi = 0$ in parameter space, all points on this line must share the same value. The ideal method is to eliminate all the extra degrees of freedom in the final linear system. However, most software packages are not designed to facilitate this implementation. Practical implementations include: do-nothing condition (leave all degrees of freedom free), Lagrange multiplier to constrain the degrees of freedom, slip boundary condition, homogeneous Dirichlet boundary condition (which is reasonable to obtain the rotation around $z$ axis), or penalty method: penalising the tangential derivatives so that the velocity is constant along the boundary. It is generally not obvious what boundary conditions is better. We will comment below on different boundaries based on our numerical results.

\begin{figure}[h!]
\centering   \includegraphics[width=0.5\linewidth]{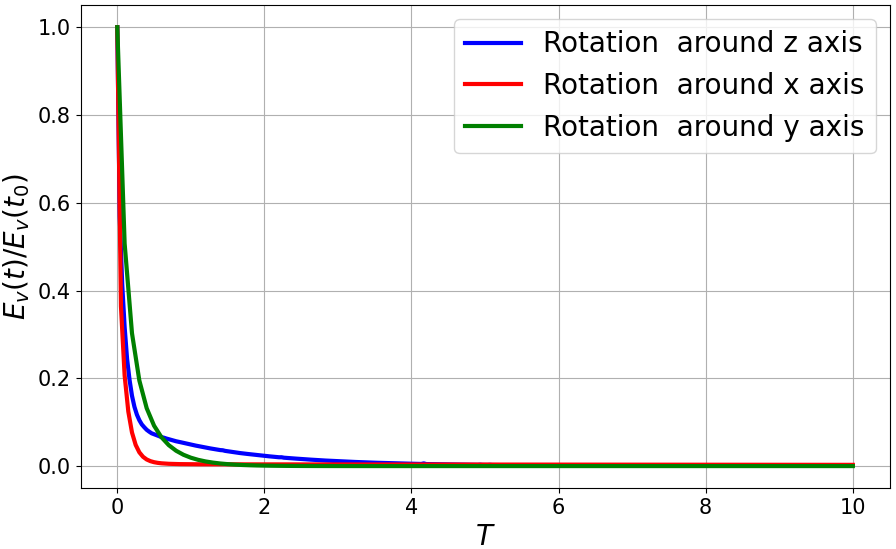}
\caption{The evolution of the viscous energy on the sphere, corresponding to three different initial conditions, each of which converges to one of the three Killing vector fields on the sphere.}
\label{fig_viscousenergy}
\end{figure}

Visually, the viscous energy quickly decrease to zero for all three rotations as shown in Figure \ref{fig_viscousenergy}. However, the following convergence study reveals the quantitative difference.

\subsubsection*{\bf Convergence study of the Killing field}
\begin{figure}[h!]
\centering   \includegraphics[width=0.32\linewidth]{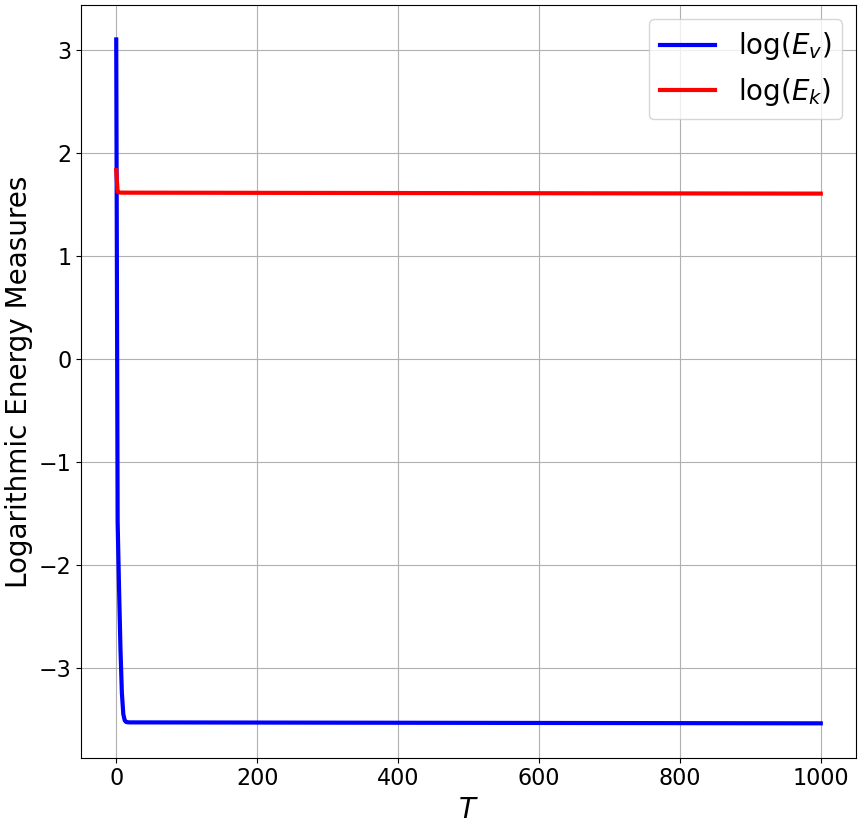}
\includegraphics[width=0.33\linewidth]{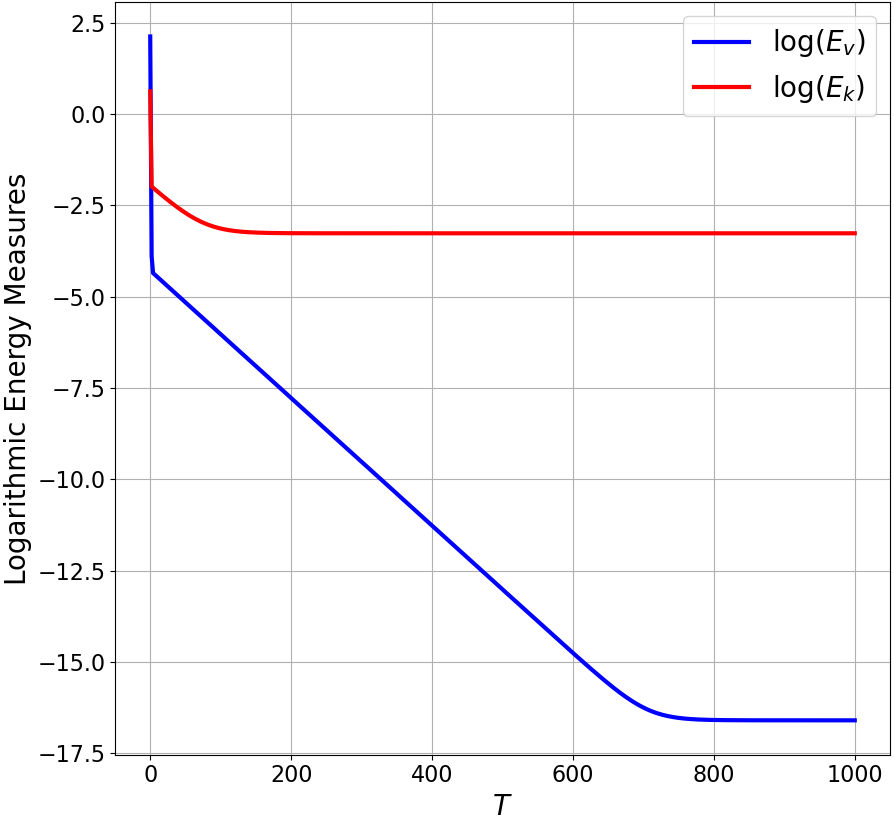}
\includegraphics[width=0.32\linewidth]{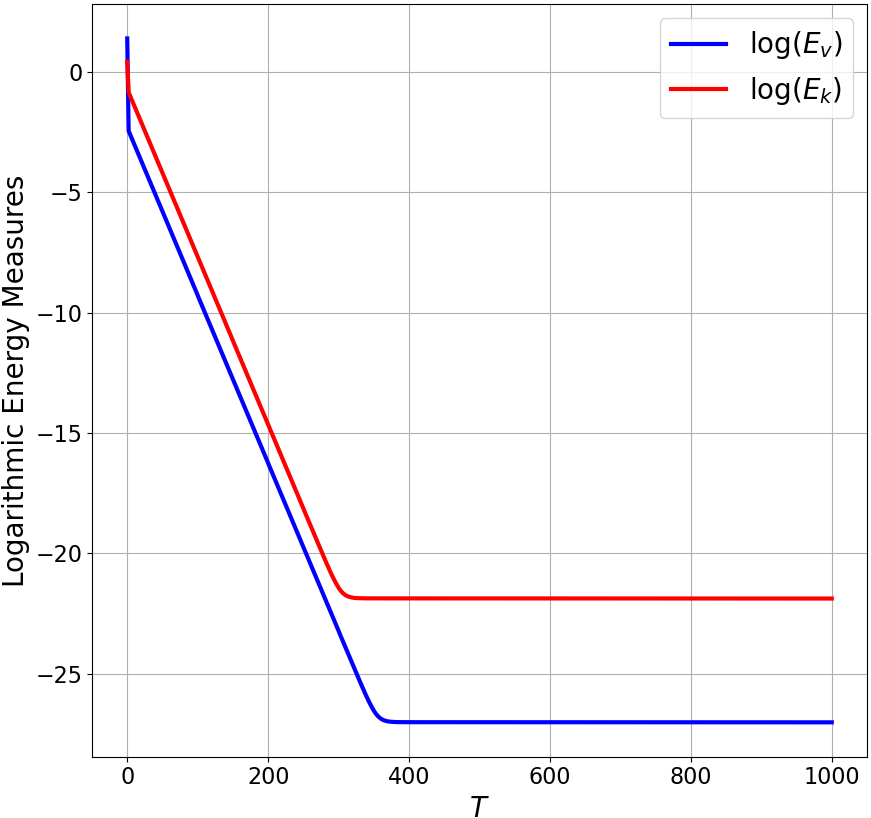}
\caption{The energy evolution for three different simulations whose steady-state solutions correspond to three Killing fields. From left to right: rotation around the $z$-, $x$-, and $y$-axes, corresponding to Figures~\ref{fig_ns_sphere_killing1}, \ref{fig_ns_sphere_killing2}, and \ref{fig_ns_sphere_killing3}, respectively.}
\label{fig_viscousenergy_longrun}
\end{figure}

We investigate whether the Killing field remains stable under long-term numerical evolution. Using the same initial velocity profiles, we run the simulations up to $t = 10^3$. The viscous
and the kinetic energy are shown in Figure~\ref{fig_viscousenergy_longrun}, from which it can be seen that the results of the $z$-rotation (left panel of Figure~\ref{fig_viscousenergy_longrun}) and the $y$-rotation (right panel of Figure~\ref{fig_viscousenergy_longrun})  are not satisfactory: The $z$-rotation is highly stable; however, the viscous energy remains at a magnitude of $10^{-3}$. The velocity of the $y$-rotation gradually decays to zero over time.

As mentioned above, it is generally not obvious what the appropriate boundary conditions should be at $\phi = 0$ and $\phi = \pi$, corresponding to the north and south poles. However, for rotational flow about the $z$-axis, we may simply impose homogeneous boundary conditions in the above tests. For the other two cases, we adopt the do-nothing (zero-normal-stress) boundary condition, which allows the fluid to flow in and out. The numerical results suggest that this choice is acceptable, at least for the $x$-rotation. However, it appears that a method is needed to enforce the consistency of the velocity at $\phi = 0$ and $\phi = \pi$. Imposing a penalty condition to force $\partial_su^i = 0$ is a simple approach ($s$ is the tangential direction at the boundary). 

\begin{figure}[h!]
\centering   \includegraphics[width=0.45\linewidth]{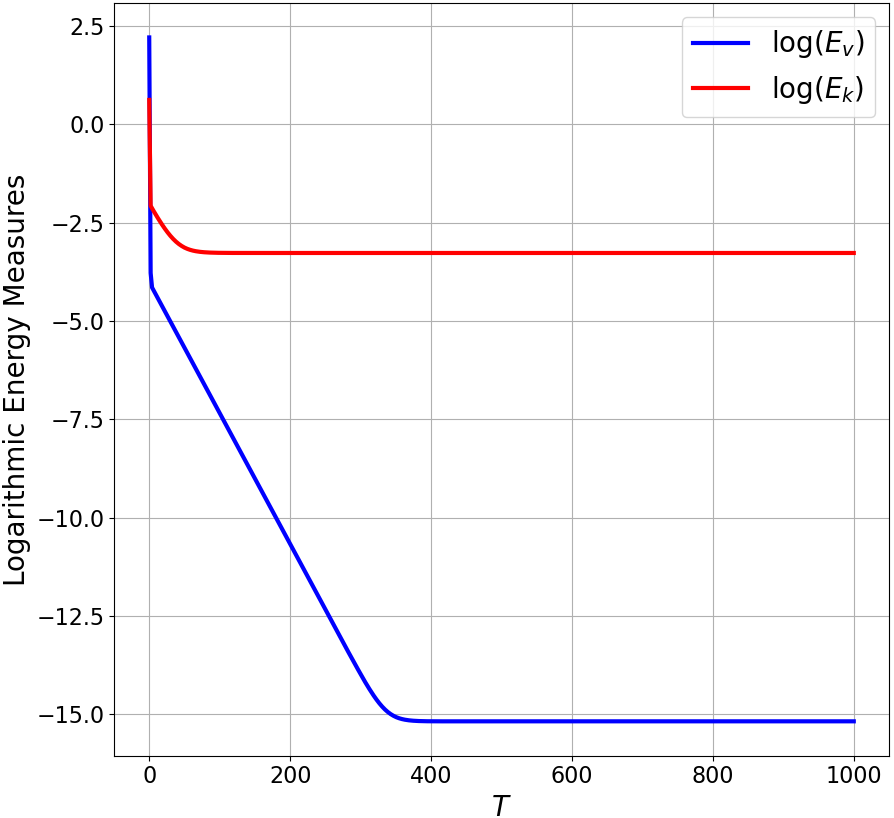}
\includegraphics[width=0.44\linewidth]{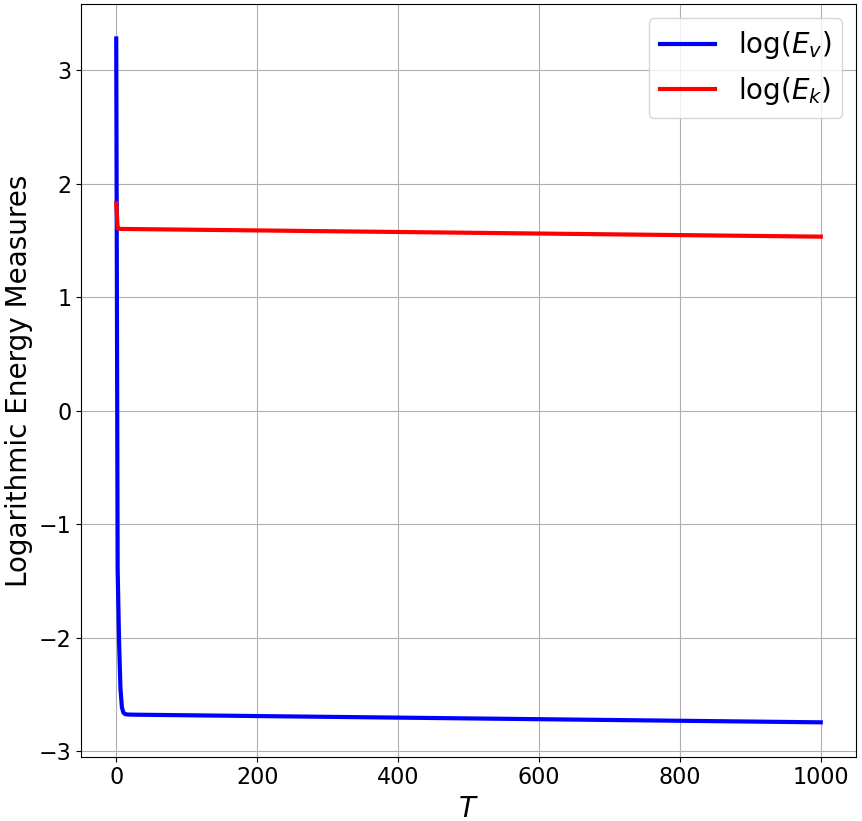}
\caption{$x-$rotation (left) and $z-$rotation (right) using an unstructured mesh.}
\label{fig_viscousenergy_longrun_unstructured}
\end{figure}

To improve the results for the $z$-rotation and the $y$-rotation, we conducted several additional tests. For the latter, we tried a finer mesh, increasing the magnitude of the initial velocity profile, and applying different penalty parameters to enforce consistency of the velocity at $\phi = 0$ and $\phi = \pi$. However, all these attempts led to a trivial zero steady state. We also ran the simulations using an unstructured mesh. For the $x$-rotation, the results are very similar to those obtained with the structured mesh. However, the $z$-rotation becomes slightly worse compared with the structured mesh previously - the viscous energy stays even larger, as shown in Figure~\ref{fig_viscousenergy_longrun_unstructured}.

\subsubsection*{\bf The related eigenvalue problem}
By definition (\ref{killing_equi}), the Killing vector fields can also be found by solving the eigenvalue problem
\[
\left(\Delta_H u\right)^i = \lambda u^i,
\]
or FEM weak form
\[
4\int_\Omega\sqrt{|g|} \,\epsilon^{ij}(u) \epsilon_{ij}(v) d\Omega = \lambda \int_\Omega\sqrt{|g|} \, u^i v_i d\Omega,
\]
corresponding to the eigenvalue $\lambda=0$.

This can be easily implemented in FreeFEM++ using EigenValue() as follows:
\begin{lstlisting}
//Defin FE space Rh and weak forms

matrix A= Hodge(Rh,Rh,solver="SPARSESOLVER",factorize=0); 
matrix B= bf(Rh,Rh,solver=CG,eps=1e-20); 

real sigma = 1.e-12;
int nev=10;  // number of computed eigen valeu close to sigma

real[int] ev(nev); // to store nev eigein value
Rh[int] [eu1,eu2](nev);   // to store nev eigen vector

int k=EigenValue(A,B,sym=true,sigma=sigma,value=ev,vector=eu1,tol=1e-16,maxit=0,ncv=30);
\end{lstlisting}

\begin{figure}[h!]
\centering   \includegraphics[width=0.5\linewidth]{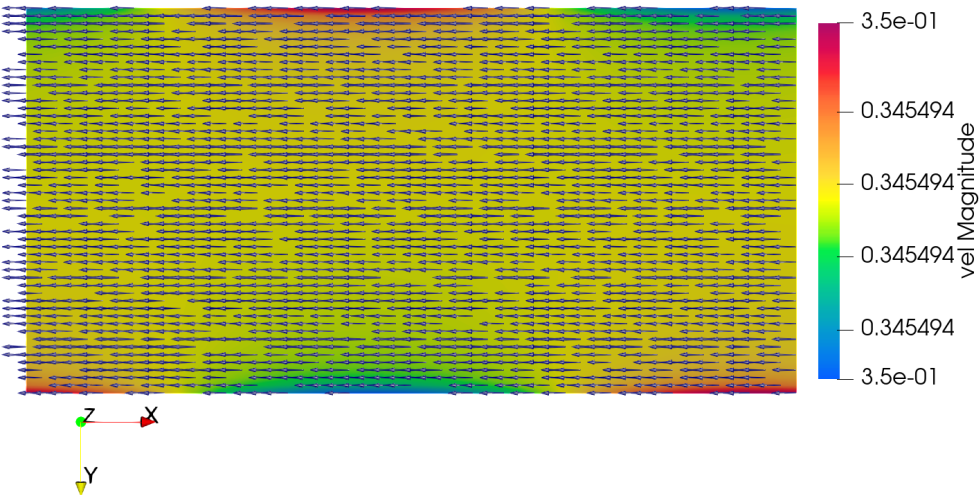}
\includegraphics[width=0.4\linewidth]{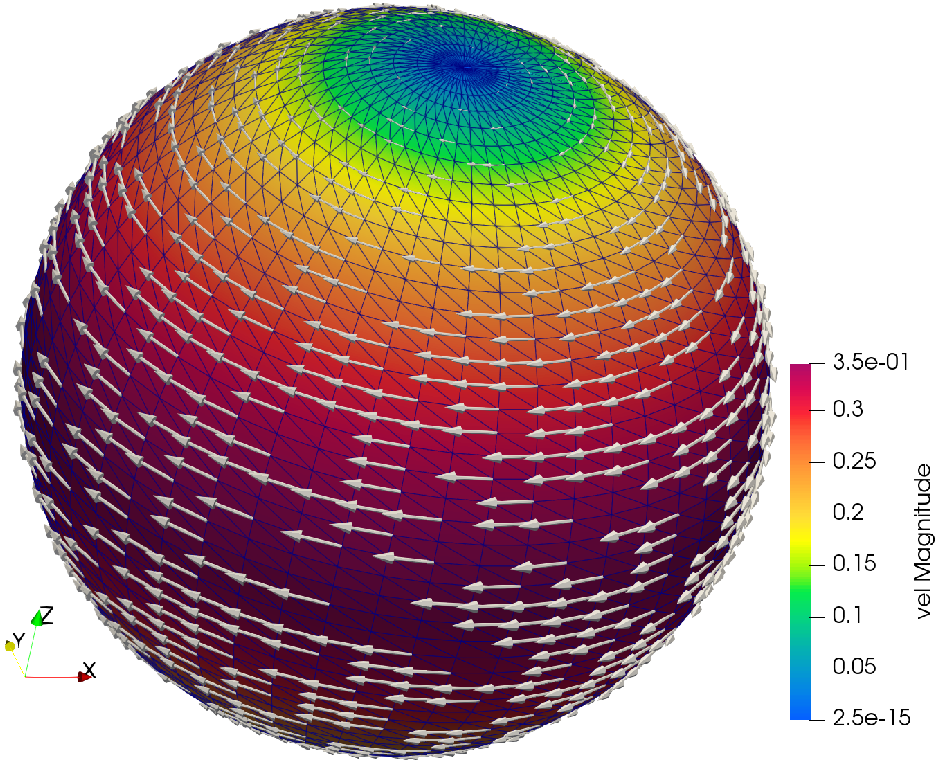}
\includegraphics[width=0.5\linewidth]{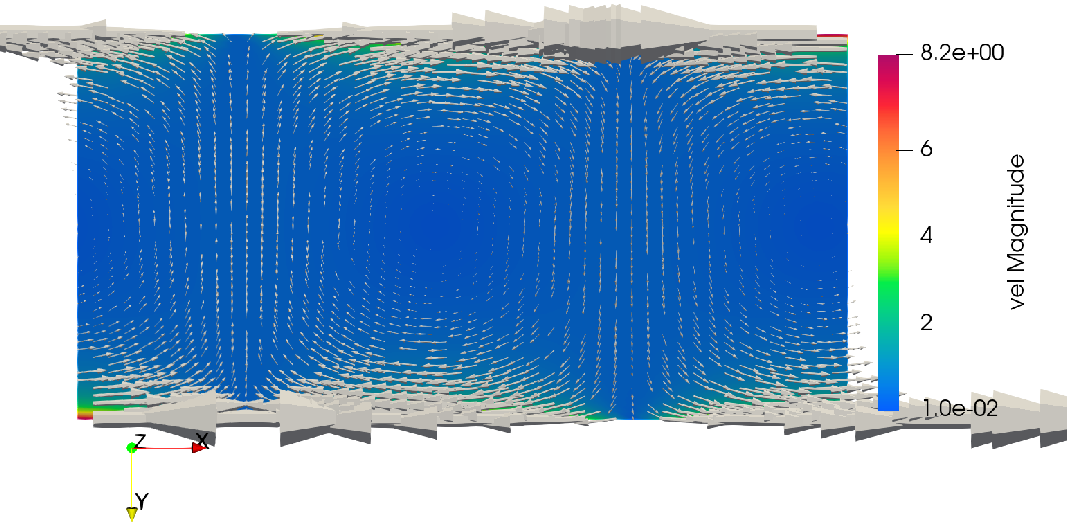}
\includegraphics[width=0.4\linewidth]{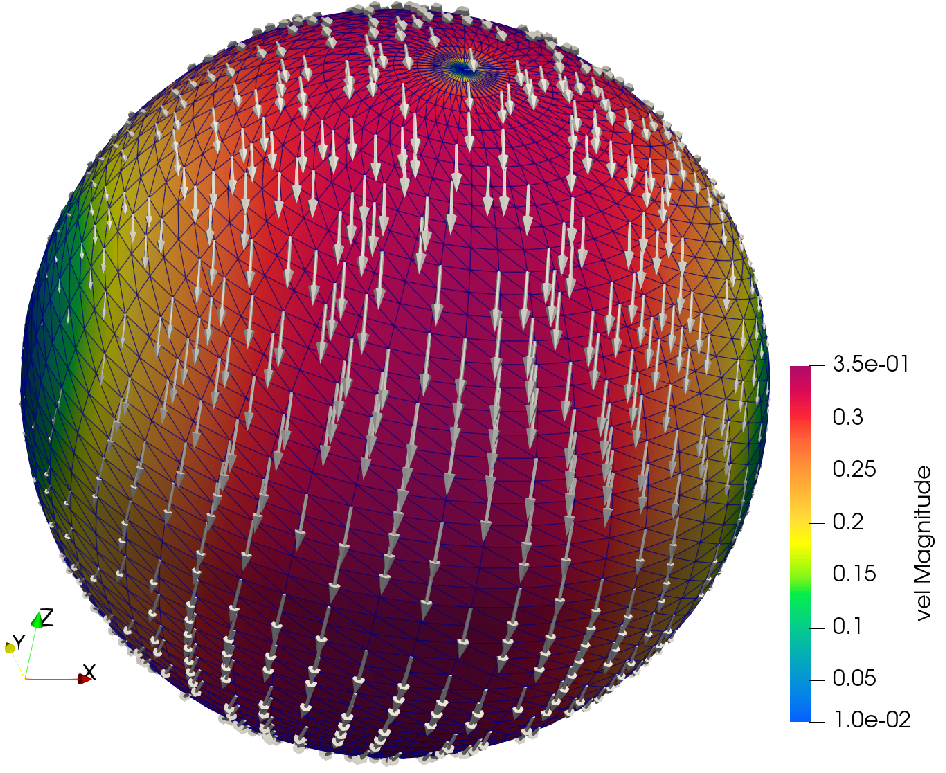}
\caption{The first and second eigenvector fields of the Hodge operator on a sphere.}
\label{fig_eigenvectors_sphere12}
\end{figure}

\begin{figure}[h!]
\centering   \includegraphics[width=0.5\linewidth]{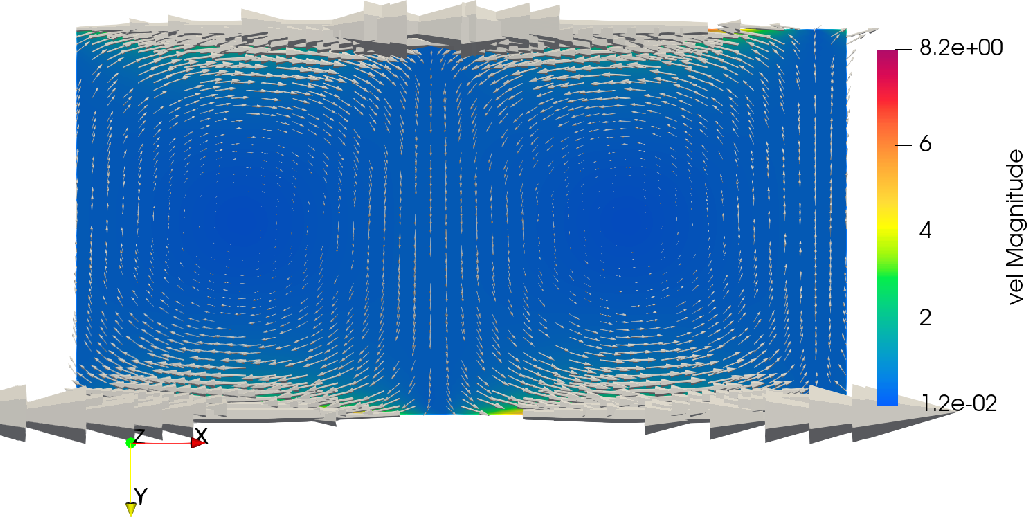}
\includegraphics[width=0.4\linewidth]{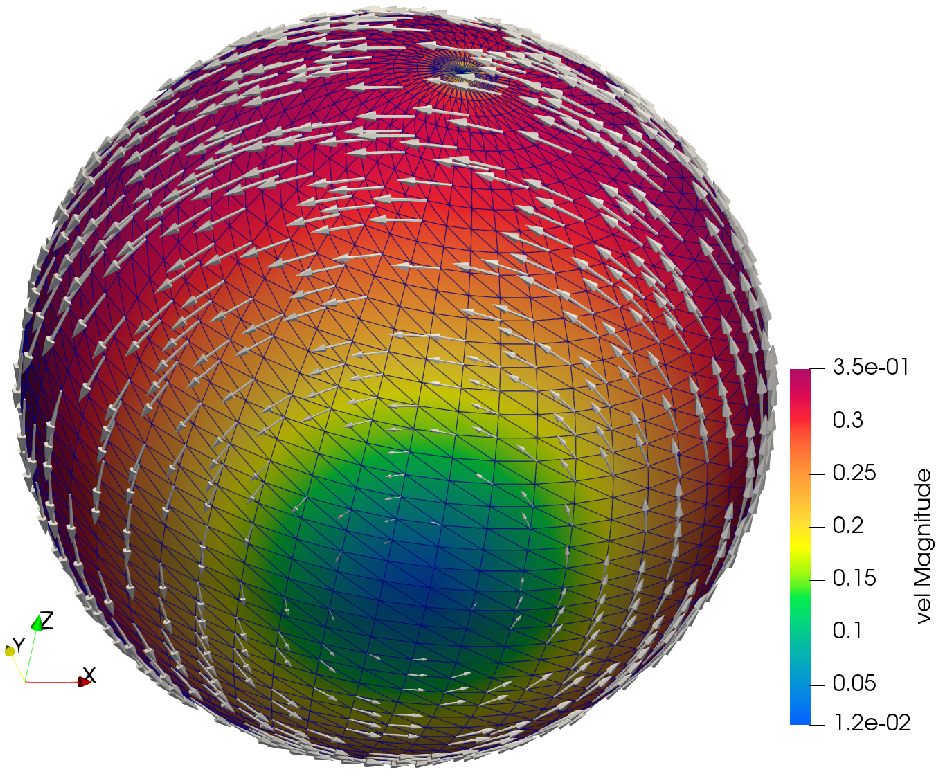}
\includegraphics[width=0.5\linewidth]{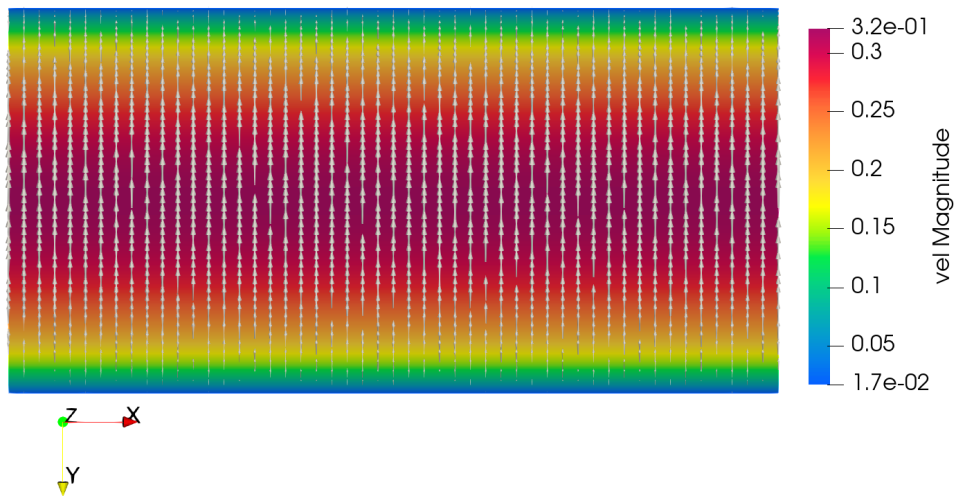}
\includegraphics[width=0.4\linewidth]{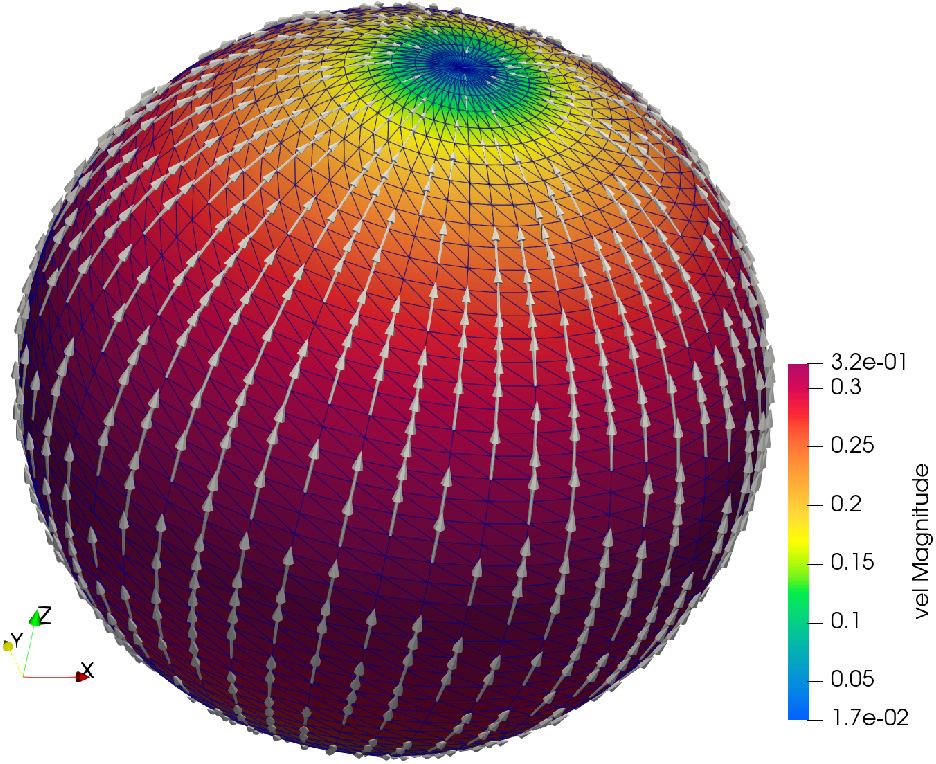}
\caption{The third and fourth eigenvector fields of the Hodge operator on a sphere.}
\label{fig_eigenvectors_sphere34}
\end{figure}

The first four eigenvector fields are displayed in Figures~\ref{fig_eigenvectors_sphere12} and~\ref{fig_eigenvectors_sphere34}. Visually, the first three correspond to the three Killing vector fields on the sphere. However, there are noticeable errors in the second ($x$-rotation) and the third ($y$-rotation), which can be observed in the parameter space near $\phi=0$ and $\phi=\pi$ - we expect the velocity is same along these two boundaries. This should either because of the inconsistent boundary conditions or the singularities at the pole, or both.

From the eigenvalues and the corresponding viscous energy in Table~\ref{table_eigenvalues}, it can be seen that the second and third solutions are incorrect. We expect the nullspace associated with the zero eigenvalue to have dimension three, corresponding to the three Killing vector fields on the sphere, with viscous energy equal to~0.

\begin{table}[h!]
\centering
\begin{tabular}{c|c|c}
\hline
 NO. of eigenvalues   &  Eigenvalue   &  Viscous Energy\\
\hline
1   &  1.004778008422281e-12 & 6.373011057817148e-24 \\
\hline
2   &  0.0459862220690389 & 7196.606203440489 \\
\hline
3   &  0.04598622206938586 & 7196.606203440624 \\
\hline
4   &  1.058483964387766 & 625.430579346591 \\
\hline
\end{tabular}
\caption{The first four eigenvalues of the Hodge operator on a sphere and the viscous energy of the corresponding eigenvectors.}
\label{table_eigenvalues}
\end{table}

\subsubsection*{\bf Comparison with the surface FEM}
The surface FEM defines all quantities in the ambient Euclidean space. For example, the velocity vector is defined as $u=(u_x,u_y,u_z)$ rather than using the local representation $u=u^i\partial_i$ ($i=1,2$) in terms of the local basis $\partial_i$.

The implementation of the scalar Laplacian is straightforward. In FreeFEM++, for a scalar $u$, if the following FE space is defined on a surface mesh \texttt{ThS},
\begin{lstlisting}
meshS ThS=square3(nx,ny,[torex,torey,torez],removeduplicate=true);
fespace Vh(ThS,P1);
Vh u;
macro GradS(u) [dx(u),dy(u),dz(u)] // EOM
\end{lstlisting}
then, the quantity \texttt{GradS(u)} is already tangential to the surface. Therefore,
\[
\int_M \nabla_\Gamma u \cdot \nabla_\Gamma v \, dM
=
\mathrm{int2d}(\mathrm{ThS})\bigl(\mathrm{GradS}(u)' * \mathrm{GradS}(v)\bigr),
\]
where $\left(\nabla_\Gamma u\right)^i = g^{ij}\partial_j u$ is the surface gradient.

However, the situation is more complicated for the vector Laplacian. For the Hodge operator acting on a surface vector field, we must first project the vector $u=(u_x,u_y,u_z)$ onto the tangent space, by $\mathcal{P}(u)$, where 
\begin{equation}\label{projection_operator_in_ambient}
\mathcal{P} = I - \tau\otimes \tau,
\end{equation}
with $\tau$ being the normal of $\partial M$ and $I$ being the identity tensor in the ambient space $\mathbb{R}^3$. We then take the gradient $\hat{\nabla} \mathcal{P}(u)$ in $\mathbb{R}^3$, and finally project the result back onto the tangent space, giving
\begin{equation}\label{projection_computation}
\mathcal{P}\left(\hat{\nabla} (\mathcal{P}u)\right)\mathcal{P}= \mathcal{P}\left(\hat{\nabla} u\right)\mathcal{P} - \left(u\cdot \tau\right)\left(\mathcal{P}\left(\hat{\nabla}\tau\right)\mathcal{P}\right),
\end{equation}
which can be used for the implementation in FreeFEM++. The right $\mathcal{P}$ restricts derivative to tangential directions, and the left $\mathcal{P}$ remove any normal component produced by differentiation. To avoid computing the gradient of the normal $\hat{\nabla}\tau$ numerically, one can provide a formula. This is simple for a sphere: $\mathcal{P}\hat{\nabla}\tau\mathcal{P} = \mathcal{P}$. 

Using the surface FEM, the first ten eigenvalues and the viscous energy of the corresponding eigenvectors are presented in Table~\ref{table_eigenvalues_surfaceFEM}. This contrasts with the results obtained from the intrinsic formulation in Table~\ref{table_eigenvalues}, where the second and third eigenvalues cannot be resolved due to the singularity in the parameter space. 

\begin{figure}[h!]
\centering
\includegraphics[width=0.45\linewidth]{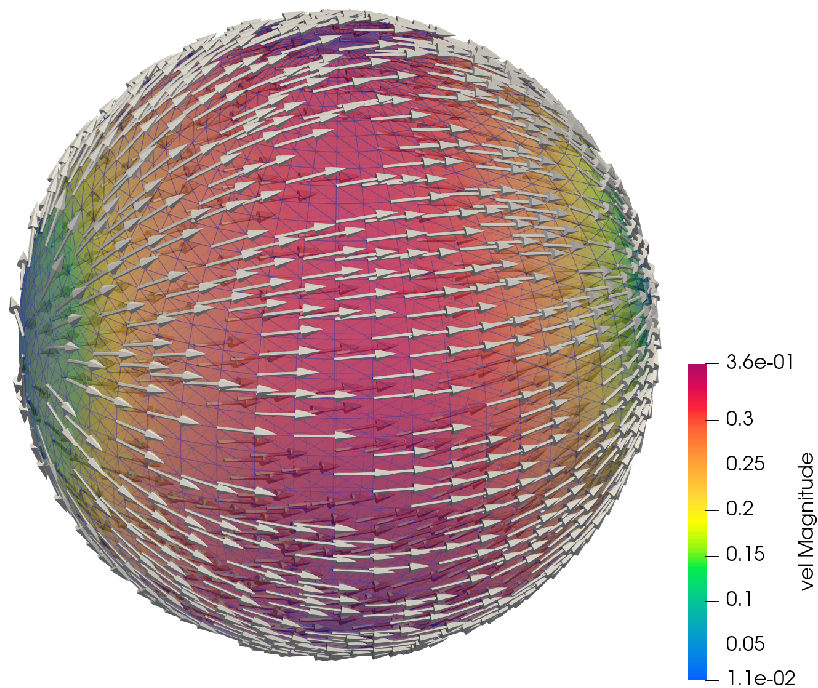}
\includegraphics[width=0.45\linewidth]{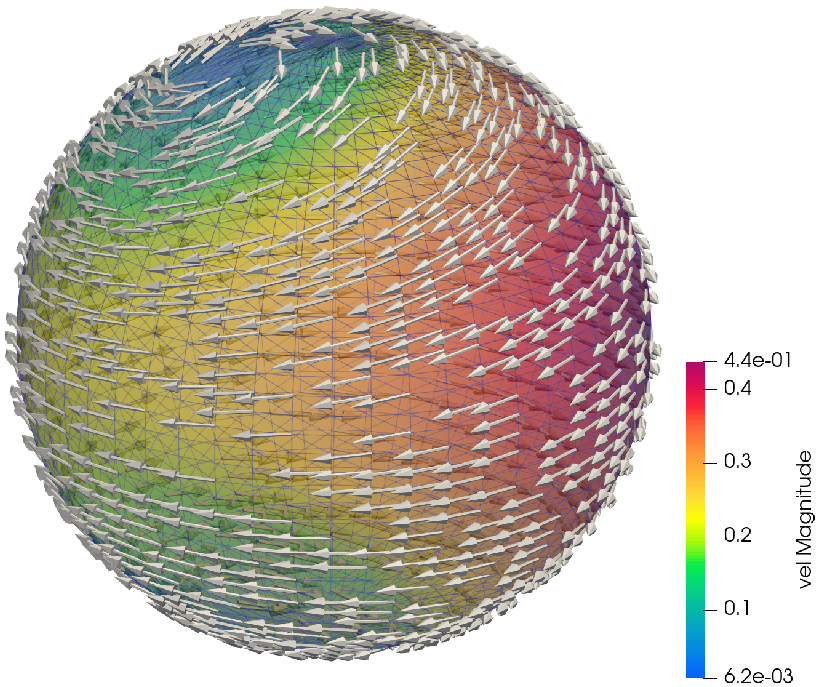}
\caption{The second and sixth modes of the Hodge operator on a sphere, computed using the surface FEM with the same mesh as in the previous intrinsic formulation.}
\label{fig_sphere_modes}
\end{figure}

However, the first six modes are mixed with the Killing vector fields: in this test, the fourth to sixth modes in Table~\ref{table_eigenvalues_surfaceFEM} correspond to the Killing vector fields, while the first three are source–sink modes. The sixth mode is visually not symmetric, as shown in Figure~\ref{fig_sphere_modes}.

Another downside is that the surface FEM is significantly slower because it involves more degrees of freedom. In addition, the magnitude of the Eigenvalue and the viscous energy can only be reduced to $10^{-5}$. This is the results of using a $P2$ mesh, while the magnitude can only be resolved to about $10^{-4}$ when $P1$ elements are used on the same mesh. 

\begin{figure}[h!]
\centering
\includegraphics[width=0.4\linewidth]{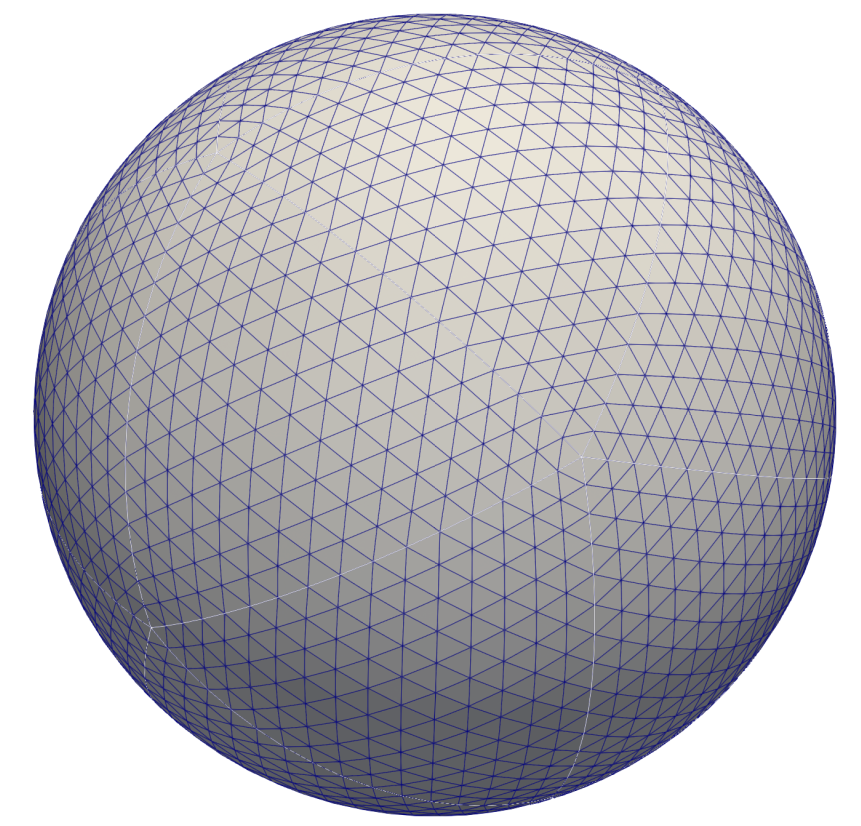}
\includegraphics[width=0.45\linewidth]{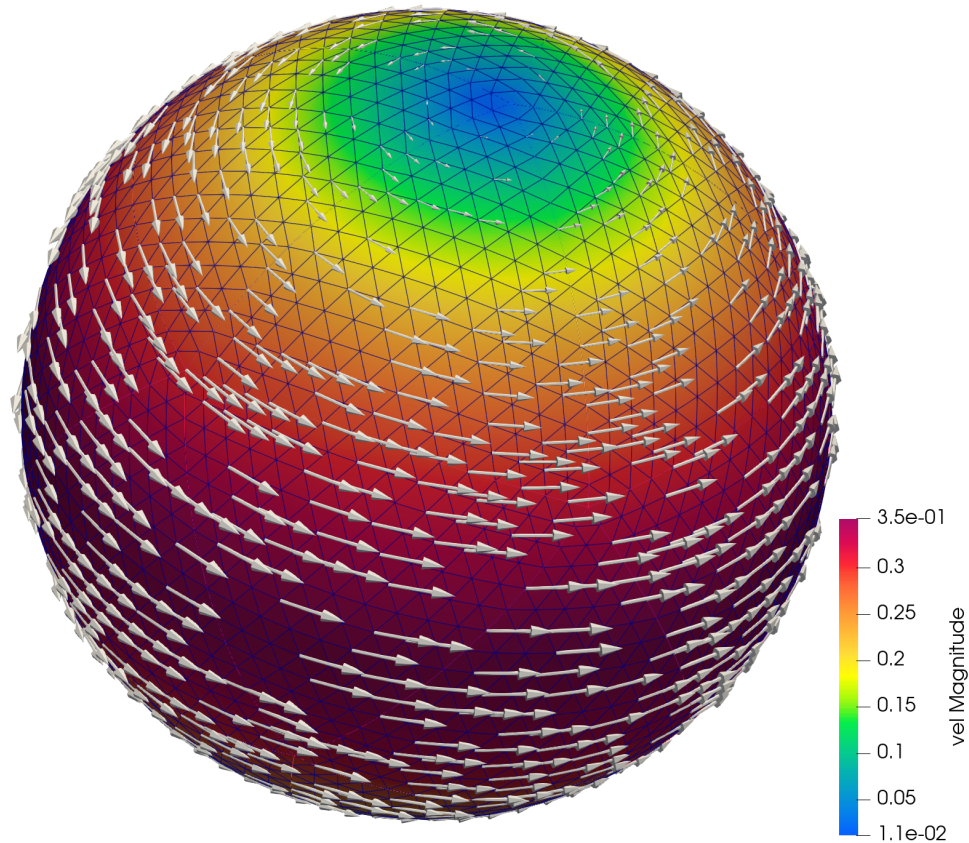}
\caption{A spherical mesh constructed from an icosahedron, with 2560 vertices and 5120 triangles (compared with 2452 vertices and 4900 triangles in the mesh previously used in the intrinsic formulation).}
\label{fig_icosahedron}
\end{figure}

Another limitation of the surface FEM is its dependence on accurate surface normals. For example, we find that defining the FEM variables as $Nx = Ns.x$, $Ny = Ns.y$, and $Nz = Ns.z$ is preferable to directly using $Ns.x$, $Ns.y$, and $Ns.z$. However, these normal fields must be represented using $P1$ finite elements. Another interesting observation is that the results deteriorate on the ``nicer'' mesh, despite having essentially the same mesh resolution, as shown in Figure~\ref{fig_icosahedron} and Table~\ref{table_eigenvalues_surfaceFEM_nicermesh}. It can be seen that both the eigenvalue errors and the viscous energy are larger than the corresponding values reported in Table~\ref{table_eigenvalues_surfaceFEM}.

\begin{table}[h!]
\centering
\begin{tabular}{c|c|c}
\hline
 NO. of eigenvalues   &  Eigenvalue   &  Viscous Energy\\
\hline
1   &  3.768457841967789e-06 & 1.884228943722436e-06 \\
\hline
2   &  4.970081660977566e-06 & 2.485040809719386e-06 \\
\hline
3   &  5.320441312495307e-06 & 2.660220668233715e-06 \\
\hline
4   &  1.069769826658529e-05 & 5.34884914638261e-06 \\
\hline
5   &  1.205550232932085e-05 & 6.027751138281373e-06 \\
\hline
6   &  1.596970336732208e-05 &7.984851686738948e-06  \\
\hline
7&3.880736614661474 &1.940368309839571\\
\hline
8&3.902819347473074 &1.951409674646688\\
\hline
9&3.927018931631206 &1.963509468450154\\
\hline
10&3.944241249316089 &1.972120625534393\\
\hline
\end{tabular}
\caption{The first ten eigenvalues of the Hodge operator on a sphere and the viscous energy of the corresponding eigenvectors, computed using the surface FEM with the same mesh as in the previous intrinsic formulation.}
\label{table_eigenvalues_surfaceFEM}
\end{table}

\begin{table}[h!]
\centering
\begin{tabular}{c|c|c}
\hline
 NO. of eigenvalues   &  Eigenvalue   &  Viscous Energy\\
\hline
1   &  0.0001287371151975334 & 6.436855760117211e-05 \\
\hline
2   &  0.000175691252156582 & 8.784562606817302e-05 \\
\hline
3   &  0.0002159611450274687 & 0.0001079805724975914 \\
\hline
\end{tabular}
\caption{The first three eigenvalues of the Hodge operator on a sphere and the viscous energy of the corresponding eigenvectors (corresponding to the Killing vector fields), using a ``nicer" mesh shown in Figure \ref{fig_icosahedron}.}
\label{table_eigenvalues_surfaceFEM_nicermesh}
\end{table}

To implement the time-dependent NS equation using surface FEM, the divergence-free equation needs to be implemented. The surface divergence of $u=(u_x, u_y, u_z)$ is given by:
\begin{equation}
\begin{split}
\nabla_\Gamma\cdot u & = \text{tr}\left(\nabla_\Gamma u\right) \\
& = \text{tr}\left(\mathcal{P}\left(\hat{\nabla} u\right)\mathcal{P} \right) - \left(u\cdot \tau\right)\text{tr}\left(\mathcal{P}\left(\hat{\nabla}\tau\right)\mathcal{P}\right)\\
& = \text{tr}\left(\mathcal{P}\left(\hat{\nabla} u\right)\mathcal{P} \right) - \left(u\cdot \tau\right)\text{tr}\left(\hat{\nabla}\tau\right)\\
& = \text{tr}\left(\mathcal{P}\left(\hat{\nabla} u\right)\mathcal{P} \right) + 2 \left(u\cdot \tau\right) H
\end{split}
\end{equation}
based on equation (\ref{projection_computation}). In the above $H$ is the mean curvature, and $\hat{\nabla}\tau = - 2H$ due to
\[
\hat{\nabla}\tau = \partial_j \tau \otimes dx^j = -h^i_j (\partial_i \otimes dx^j),
\]
and $H = h_m^m/2$.

\begin{figure}[h!]
\centering   \includegraphics[width=0.3\linewidth]{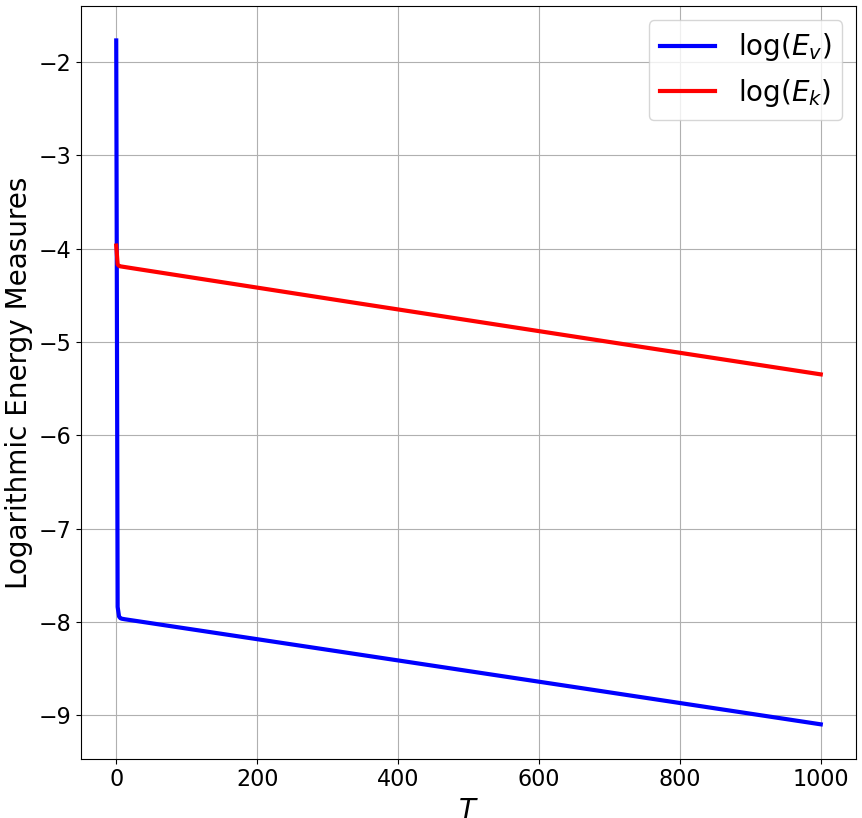}
\includegraphics[width=0.3\linewidth]{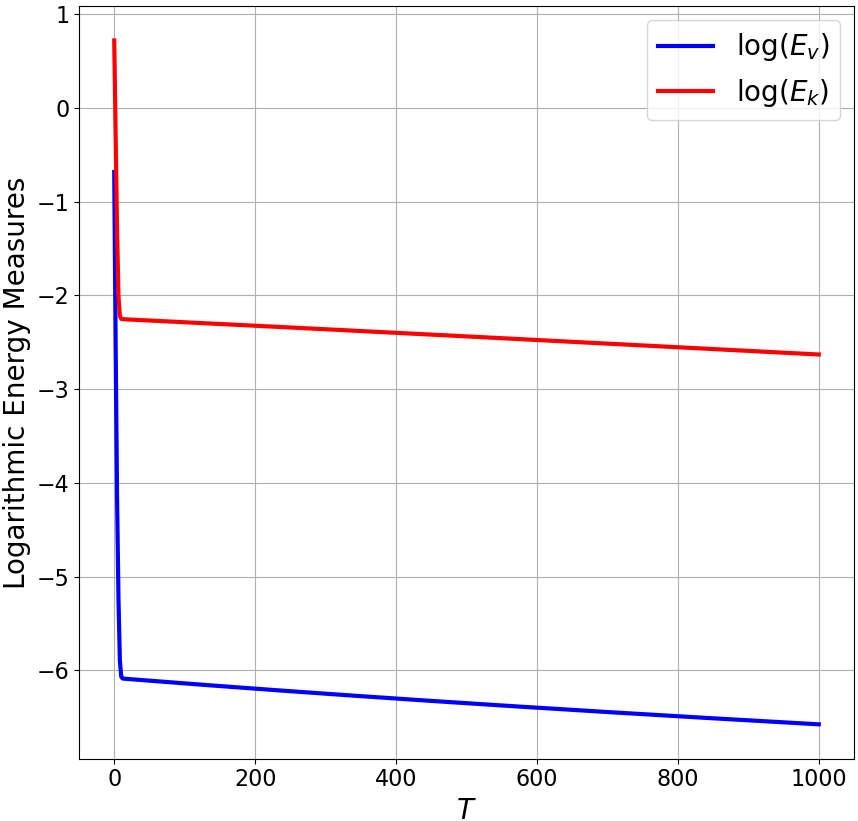}
\includegraphics[width=0.3\linewidth]{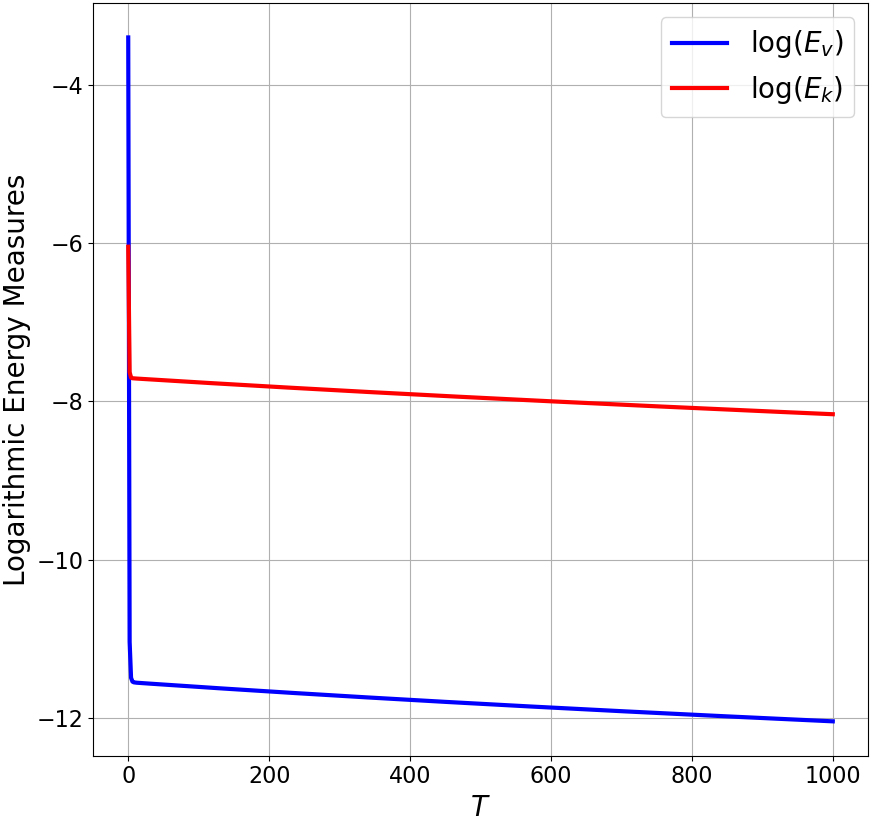}
\caption{Energy evolution for three different initial velocity profiles using the surface FEM.}
\label{fig_viscousenergy_longrun_surfaceFEM}
\end{figure}

In FreeFEM++, function trace() can be directly used to compute the divergence:
\begin{lstlisting}
// Surface gradient and divergence on a sphere
macro GradS(u) (Pr*Grad(u)*Pr - (vect(u)'*Nor)*Pr) 
macro div(u) trace(GradS(u)) //  
\end{lstlisting}

We use the same initial velocity profile as in the previous tests and examine whether this initial velocity converges to a steady state corresponding to the Killing fields on the sphere. Note that the velocity must be mapped onto the sphere in order to apply the surface finite element method. The positive observation is that the solution does converge to the Killing vector fields. However, the magnitude of viscous energy is large, and both the kinetic and viscous energy continue to decay over time, likely due to numerical viscosity, as shown in Figure~\ref{fig_viscousenergy_longrun_surfaceFEM}. Another note is that in order to obtain the Killing vector fields, it is important to project the solution onto the tangent space during the time iteration.

\subsection{Torus and helicoid}\label{sec_torus}
Based on the previous numerical experiments, the intrinsic FEM formulated in the parameter space appears to have advantages in both efficiency and accuracy, provided that no coordinate singularities are present. This will be further investigated and verified on two geometries in this section: a torus and a helicoid, both of which admit singularity-free parameterisations.

\subsubsection{Torus}
Consider the torus $M = T_{R,r} \subset \mathbb{R}^3$ with major radius $R$ and minor radius $r$, parameterised by
\begin{equation}\label{torus_parameterised}
x = (R + r\cos\theta)\cos\phi, \quad
y = (R + r\cos\theta)\sin\phi, \quad
z = r\sin\theta,
\qquad
\theta,\phi \in [0,2\pi).
\end{equation}

The components of the metric tensor are:
\[
g_{\theta\theta}=r^2,\quad g_{\phi\phi}=(R + r\cos\theta)^2, \quad g_{\theta\phi}=g_{\phi\theta}=0.
\]

Since $g_{\theta\theta}$ and $g_{\phi\phi}$ depend only on $\theta$, the nonzero components of the Christoffel symbols are:
\[
\Gamma_{\phi\phi}^\theta
= -\frac{1}{2} g^{\theta\theta} \partial_\theta g_{\phi\phi}
= -\frac{1}{2r^2}\big(2(R+r\cos\theta)(-r\sin\theta)\big)
= \frac{(R+r\cos\theta)\sin\theta}{r},
\]
\[
\Gamma_{\theta\phi}^\phi = \Gamma_{\phi\theta}^\phi
= \frac{1}{2} g^{\phi\phi} \partial_\theta g_{\phi\phi}
= \frac{1}{2(R+r\cos\theta)^2}
\big(2(R+r\cos\theta)(-r\sin\theta)\big)
= -\frac{r\sin\theta}{R+r\cos\theta}.
\]

\begin{figure}[h!]
\centering   
\includegraphics[width=0.7\linewidth]{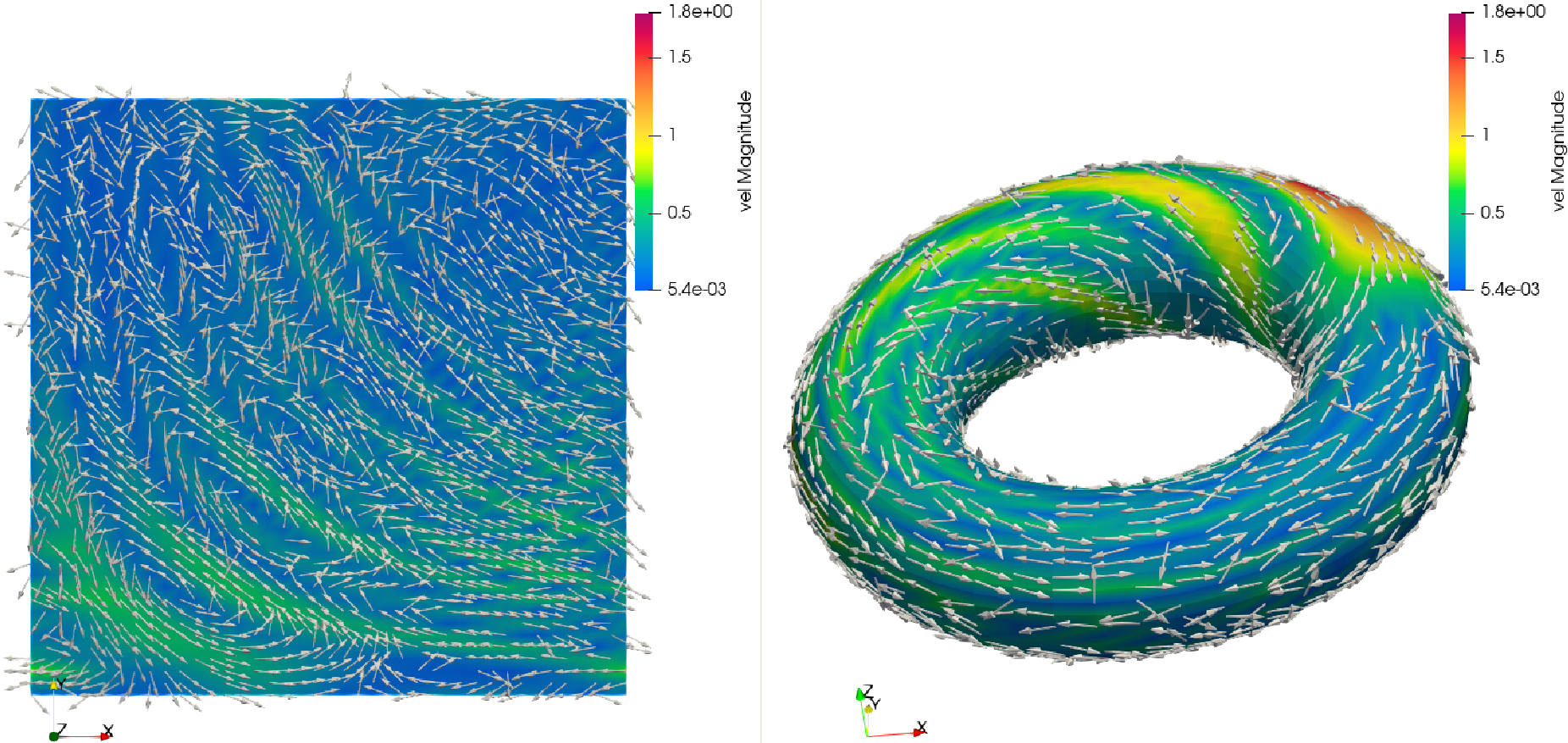}
\includegraphics[width=0.7\linewidth]{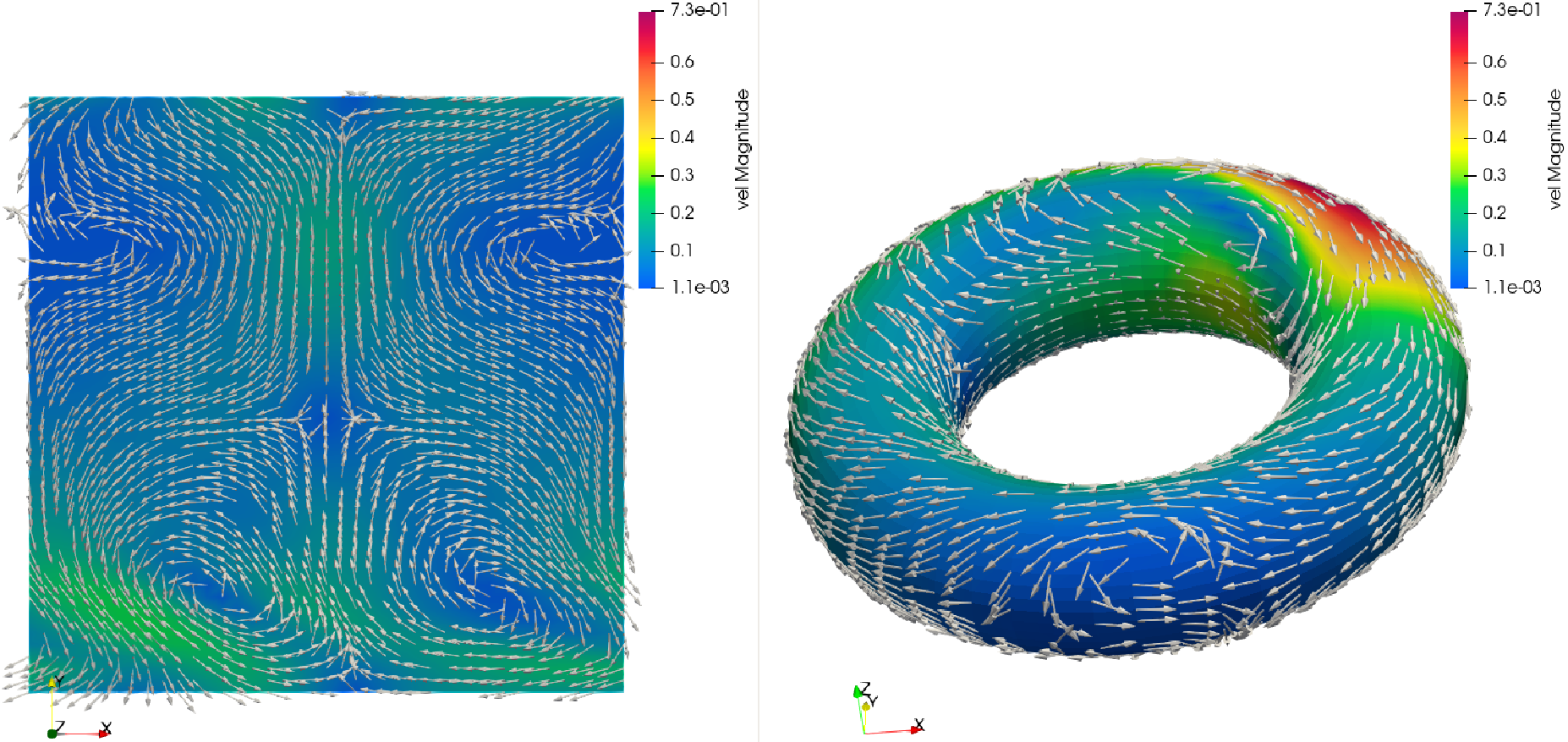}
\includegraphics[width=0.7\linewidth]{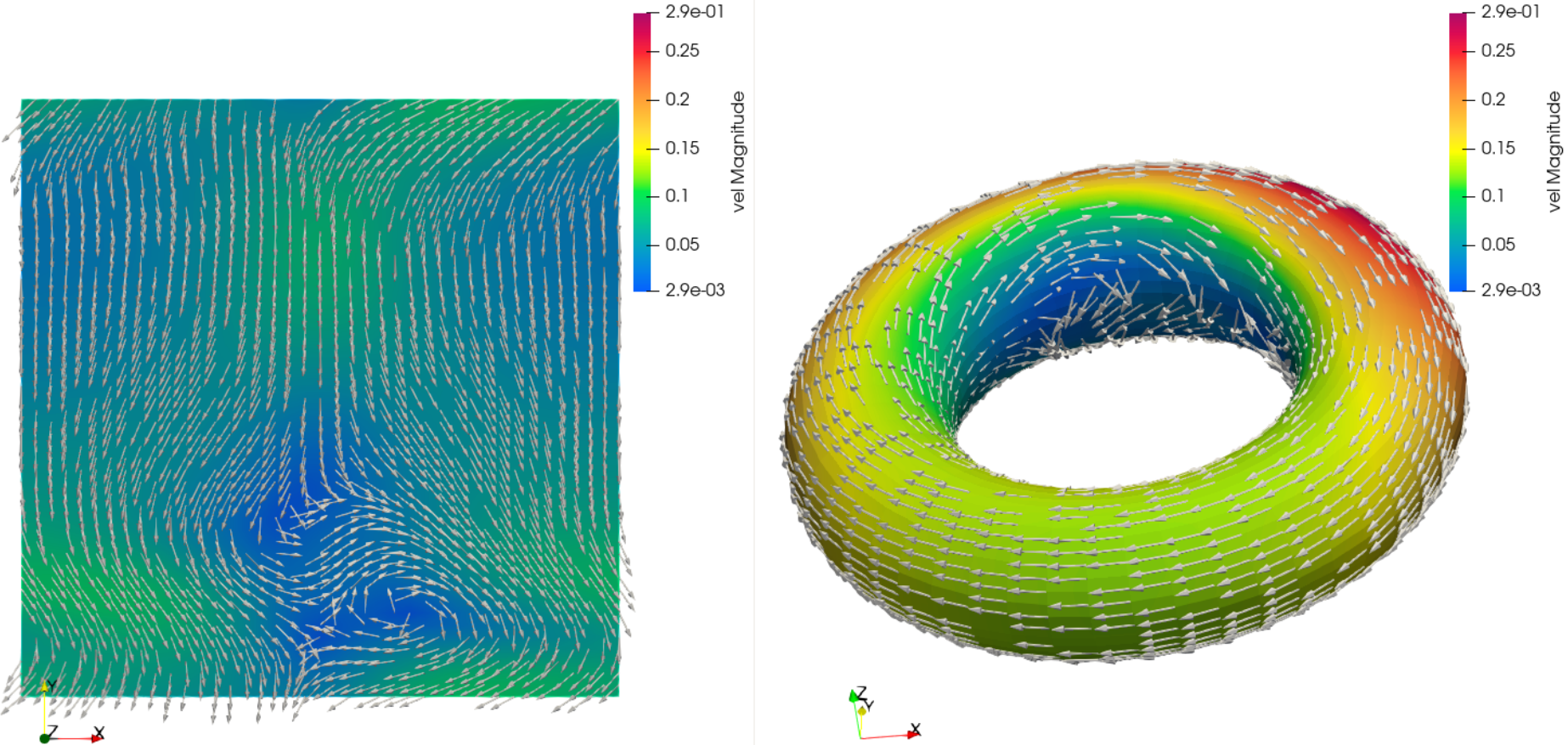}
\includegraphics[width=0.7\linewidth]{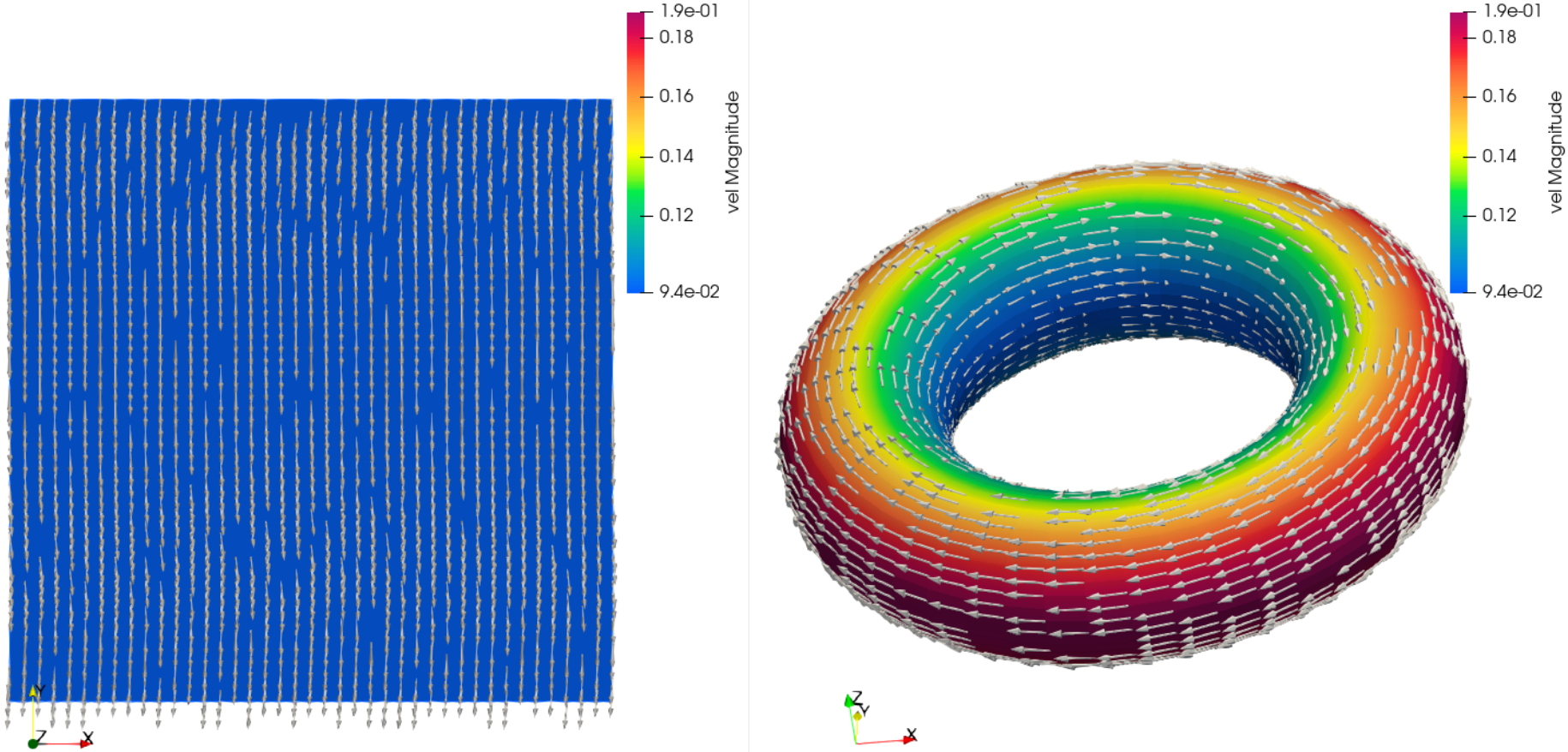}
\caption{The initial flow $u = (\sin\theta \sin\phi,\,-\sin(\theta\phi)\cos(\theta\phi))$ converges to the steady-state solution corresponding to the Killing field representing rotation about the symmetry axis, with $\rho = 1$ and $\mu = 1$. From top to bottom, the figures correspond to $t = 0$, $t = 0.1$, $t = 0.5$, and $t = 3$.}
\label{fig_ns_torus}
\end{figure}

\begin{figure}[h!]
\centering   
\includegraphics[width=0.4\linewidth]{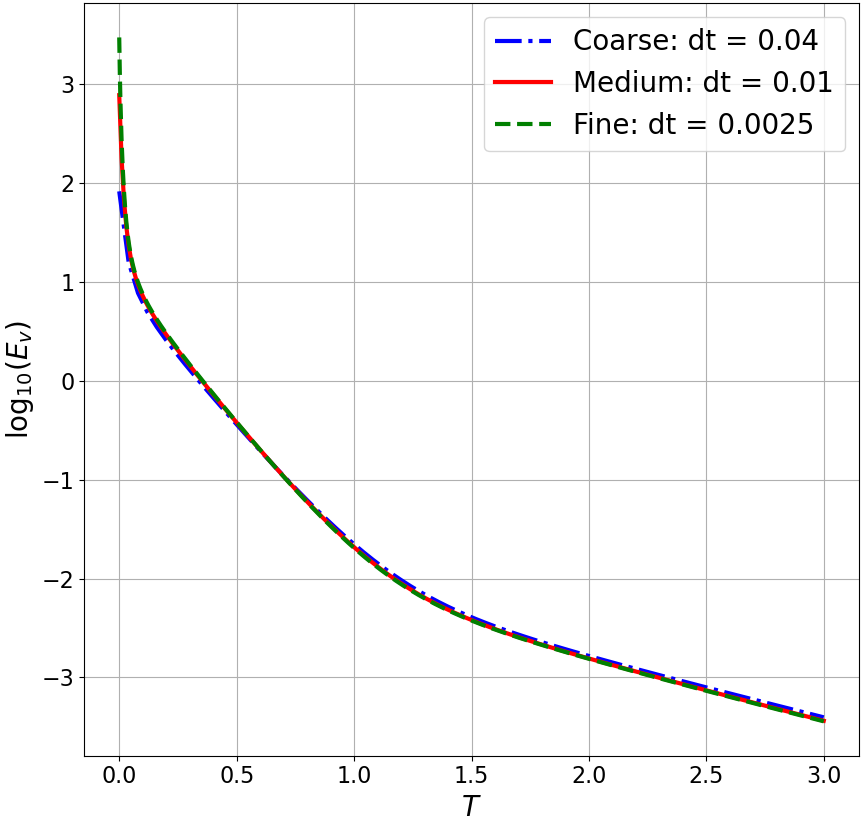}
\includegraphics[width=0.4\linewidth]{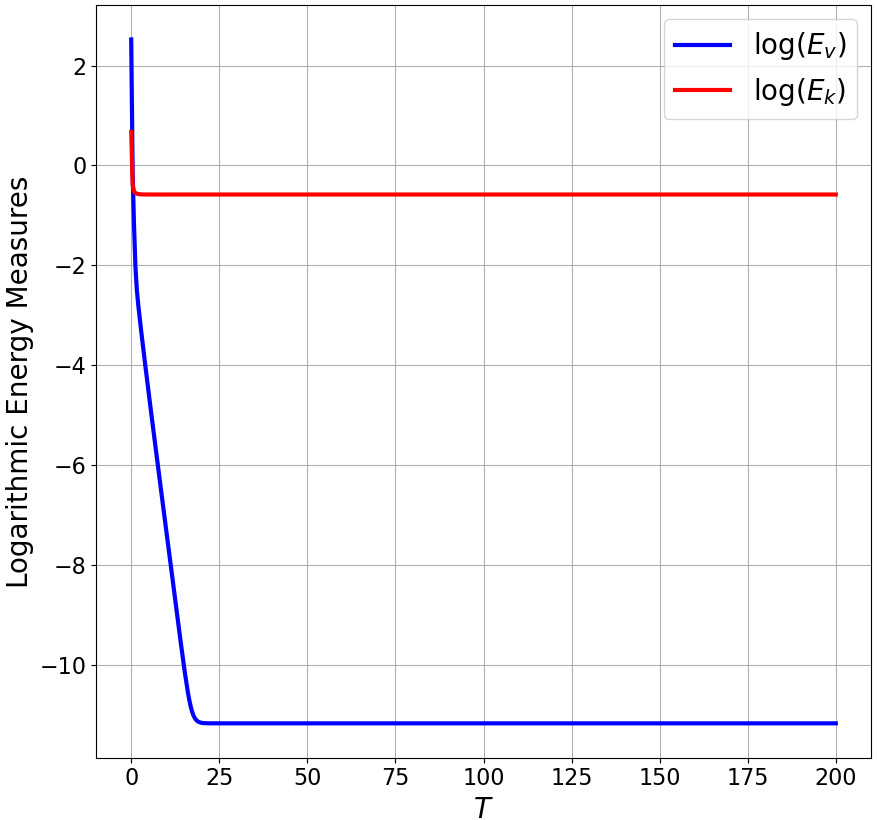}
\caption{Fluid flow on a torus simulated on three different meshes (left) and extended to a long-time simulation up to $t = 200$ (right) using the medium-sized mesh.}
\label{convergence_torus}
\end{figure}

As expected, a rotational Killing vector field is obtained as the steady-state solution of the NS equation, as shown in Figures~\ref{fig_ns_torus}, with geometrical parameters: $R=3$, $r=1$ and $\theta \in [0,2\pi)$. The computational domain is $(\theta, \phi) \in [0, 2\pi] \times [0, 2\pi]$ with periodic boundary conditions at $\theta=0, 2\pi$. The steady-state solution is very stable as shown in Figure \ref{convergence_torus}.

\subsubsection{ Helicoid}
The helicoid can be parameterised as:
\[
x(\xi,\eta) = \eta\cos \xi,\quad y(\xi,\eta) = \eta\sin \xi,\quad z(\xi,\eta) = \alpha \xi,\quad \xi, \eta \in \mathbb{R}, \quad \alpha \in \mathbb{R}.
\]

The components of the metric tensor are:
\[
g_{\xi\xi} = \eta^2+\alpha^2,\quad g_{\eta\eta}=1, \quad g_{\xi\eta}=g_{\eta\xi}=0.
\]

The non-zero Christoffel symbols are:
\[
\Gamma^\xi_{\xi\eta} = \Gamma^\xi_{\eta\xi} = \frac{\eta}{\eta^2 + \alpha^2}, \quad
\Gamma^\eta_{\xi\xi} = -\eta.
\]

\begin{figure}[h!]
\centering   
\includegraphics[width=1\linewidth]{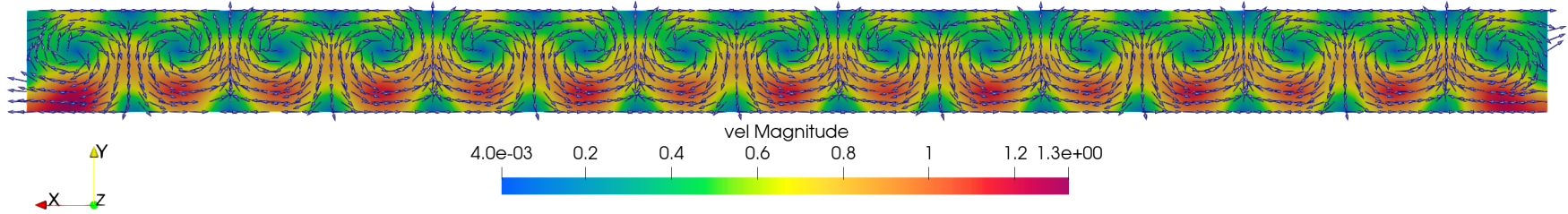}
\includegraphics[width=1\linewidth]{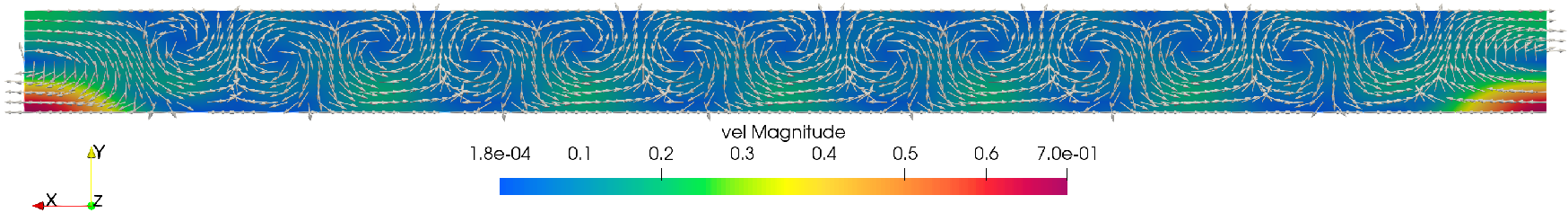}
\includegraphics[width=1\linewidth]{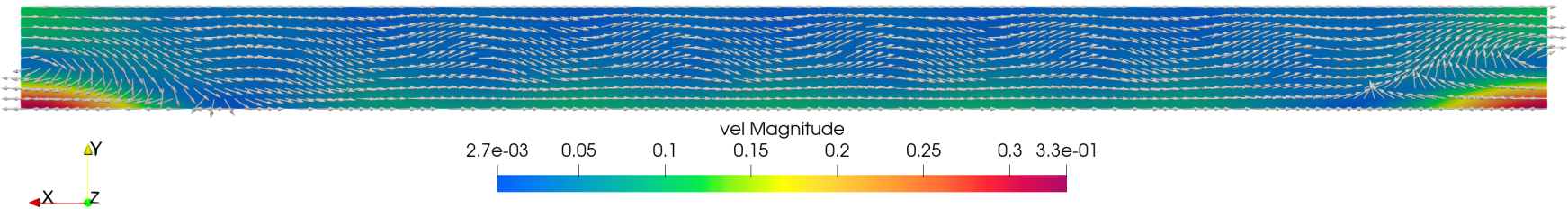}
\includegraphics[width=1\linewidth]{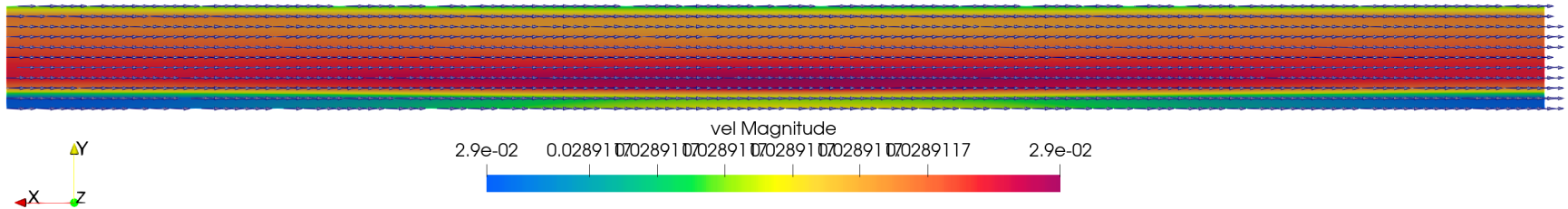}
\caption{The initial flow in the parameter space converges to the steady-state solution corresponding to the translational Killing vector field. From left to right and top to bottom, the figures correspond to $t = 0$, $t = 0.1$, $t = 0.2$, and $t = 5$.}
\label{fig_ns_helicoid_parameter}
\end{figure}

For the simulation, let $\xi \in [0, 15]$ be periodic, and let $\eta \in [0, 1]$ where slip boundary conditions is applied. The parameter is set to $\alpha = 0.2$. The initial flow profile is given by
\[
u(\xi,\eta) = \left(\sin(\pi \xi)\cos(\pi \eta), -\cos(\pi \xi)\sin(\pi \eta)\right),
\]
as shown in the first panel of Figure~\ref{fig_ns_helicoid_parameter}. 

\begin{figure}[h!]
\centering   
\includegraphics[width=0.45\linewidth]{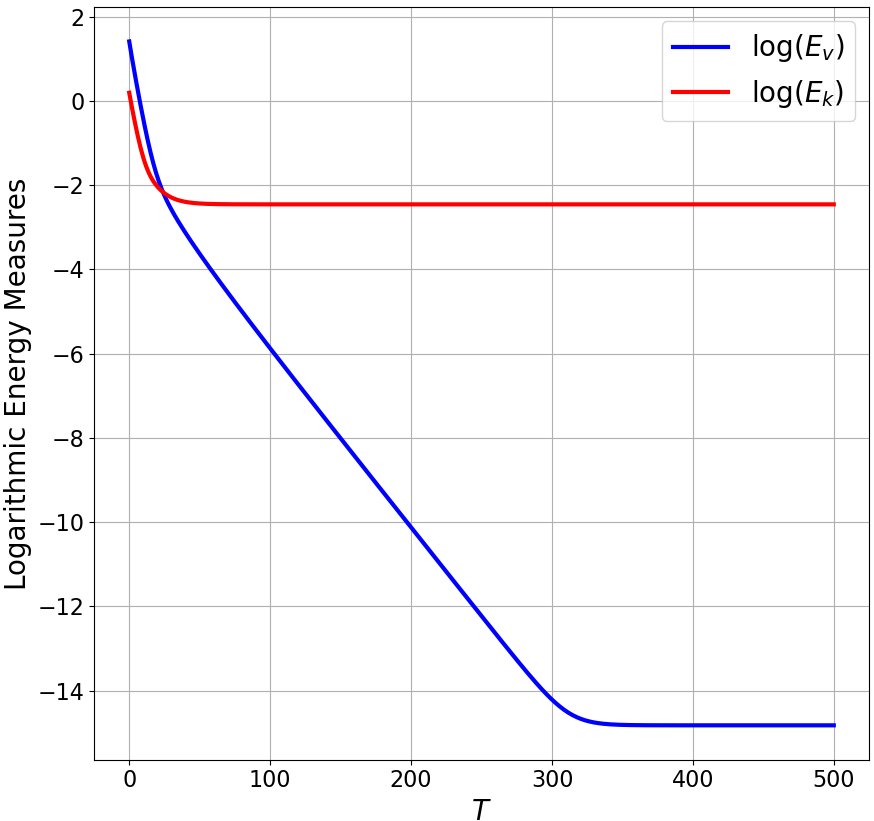}
\caption{Evolution of the viscous and kinetic energy for fluid on the helicoid.}
\label{convergence_helicoid}
\end{figure}

This initial velocity converges to a steady-state solution, which corresponds to the translational Killing vector field on the helicoid. The convergence of the viscous and kinetic energies is plotted in Figure~\ref{convergence_helicoid}, where the viscous energy decays to zero while the kinetic energy remains stable. 

\begin{figure}[h!]
\centering   
\includegraphics[width=0.4\linewidth]{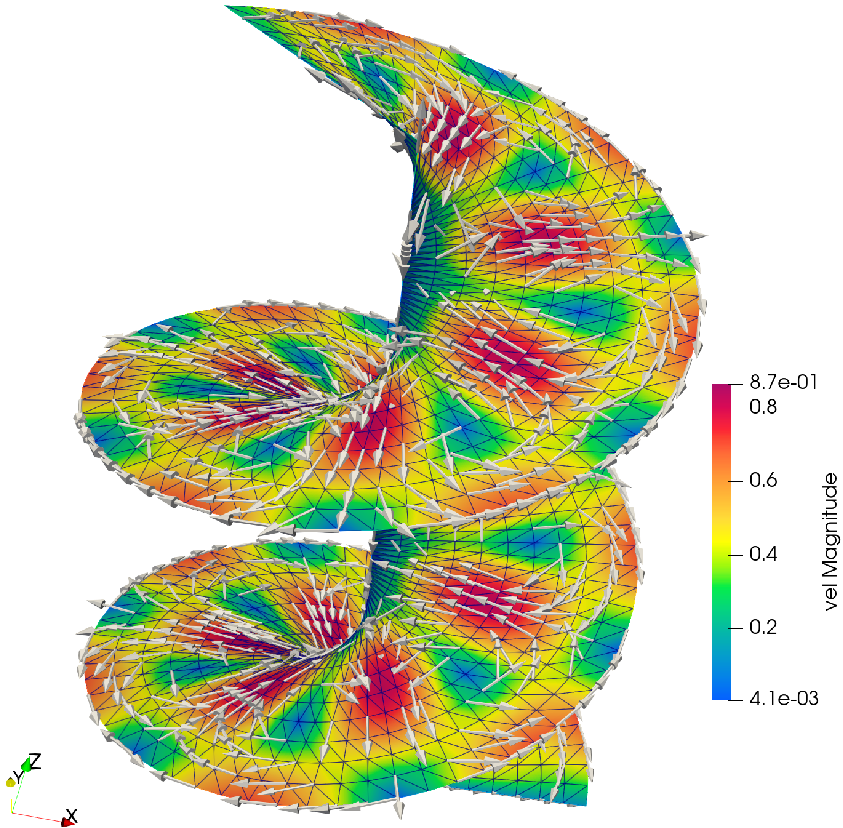}
\includegraphics[width=0.4\linewidth]{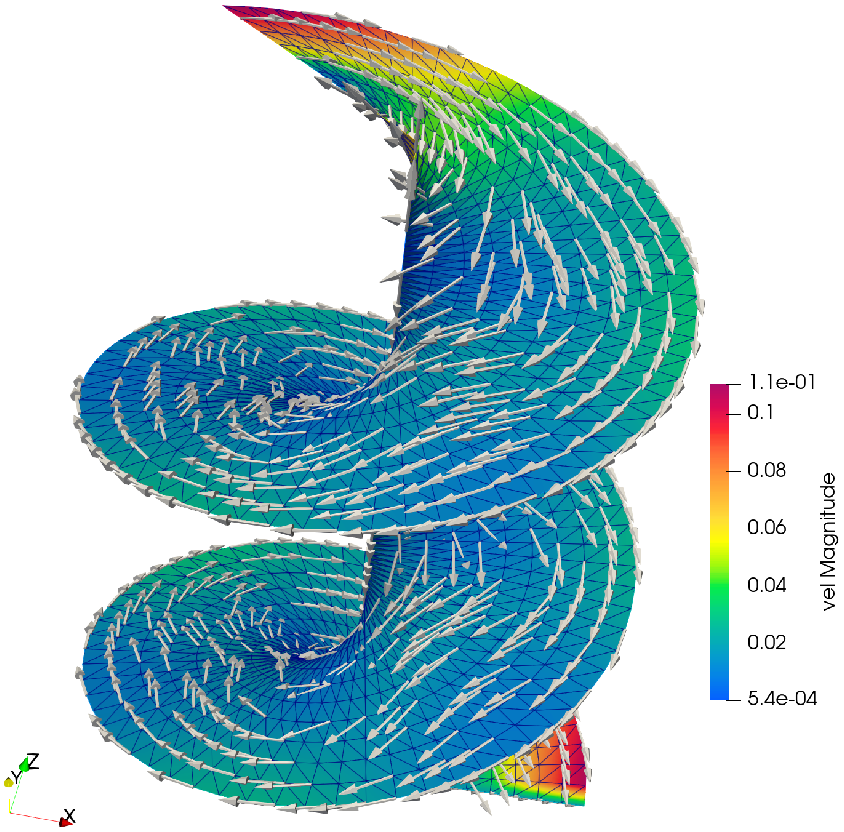}
\includegraphics[width=0.4\linewidth]{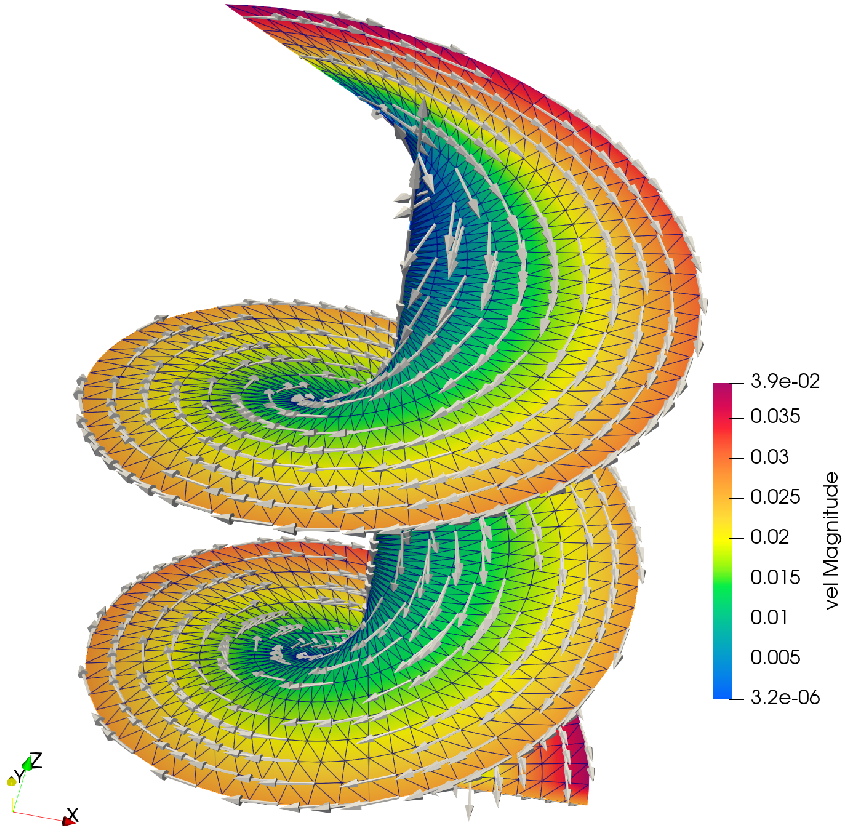}
\includegraphics[width=0.4\linewidth]{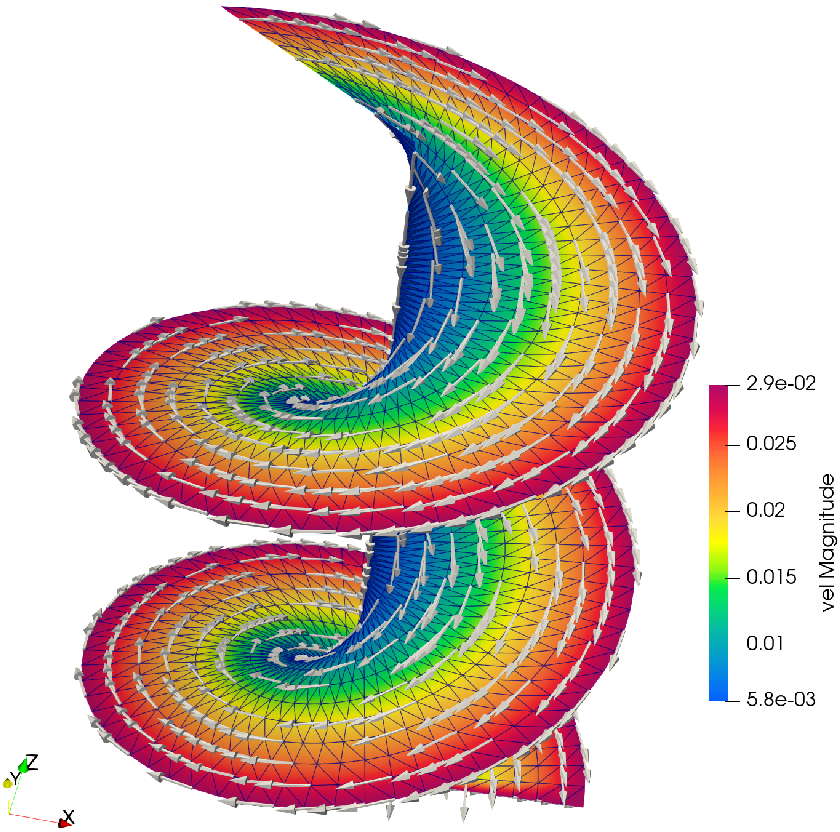}
\caption{The initial flow on the helicoid converges to the steady-state solution corresponding to the translational Killing vector field. From left to right and top to bottom, the figures correspond to $t = 0$, $t = 0.2$, $t = 0.5$, and $t = 5$.}
\label{fig_ns_helicoid}
\end{figure}

The convergence of the velocity field is illustrated in Figure~\ref{fig_ns_helicoid_parameter} (in parameter space) and Figure~\ref{fig_ns_helicoid} (for the helicoid mapped from the parameter space $\Omega = [0,15]\times[0,1]$ to $\mathbb{R}^3$).

\subsection{Flow in a spherical shell with Schwarzschild spatial metric}\label{sec_schwarzschild}

Using the parameterisation described in~(\ref{sphere_parameterised}), where $\theta\in[0,2\pi)$ is the longitude and $\phi\in[0,\pi]$ is the colatitude (so that $\pi/2-\phi$ is the latitude), the Schwarzschild spatial metric can be written as
\begin{equation}
ds^2
=
\left(1-\frac{\mathrm{M}}{r}\right)^{-1}dr^2
+r^2\left(d\phi^2+\sin^2\phi\,d\theta^2\right).
\end{equation}
Here, $\mathrm{M}=2G\mathcal{M}/c^2$ is the Schwarzschild radius, where $\mathcal{M}$ is the mass of the central body, $G$ is the gravitational constant, and $c$ is the speed of light.

The nonzero Christoffel symbols are:
\begin{align*}
\Gamma^r_{rr} &= \frac{M}{2r(r - M)}, \\
\Gamma^r_{\phi\phi} &= -(r - M), \\
\Gamma^r_{\theta\theta} &= -(r - M)\sin^2\phi, \\
\Gamma^\phi_{r\phi} &= \Gamma^\phi_{\phi r} = \frac{1}{r}, \\
\Gamma^\phi_{\theta\theta} &= -\sin\phi \cos\phi, \\
\Gamma^\theta_{r\theta} &= \Gamma^\theta_{\theta r} = \frac{1}{r}, \\
\Gamma^\theta_{\phi\theta} &= \Gamma^\theta_{\theta\phi} = \cot\phi.
\end{align*}

Due to the large scale of the numerical discretisation -- $60{,}000$ nodes, $300{,}000$ tetrahedra, and approximately $2{,}000{,}000$ degrees of freedom -- we perform the numerical simulations in parallel using FreeFEM++ and PETSc. The resulting algebraic system is solved using FGMRES with a Schur complement preconditioner. For the full sphere, which contains coordinate singularities at $\phi = 0$ and $\phi = \pi$, we find that the iterative solver struggles to converge. To avoid this issue, we exclude the singular points and consider the computational domain
\[
(\theta, \phi, r)\in [0,2\pi]\times[\pi/4,3\pi/4]\times[2,3],
\]
with periodic boundary conditions at $\theta = 0$ and $\theta = 2\pi$, and slip boundary conditions at $\phi = \pi/4$, $\phi = 3\pi/4$, $r = 2$, and $r = 3$.

\begin{figure}
\centering
\includegraphics[width=1\linewidth]{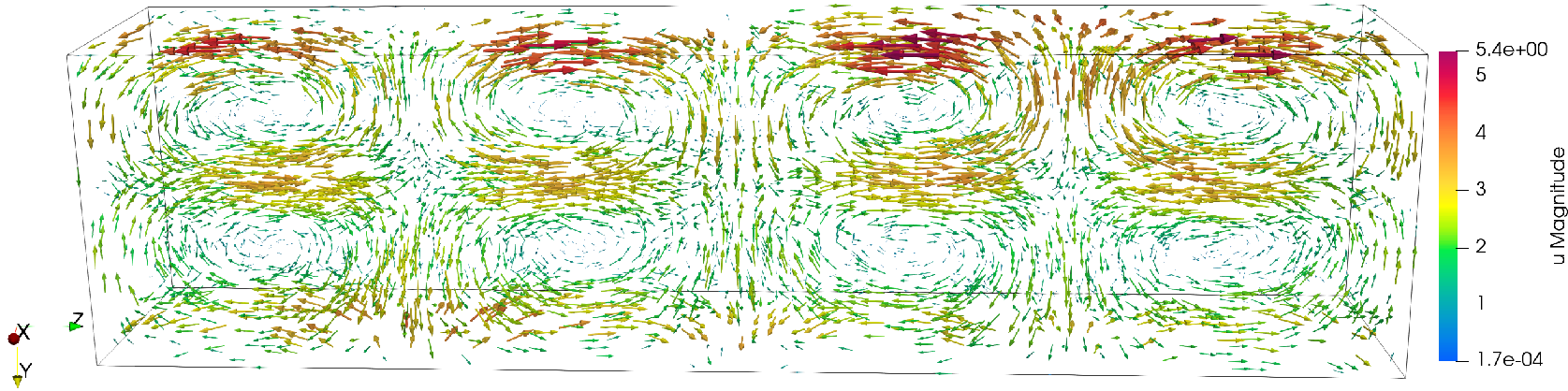}
\includegraphics[width=0.55\linewidth]{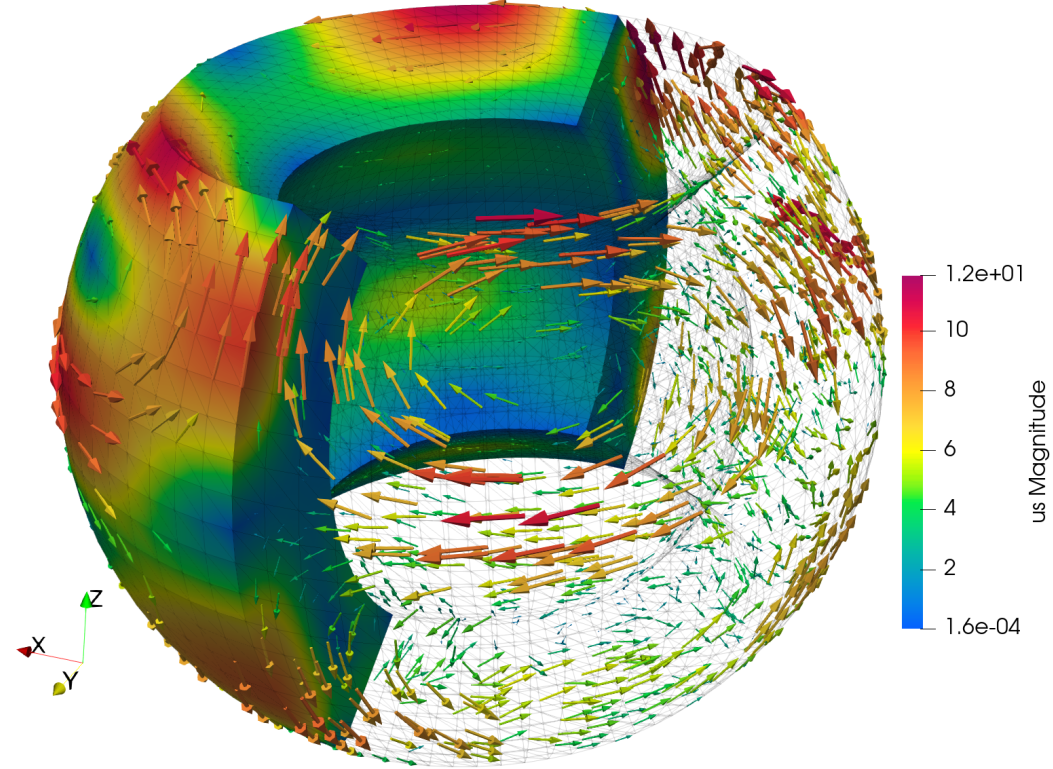}
\caption{Initial velocity field defined by~(\ref{schwarz_initial}). The top panel shows the velocity field in the parameter space, while the bottom panel shows the corresponding velocity field on the manifold.}
\label{fig_schwartschild0}
\end{figure}

We consider the following initial velocity field with periodic features:
\begin{equation}\label{schwarz_initial}
\begin{split}
u_\theta &= \Bar{u}\sin(2\theta)\cos(2\phi)\sin(2r),\\
u_\phi &= \Bar{u}\cos(\theta)\sin(\phi)\cos(r),\\
u_r &= \Bar{u}\cos(r)\sin(r)\sin(2\theta\phi),
\end{split}
\end{equation}
as shown in Figure~\ref{fig_schwartschild0}.

\begin{figure}
    \centering
    \includegraphics[width=1\linewidth]{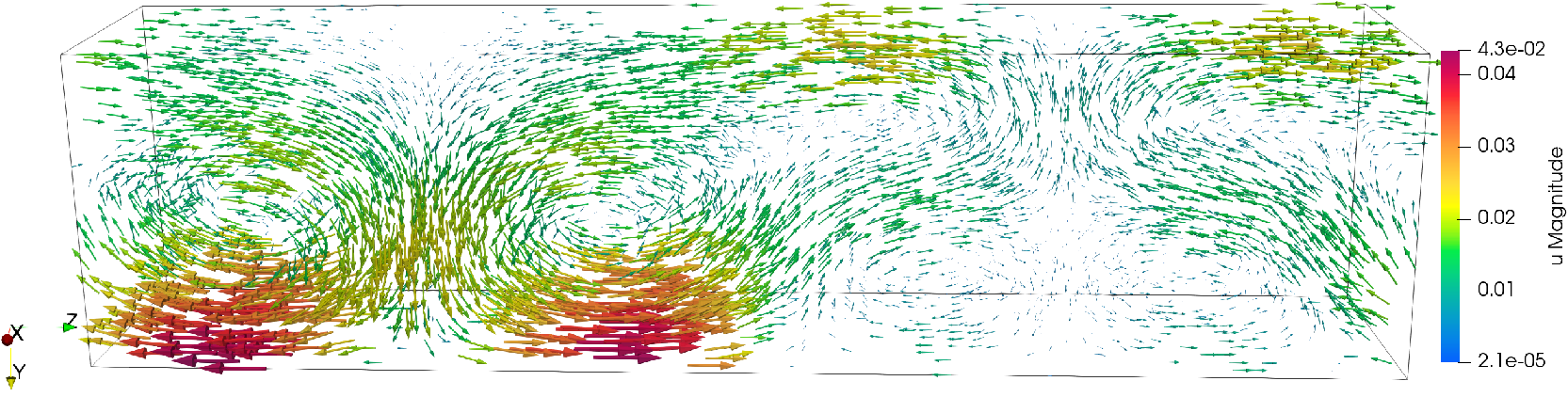}
    \includegraphics[width=1\linewidth]{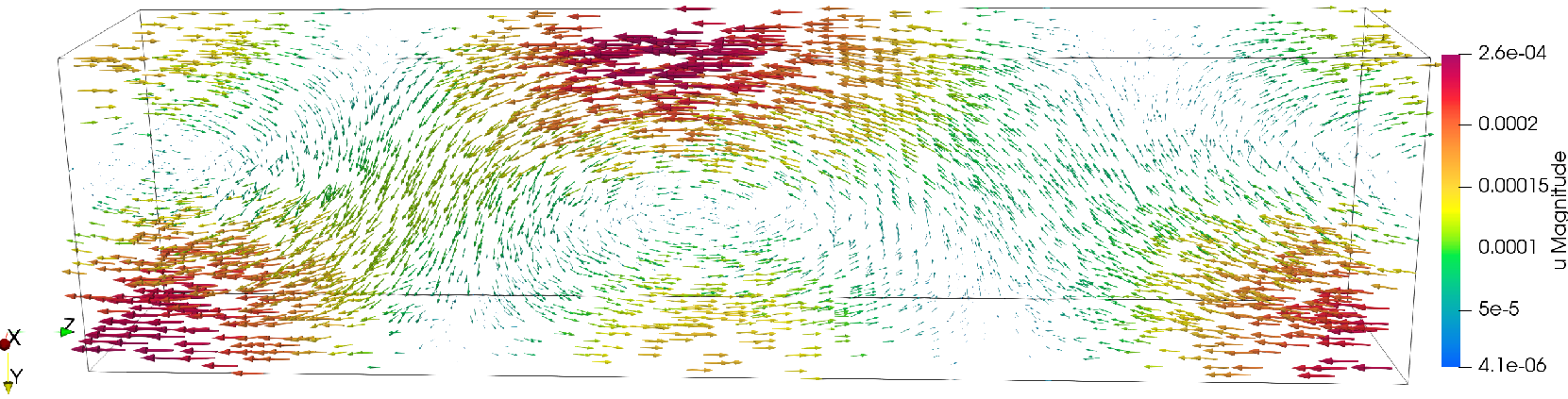}
    \includegraphics[width=1\linewidth]{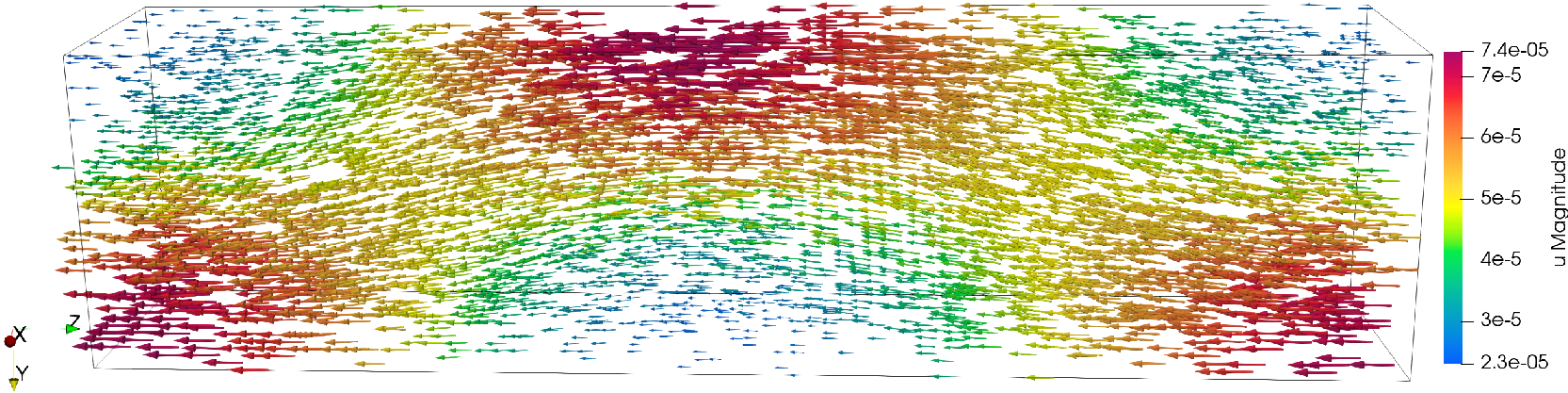}
    \includegraphics[width=1\linewidth]{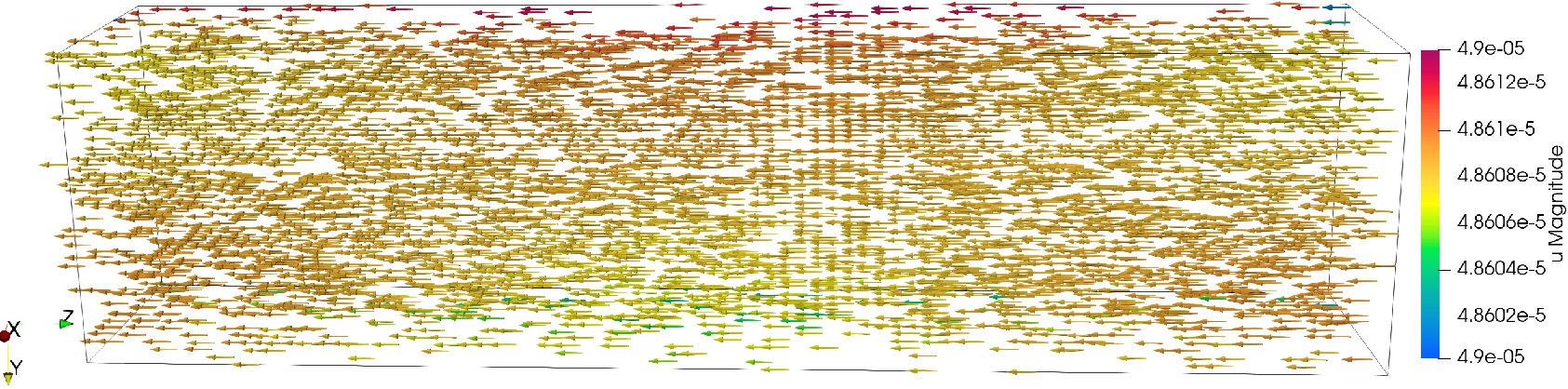}
\caption{The initial velocity field defined by~(\ref{schwarz_initial}) converges to a steady-state solution corresponding to a rotational Killing vector field, shown in the parameter space.}
\label{fig_schwartschild1}
\end{figure}

\begin{figure}
    \centering
    \includegraphics[width=0.45\linewidth]{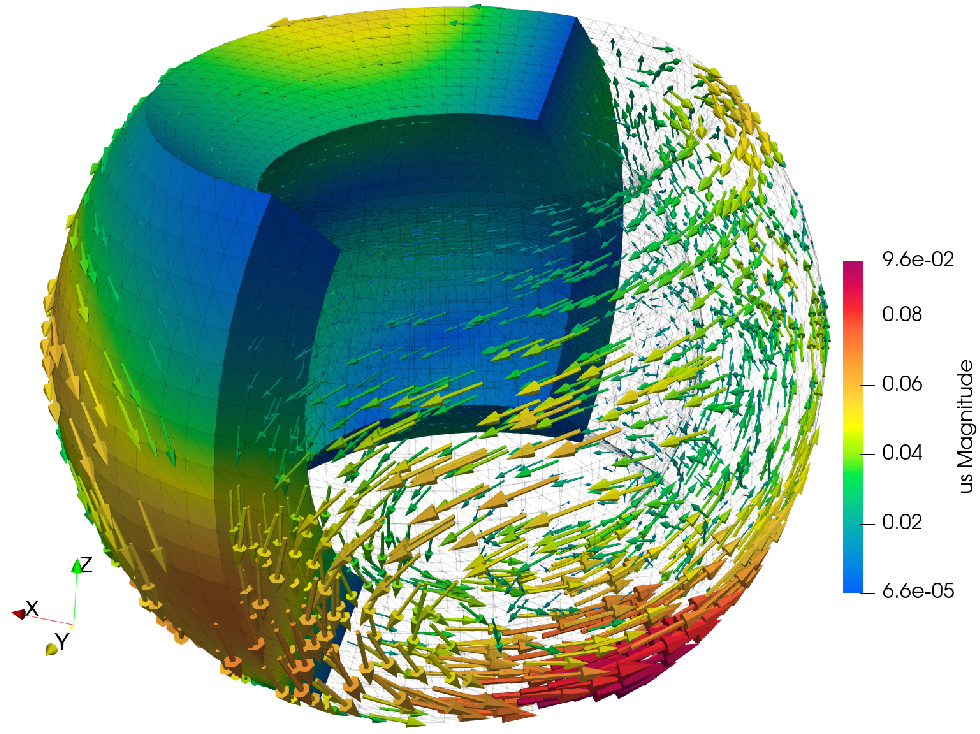}
    \includegraphics[width=0.45\linewidth]{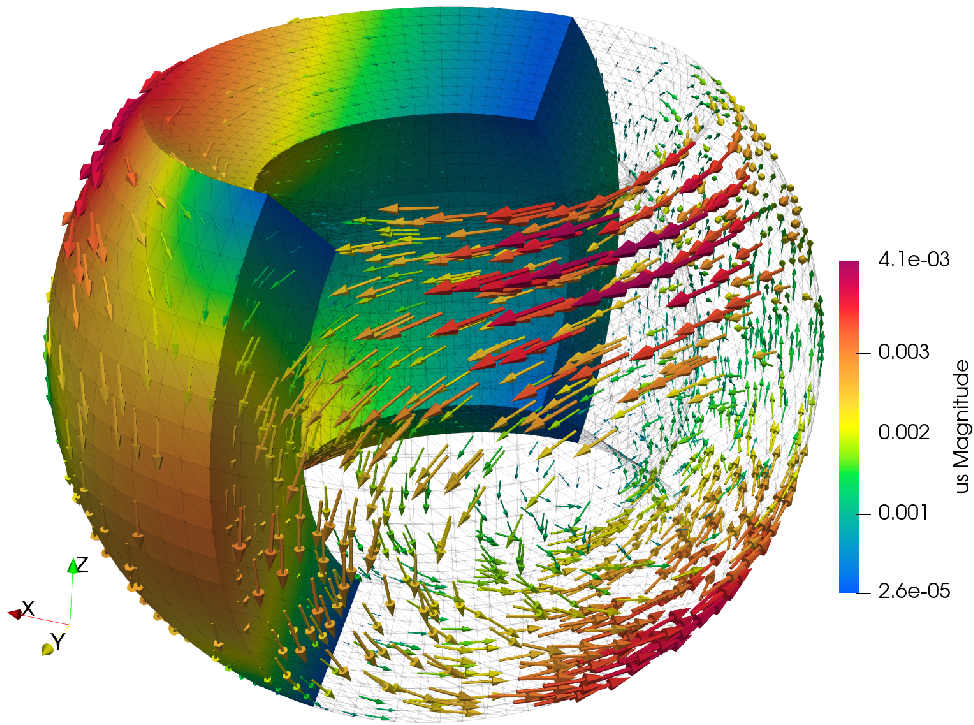}
    \includegraphics[width=0.45\linewidth]{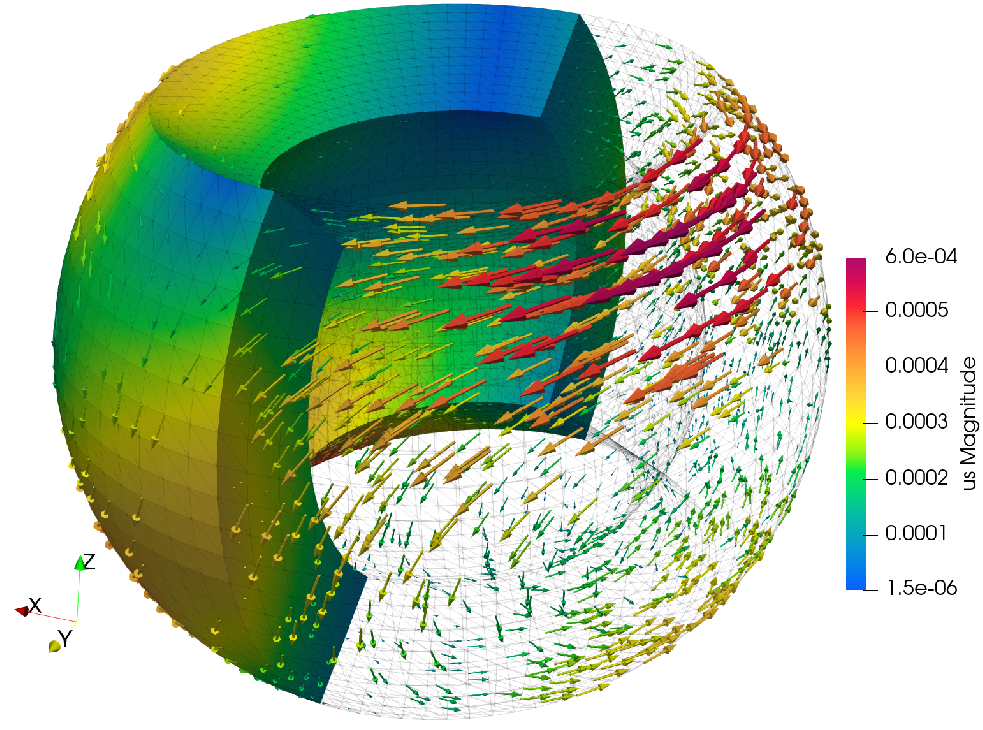}
    \includegraphics[width=0.45\linewidth]{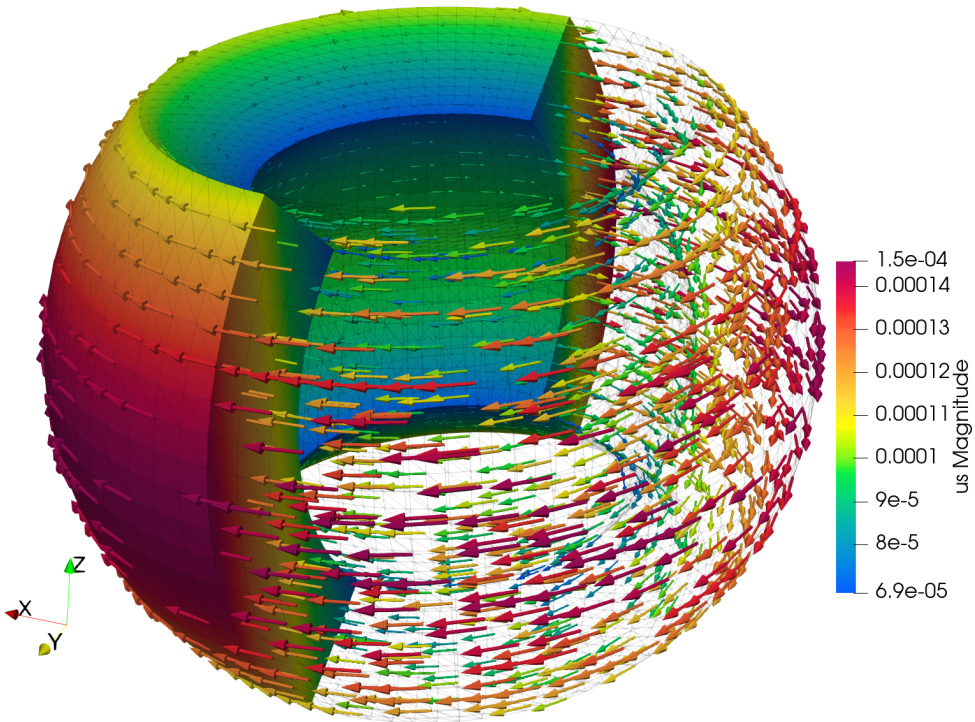}
\caption{The initial velocity field defined by~(\ref{schwarz_initial}) converges to a steady-state solution corresponding to a rotational Killing vector field, shown on the manifold.}
\label{fig_schwartschild2}
\end{figure}

\begin{figure}[h!]
\centering   
\includegraphics[width=0.5\linewidth]{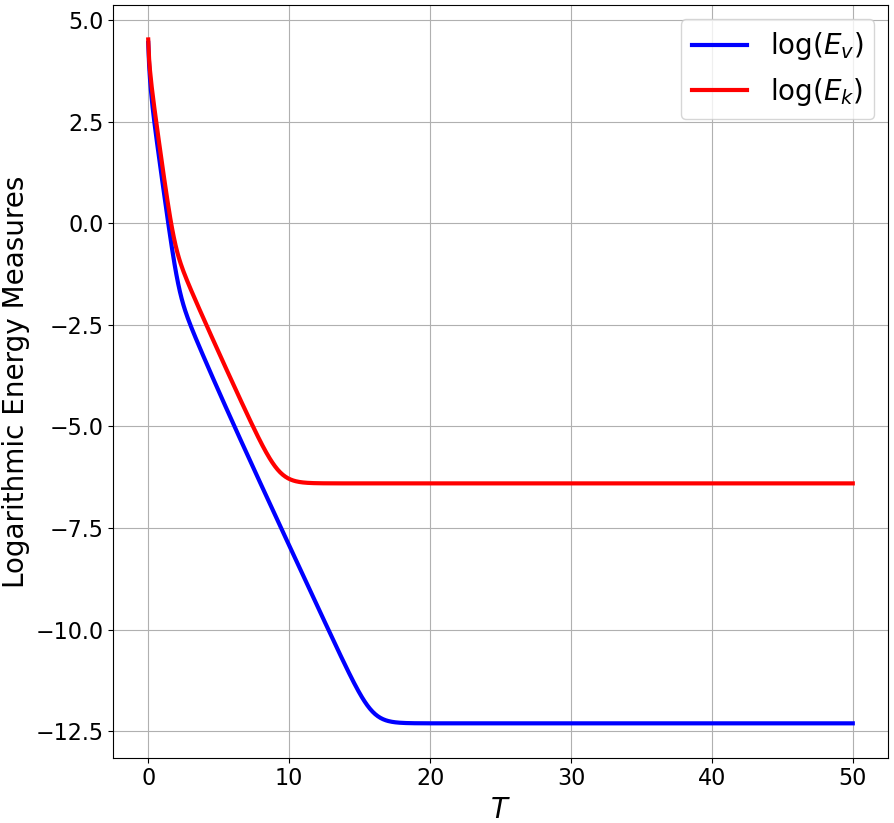}
\caption{Evolution of the kinetic and viscous energies for fluid flow in a spherical shell equipped with the Schwarzschild spatial metric.}
\label{convergence_schwarzschild_energy}
\end{figure}

Again, the flow evolves towards the steady-state solution corresponding to a rotational Killing vector field, as shown in Figures~\ref{fig_schwartschild1} and~\ref{fig_schwartschild2}. The steady-state solution is also observed to be stable, as demonstrated by the energy evolution in Figure~\ref{convergence_schwarzschild_energy}.

\section{Conclusion}\label{sec_conclusion}

This paper has developed and validated an intrinsic finite element formulation for the incompressible Navier--Stokes equations on Riemannian manifolds. The method is assembled directly in the parameter space, with the metric tensor, inverse metric, and Christoffel symbols carrying all geometric information. We derived the corresponding weak formulation and proved that the backward Euler time discretisation is energy-stable. The numerical experiments employed Killing vector fields as the primary benchmark, since these fields are steady-state solutions of the NS equations and form the null space of the corresponding strain-rate operator.

The results demonstrate that, for geometries admitting regular parameterisations, the intrinsic formulation offers significant advantages in both efficiency and accuracy over the surface FEM. The intrinsic formulation is also geometrically transparent, since all terms can be traced directly to the Riemannian metric and the Levi--Civita connection. Moreover, the method extends naturally to higher-dimensional manifolds. The Schwarzschild-shell example demonstrates that essentially the same formulation can be applied to a three-dimensional Riemannian manifold. For regular geometries such as the torus and helicoid, the simulations exhibit stable convergence towards the expected Killing fields, with the viscous energy decaying to zero while the kinetic energy remains constant in the long-time regime.

Although the surface FEM has become increasingly popular in recent years, the findings of this paper also reveal several of its limitations. High-quality meshes are important for achieving accurate solutions, and the surface normals must be approximated sufficiently accurately to attain the same level of accuracy as the intrinsic formulation. In the present computations, the surface finite element method is also more computationally expensive than the intrinsic formulation and may exhibit excessive numerical dissipation.

Overall, the intrinsic and embedded surface FEM formulations should be viewed as complementary rather than competing frameworks. The intrinsic parameter-space method is particularly attractive when a regular parameterisation is available or when the problem is posed on a higher-dimensional Riemannian manifold. In contrast, the surface FEM is likely to be preferable when no global parameterisation exists or when coordinate singularities cannot be avoided. Future work will focus on evolving-surface problems using the intrinsic formulation, in which the metric tensor is evolved directly instead of updating the computational mesh. Such an approach could provide a further advantage of the proposed intrinsic formulation.

\section*{Declarations}

\begin{itemize}
    \item Funding: This research received no external funding.

    \item Conflict of interest/Competing interests: The author declares that there are no competing interests.

    \item Ethics approval and consent to participate: Not applicable.

    \item Consent for publication: Not applicable.

    \item Data availability: The data supporting the findings of this study, together with the corresponding FreeFEM source code, are publicly available at
    \url{https://yongxingwang.github.io/surfacefluids/}.

    \item Materials availability: Not applicable.

    \item Code availability: The FreeFEM source code used to generate the numerical results presented in this paper is publicly available at
    \url{https://yongxingwang.github.io/surfacefluids/}.

    \item Author Contributions: The author solely conceived the study, developed the mathematical and numerical formulations, performed the theoretical analysis, implemented the numerical methods, conducted the computational experiments, analysed the results, and wrote, reviewed, and revised the manuscript.
\end{itemize}

\appendix
\section{Geometric preliminaries and covariant derivatives}\label{sec_notations}
Let $(M,g)$ be a compact, smooth, and oriented Riemannian manifold, with boundary $\partial M$. Let $(\cdot, \cdot)_g$ denote the Riemannian metric on $M$. For each point $p \in M$, denote by $T_pM$ the tangent space and by $T_p^*M$ the cotangent space at $p$.

The disjoint union of all tangent spaces and cotangent spaces over $M$ forms the tangent bundle $TM$ and the cotangent bundle $T^*M$, respectively.  
Both $TM$ and $T^*M$ are smooth vector bundles over $M$: each fiber $T_pM$ and $T_p^*M$ is a vector space, but the bundles themselves do not carry a natural vector space structure globally, since vectors belonging to different fibers cannot be added.

A smooth section of the tangent bundle is a smooth vector field; the space of smooth sections is denoted by $\Gamma( TM)$. Thus, for $X \in \Gamma( TM)$, one has $X(p) \in T_pM$ for every $p \in M$. Similarly, the space of smooth sections of the cotangent bundle is denoted by $\Gamma( T^*M)$; its elements are smooth one-forms $\omega$, with $\omega(p) \in T_p^*M$ for each $p \in M$.

$C^k(TM)$ denotes $k-$times continuously differentiable vector fields. $C(TM)=C^0(TM)$, and $\Gamma(TM)=C^\infty( TM)$. 

Let $T_n^mM := \overbrace{TM\otimes TM \cdots \otimes TM}^m \otimes \underbrace{T^*M\otimes T^*M \cdots \otimes T^*M}_n$ denote the $(m,n)-$tensor bundle of $M$. With this notation $TM=T^1M$ and $T^*M=T_1M$. For example, $T_2^1M$ is a $(2,1)-$tensor bundle over $M$, and its fiber at a point $p\in M$ is $(T^2_1M)_p=T_pM\otimes T_pM\otimes T_p^*M$. Notions $\Gamma( T^m_nM)$ and $C^k( T^m_nM)$ have the similar meaning as defined above.

In local coordinates, the coordinate basis of the tangent space $T_pM$ is given by
\[
{\bf e}_k \equiv \partial_k \equiv   \frac{\partial}{\partial x^k},
\]
and the corresponding dual basis of the cotangent space $T_p^*M$ is $\{dx^k\}$. As an example, for a tensor $V\in \Gamma( T^2_1 M)$, the local representation of $V$ is given by:
\[
V=V^{ij}_k \partial_i\otimes\partial_j\otimes dx^k.
\]

Covariant differentiation of a vector field $v\in\Gamma(TM)$ with respect to the $j$th coordinate extends the ordinary partial derivative by accounting for the variation of the basis vectors:
\begin{equation}\label{partial_covariant}
\nabla_j (v^i \mathbf{e}_i)
=
\partial_j v^i\, \mathbf{e}_i
+
v^i\,\partial_j \mathbf{e}_i.
\end{equation}

The derivative of a basis vector can, in general, be expressed as a linear combination of the basis vectors:
\begin{equation}\label{derivative_basis_tangent}
\partial_j \mathbf{e}_i
=
\Gamma_{ij}^k \mathbf{e}_k,
\end{equation}
where $\Gamma_{ij}^k$ are called the Christoffel symbols (or connection coefficients). A natural question is: what conditions are required to determine the connection coefficients uniquely? The Levi--Civita connection provides a unique connection by requiring that it be torsion-free and compatible with the Riemannian metric:

\begin{itemize}
\item Torsion-free:
\[
\Gamma_{ij}^k = \Gamma_{ji}^k
\;\Longleftrightarrow\;
\nabla_i \partial_j = \nabla_j \partial_i
\quad
\left(
\nabla_{\mathbf{e}_i} \mathbf{e}_j
=
\nabla_{\mathbf{e}_j} \mathbf{e}_i
\right).
\]

\item Metric compatibility:
\[
\nabla_w (u \cdot v)
=
v\cdot\nabla_w u 
+
u \cdot \nabla_w v,
\qquad
\text{where }
u \cdot v = (u,v)_g .
\]
\end{itemize}

The metric compatibility is equivalent to
\begin{equation}\label{metric_compatibility}
\partial_k g_{ij}
=
\Gamma_{ki}^l g_{jl}
+
\Gamma_{kj}^l g_{il}.
\end{equation}

Indeed, taking \(u = \mathbf{e}_i\), \(v = \mathbf{e}_j\), and \(w = \mathbf{e}_k\), we have
\[
\nabla_k (\mathbf{e}_i \cdot \mathbf{e}_j)
=
\nabla_k \mathbf{e}_i \cdot \mathbf{e}_j
+
\mathbf{e}_i \cdot \nabla_k \mathbf{e}_j,
\]
which yields \ref{metric_compatibility}.

The torsion-free condition and the metric-compatibility allows us to determine the connection coefficients.
From~\eqref{metric_compatibility}, we can write two additional equations:
\begin{equation}\label{metric_com1}
\partial_i g_{kj}
=
\Gamma_{ik}^l g_{jl}
+
\Gamma_{ij}^l g_{kl},
\end{equation}
\begin{equation}\label{metric_com2}
-\partial_j g_{ik}
=
-\Gamma_{ji}^l g_{kl}
-
\Gamma_{jk}^l g_{il}.
\end{equation}

Using the torsion-free condition, the sum of equations
\eqref{metric_compatibility}, \eqref{metric_com1}, and \eqref{metric_com2}
yields
\[
\partial_k g_{ij}
+
\partial_i g_{kj}
-
\partial_j g_{ik}
=
2 \Gamma_{ki}^l g_{jl},
\]
which allows us to compute the connection coefficients:
\begin{equation}\label{christoffel_symbol}
\Gamma_{ki}^l
=
\frac{1}{2} g^{lj}
\left(
\partial_k g_{ij}
+
\partial_i g_{kj}
-
\partial_j g_{ik}
\right).
\end{equation}

Another immediate consequence of metric compatibility is that
\(\nabla g = 0\).
This follows from a direct computation:
\begin{equation*}
\begin{split}
\nabla g
&=
\nabla_k
\left(
g_{ij} \, dx^i \otimes dx^j
\right)
\otimes dx^k
\\
&=
\partial_k g_{ij}
\left(
dx^i \otimes dx^j \otimes dx^k
\right)
+
g_{ij}
\left(
\nabla_k dx^i \otimes dx^j \otimes dx^k
\right)
+
g_{ij}
\left(
dx^i \otimes \nabla_k dx^j \otimes dx^k
\right)
\\
&=
\left(
\partial_k g_{lm}
-
\Gamma_{kl}^i g_{im}
-
\Gamma_{km}^j g_{lj}
\right)
dx^l \otimes dx^m \otimes dx^k
= 0 .
\end{split}
\end{equation*}

\section{Bochner and Hodge Laplacians}\label{sec_hodge_laplacian}

Let $\nabla$ be the Levi-Civita connection on $M$. For $u\in C^1(TM)$, the covariant derivative $\nabla u\in C(T_1^1M)$ is given in the local coordinates by:
\begin{equation}
\begin{split}
\nabla u &= \nabla_j u\otimes dx^j = \nabla_j (u^i\partial_i) \otimes dx^j \\
&= \left(\partial_ju^i\partial_i + u^i\nabla_j\partial_i\right) \otimes dx^j\\
&=  \left(\partial_ju^k\partial_k + u^i\Gamma_{ij}^k\partial_k\right) \otimes dx^j\\
&=  \left(\partial_ju^k + u^i\Gamma_{ij}^k\right)\partial_k \otimes dx^j.
\end{split}
\end{equation}

A conventional short-hand notation for the components of the covariant derivative is: $u^k_{;j}=\partial_ju^k+u^i\Gamma_{ij}^k=\nabla_j u^k$. While the partial derivative is: $u^k_{,j}=\partial_ju^k$. This short-hand notation needs to be clarified before being used for high-order derivatives: for example, if $v\in C^1\left(T^1_1M\right)$, then
\begin{equation}\label{nabla_v}
\begin{split}
\nabla v &= \nabla_j v\otimes dx^j = \nabla_j (v^i_k\partial_i\otimes dx^k) \otimes dx^j \\
&= \left(\nabla_j(v^i_k\partial_i)\otimes dx^k + (v^i_k\partial_i)\otimes \nabla_j(dx^k)\right) \otimes dx^j\\
&= \left(\nabla_j(v^i_k\partial_i)\otimes dx^k - (v^i_k\partial_i)\otimes \Gamma_{mj}^k dx^m\right) \otimes dx^j\\
&= \left(\nabla_j(v^i_m\partial_i) - (v^i_k\partial_i) \Gamma_{mj}^k \right) \otimes dx^m\otimes dx^j.
\end{split}
\end{equation}
If $v=\nabla u$, consistent notations should be: $\nabla_m u = v^i_m\partial_i$, $\nabla_m u^i = v^i_m$, $u^i_{; m} = v^i_m$, then (\ref{nabla_v}) becomes:
\begin{equation}\label{nabla_nabla_u}
\begin{split}
\nabla (v) &= \nabla (\nabla u) \\
&= \left(\nabla_j(v^i_m\partial_i) - (v^i_k\partial_i) \Gamma_{mj}^k \right) \otimes dx^m\otimes dx^j\\
&= \left(\partial_j(v^i_m) \partial_i + v^i_m \Gamma_{ij}^n \partial_n- (v^i_k\partial_i) \Gamma_{mj}^k \right) \otimes dx^m\otimes dx^j \\
&= \left(\partial_jv^n_m  + v^i_m \Gamma_{ij}^n - v^n_k \Gamma_{mj}^k \right) \partial_n\otimes dx^m\otimes dx^j\\
&= \left(\partial_ju^n_{;m}  + u^i_{;m} \Gamma_{ij}^n - u^n_{;k} \Gamma_{mj}^k \right) \partial_n\otimes dx^m\otimes dx^j.
\end{split}
\end{equation}
Now, to use notation
\begin{equation}\label{second_order_derivative_components}
  u^n_{;mj} = \partial_ju^n_{;m}  + u^i_{;m} \Gamma_{ij}^n - u^n_{;k} \Gamma_{mj}^k  
\end{equation}
for the component of the second covariant of $u\in C^1(TM)$, it can be seen from the last line in (\ref{nabla_v}) that
\[
\nabla_j(\nabla_m u)- (\nabla_k u) \Gamma_{mj}^k = u^n_{;mj} \partial_n,
\]
or
\begin{equation}
\left(\nabla_j\nabla_m - \Gamma_{mj}^k\nabla_k\right) u^n\partial_n  = u^n_{;mj} \partial_n.   
\end{equation}

In other words, $u^n_{;mj}$ is \emph{not} the same as $\nabla_j\nabla_m u^n$. Instead, $u^n_{;mj}$ is the component of second-order covariant derivative $\nabla(\nabla u)$ of $u$. While $\nabla_j(\nabla_m u)$ means: first, compute $\nabla_m u$, then throw the basis $dx^m$, finally compute $\nabla_j(\nabla_m u)$. Consiering the basis:
\begin{equation}\label{second_order_derivative}
\begin{split}
\nabla_j(\nabla_m u \otimes dx^m) &= \nabla_j(\nabla_m u) \otimes dx^m + \nabla_m u \otimes \nabla_j(dx^m)  \\
& = \nabla_j(\nabla_m u) \otimes dx^m - \nabla_m u \otimes \Gamma_{jn}^m dx^n\\
& = \left(\nabla_j(\nabla_n u)  - \nabla_m u  \Gamma_{jn}^m \right)\otimes dx^n
=u_{;nj}^k\partial_k \otimes dx^n.
\end{split}
\end{equation}

Based on the above derivation, $u^n_{;mj}$ represent components of a tensor (components), while $\nabla_j\nabla_m(u^n)$ is \emph{not} a tensor.

\begin{remark}
It is important to distinguish between the three notations $\nabla$, $\nabla_j$, and $\nabla_v$, as they represent different objects. Let $u, v \in \Gamma(TM)$. The covariant derivative of $u$ is a $(1,1)$-tensor given by
\[
\nabla u = (\nabla_j u^i)\,\partial_i \otimes dx^j,
\]
where $\nabla_j u^i$ denotes the components of the covariant derivative, while $\nabla_j u$ denotes the vector field corresponding to the basis one-form $dx^j$. Finally, the covariant derivative of $u$ in the direction of $v$ is the vector field
\[
\nabla_v u = v^j \nabla_j (u^i \partial_i).
\]
\end{remark}

We now compute the divergence of the strain rate tensor defined in~(\ref{viscous_strain_rate}) which yields the Hodge Laplacian (\ref{hodge_laplacian_paper}).
From~(\ref{nabla_nabla_u}), we have
\[
\nabla(\nabla u)
= \nabla\!\left(u^i{}_{;j}\,\partial_i \otimes dx^j\right)
= u^i{}_{;jk}\,\partial_i \otimes dx^j \otimes dx^k .
\]

Notice that raising indices and taking covariant derivatives commute (since $\nabla g = 0$), so
\[
\nabla\!\left(g^{mj}\left(u^i{}_{;j}\,\partial_i \otimes dx^j\right)\right)
= g^{mj}\left(u^i{}_{;jk}\,\partial_i \otimes dx^j \otimes dx^k\right)
= g^{mj} u^i{}_{;jk}\left(\partial_i \otimes \partial_m \otimes dx^k\right).
\]

Similarly, swapping the indices $i$ and $m$ gives the covariant derivative of the transpose in~(\ref{viscous_strain_rate}); the order of the basis elements remains unchanged:
\[
\nabla\!\left(g^{ij}\left(u^m{}_{;j}\,\partial_m \otimes dx^j\right)\right)
= g^{ij} u^m{}_{;jk}\left(\partial_i \otimes \partial_m \otimes dx^k\right).
\]

Contracting the indices $m$ and $k$ yields
\[
\operatorname{div}(2\epsilon(u))
= \left(g^{kj} u^i{}_{;jk} + g^{ij} u^k{}_{;jk}\right)\partial_i .
\]

The first term gives the Bochner Laplacian $\Delta_B(u)$.  
The second term is analysed as follows (it is not zero on a general manifold, although it vanishes in Euclidean space due to the divergence-free condition).  
From~(\ref{second_order_derivative_components}), we have
\begin{equation}\label{second_order_covariant_derivative}
    u^n{}_{;mj}
= \partial_j\!\left(\partial_m u^n + u^l \Gamma_{lm}^n\right)
+ \left(\partial_m u^i + u^l \Gamma_{lm}^i\right)\Gamma_{ij}^n
- \left(\partial_k u^n + u^l \Gamma_{lk}^n\right)\Gamma_{mj}^k.
\end{equation}
Setting $j = n$ gives
\begin{equation*}
\begin{split}
u^n{}_{;mn}
&= \partial_n\!\left(\partial_m u^n + u^l \Gamma_{lm}^n\right)
+ \left(\partial_m u^i + u^l \Gamma_{lm}^i\right)\Gamma_{in}^n
- \left(\partial_k u^n + u^l \Gamma_{lk}^n\right)\Gamma_{mn}^k \\
&= \partial_n \partial_m u^n
+ (\partial_n u^l)\Gamma^n_{lm}
+ (\partial_m u^i)\Gamma_{in}^n
- (\partial_k u^n)\Gamma_{mn}^k \\
&\quad + \left(\partial_n \Gamma^n_{lm}
+ \Gamma_{lm}^i \Gamma_{in}^n
- \Gamma_{lk}^n \Gamma_{mn}^k\right) u^l \\
&= \partial_m\!\left(\partial_n u^n + u^l \Gamma_{ln}^n\right)
- u^l \partial_m \Gamma_{ln}^n
+ (\partial_n u^l)\Gamma^n_{lm}
- (\partial_k u^n)\Gamma_{mn}^k \\
&\quad + \left(\partial_n \Gamma^n_{lm}
+ \Gamma_{lm}^i \Gamma_{in}^n
- \Gamma_{lk}^n \Gamma_{mn}^k\right) u^l \\
&= \partial_m (u^n{}_{;n})
+ \left(\partial_n \Gamma^n_{lm}
- \partial_m \Gamma_{ln}^n
+ \Gamma_{lm}^i \Gamma_{in}^n
- \Gamma_{lk}^n \Gamma_{mn}^k\right) u^l \\
&= R_{lm} u^l ,
\end{split}
\end{equation*}
with $ R_{kj}
= \partial_i \Gamma_{jk}^i
- \partial_j \Gamma_{ik}^i
+ \Gamma_{jk}^l \Gamma_{il}^i
- \Gamma_{ik}^l \Gamma_{jl}^i$, and the divergence-free condition $u^n{}_{;n} = 0$ in mind.  
Finally, we obtain
\begin{equation}
\operatorname{div}(D(u))
= \Delta_B(u) + g^{ij} R_{jl} u^l \partial_i,
\end{equation}
or (\ref{hodge_laplacian_paper}) after raising the index. 

\section{Lie derivative}\label{sec_lie_derivative}
We introduce the Lie derivative to facilitate the discussion of Killing vector fields in \ref{sec_killing_fields}. Let $X$ be a smooth vector field on $M$. The vector field $X$ generates a flow $\varphi_t(p)$ (corresponding to streamlines in flat space) in the following sense:
\[
\frac{d}{dt}\varphi_t(p) = X(\varphi_t(p)),
\]
where $p = \varphi_0(p)$ is the initial point -- that is, $p$ explicitly specifies the initial condition. In flat space, one usually refers to the streamlines as curves $\gamma(t)$, but one still needs an initial point to integrate and determine the curve. In other words, $\varphi_t(p)$ moves the point $p$ around on $M$.

For a tensor field $T$, the Lie derivative of $T$ along $X$ is defined as
\[
\mathcal{L}_X T = \left. \frac{d}{dt} \right|_{t=0} \varphi_t^* T,
\]
where $\varphi_t^*$ is the pullback operator --- it pulls a tensor back to $p$ so that tensors at different points can be compared. For example, if $\omega$ is a 1-form, $(\varphi_t^* \omega)_p$ is a 1-form at $p$, defined by
\[
(\varphi_t^* \omega)_p(v) = \omega_{\varphi_t(p)} \big( d\varphi_t(v) \big).
\]
Here, $d\varphi_t$ is the pushforward operator acting on vectors.

\begin{itemize}
    \item Lie derivative of a scalar $f$: $\mathcal{L}_X f= X(f) = X^i\partial_i f$.
    \item Lie derivative of a vector field $Y$: $\mathcal{L}_X Y = [X,Y]$, or in coordinate form: \[(\mathcal{L}_X Y)^i = X^j\partial_jY^i - Y^j\partial_jX^i\]
    \item Lie derivative of a 1-form $\omega$: $\left(\mathcal{L}_X \omega\right)(Y) = X\left(\omega(Y)\right) - \omega([X,Y])$, or in coordinate form: \[(\mathcal{L}_X \omega)_i = X^j\partial_j\omega_i - \omega_j\partial_iX^j\]
    \item Lie derivative of a (1,1)-tensor $T$ in coordinate form: 
    \[(\mathcal{L}_X T)^i_j = X^k\partial_k T^i_j - T^k_j\partial_kX^i + T^i_k\partial_jX^k\]   
    \item Lie derivative of a (2,0)-tensor $S$ in coordinate form: 
    \[(\mathcal{L}_X S)^{ij} = X^k\partial_k S^{ij} - S^{kj}\partial_kX^i - S^{ik}\partial_kX^j\] 
    \item Lie derivative of a (0,2)-tensor $g$ in coordinate form: 
    \begin{equation}\label{lie_derivative_g}
        (\mathcal{L}_X g)_{ij} = X^k\partial_k g_{ij} + g_{kj}\partial_iX^k + g_{ik}\partial_jX^k
    \end{equation}
\end{itemize}

\section{Killing vector fields}\label{sec_killing_fields}
A vector field $u\in C^1(TM)$ is called a Killing field if
\begin{equation}
(\nabla_vu, w)_g + (\nabla_w u, v)_g =0, v,w\in\Gamma(TM).
\end{equation}
It's not difficult to prove that $u$ is a Killing field iff 
\begin{equation}\label{killing_equi}
\nabla u + (\nabla u)^\top = (u^i_{;j} + g^{in}u^m_{;n}g_{mj})\partial_i\otimes dx^j =0.
\end{equation}

Equivalent definition of Killing fields is the Lie derivative of the metric tensor
\[
\mathcal{L}_u g =0
\]
vanishes along $u$ -- The metric does not change under the flow generated by $u$. In the coordinate form, this is (\ref{lie_derivative_g}). Substituting the metric compatibility condition
\begin{equation*}
\begin{split}
0&=\nabla_k(g_{ij}dx^i\otimes dx^j)=\partial_kg_{ij}dx^i\otimes dx^j -  g_{ij}\Gamma^i_{km}dx^m\otimes dx^j
 -  g_{ij}\Gamma^j_{km}dx^i\otimes dx^m \\
 &=\left(\partial_kg_{ij} - g_{mj}\Gamma^m_{ki} -  g_{im}\Gamma^m_{kj}\right)dx^i\otimes dx^j.
\end{split}
\end{equation*}
into (\ref{lie_derivative_g}):
\[
(\mathcal{L}_u g)_{ij} = u^k\left(g_{mj}\Gamma^m_{ki} +  g_{im}\Gamma^m_{kj}\right) + g_{kj}\partial_iu^k + g_{ik}\partial_ju^k.
\]
Swapping the indices $m$ and $k$ in the first two terms, we have:
\begin{equation*}
\begin{split}
(\mathcal{L}_u g)_{ij} &= u^m\left(g_{kj}\Gamma^k_{mi} +  g_{ik}\Gamma^k_{mj}\right) + g_{kj}\partial_iu^k + g_{ik}\partial_ju^k\\
&=g_{kj}\left(u^m\Gamma^k_{mi} + \partial_iu^k\right)
+g_{ki}\left(u^m\Gamma^k_{mj} + \partial_ju^k\right)\\
&=g_{jk}\nabla_iu^k + g_{ik}\nabla_ju^k \\
&= \nabla_i(g_{jk}u^k) + \nabla_j(g_{ik}u^k)\\
&=\nabla_i u_j + \nabla_j u_i.
\end{split} 
\end{equation*}
Since $\nabla_i u_j + \nabla_j u_i=0$ $\Leftrightarrow$ $g^{ki}\nabla_k u_j + \nabla_j u_k g^{ki}=g^{ki}\nabla_k u_j + \nabla_j u^i=0$ $\Leftrightarrow$ $g^{ki}\nabla_k u^l + g^{lj}\nabla_j u^i=0$
$\Leftrightarrow$
$g^{ki}\nabla_k u^l g_{ml}+ g_{ml}g^{lj}\nabla_j u^i=0$
$\Leftrightarrow$
$g^{ki}\nabla_k u^l g_{jl}+ \nabla_j u^i=0$
$\Leftrightarrow$
$g^{ki}u^l_{;k} g_{jl}+ u^i_{;j}=0$ $\Leftrightarrow$ (\ref{killing_equi}).

\begin{remark}
Since
\[
\nabla_i u_j
=
\frac{1}{2}\left(\nabla_i u_j + \nabla_j u_i\right)
+
\frac{1}{2}\left(\nabla_i u_j - \nabla_j u_i\right),
\]
the covariant derivative of the velocity decomposes into its symmetric and antisymmetric parts. The Killing condition implies that the symmetric part vanishes, so only the antisymmetric (rotational) part remains.
\end{remark}

\begin{lemma}
Suppose $u$, $v$ are Killing fields. Then
\begin{equation}
    \nabla_u v + \nabla_v u = -\text{grad}(u,v)_g.
\end{equation}
In particular, for a Killing filed $u$,
\begin{equation}\label{killing_convection}
    \nabla_u u = -\frac{1}{2}\text{grad}||u||^2_g.
\end{equation}
\end{lemma}

\begin{proof} From the definition of gradient, we have
\begin{equation}\label{killing_proof1}
\text{grad}(u,v)_g = g^{lk}\nabla_k(g_{ij}u^iv^j)\partial_l
=g^{lk}g_{ij}(u^i_{;k}v^j + u^iv^j_{;k})\partial_l,
\end{equation}
where metric compatibility condition is used. Since both $u$ and $v$ and Killing field: $u^i_{;k} + g^{in}u^m_{;n}g_{mk}=0$ and $v^j_{;k} + g^{jn}v^m_{;n}g_{mk}=0$, using which equation (\ref{killing_proof1}) can be expressed:
\begin{equation*}
\begin{split}
\text{grad}(u,v)_g 
&=-g^{lk}g_{ij}(v^jg^{in}u^m_{;n}g_{mk} + u^ig^{jn}v^m_{;n}g_{mk})\partial_l\\
&=-\left(\delta^l_m\delta^n_jv^ju^m_{;n} +\delta^l_m\delta^n_iu^iv^m_{;n}\right)\partial_l\\
&=-(v^ju^l_{;j} + u^iv^l_{;i})\partial_l \\
&= - v^j\nabla_j(u^l\partial_l) - u^i\nabla_i(v^l\partial_l) = -(\nabla_vu+\nabla_uv).
\end{split}
\end{equation*}
\end{proof}

For any Killing field $u$, after lowering or raising indices using the metric, we conclude that the strain-rate tensor satisfies $\epsilon(u)=0$. Hence, if $u$ is a Killing field, the stationary Navier--Stokes equations
\[
\rho \nabla_u u
-
2\mu\,\operatorname{div}(\epsilon(u))
+
\operatorname{grad}(\pi)
=
0,
\qquad
\operatorname{div}(u)=\nabla_i u^i=0,
\]
reduces to
\[
\rho \nabla_u u + \operatorname{grad}(\pi)=0.
\]
Alternatively, using (\ref{killing_convection}) we have
\[
-\frac{\rho}{2}\operatorname{grad}\|u\|_g^2
+
\operatorname{grad}(\pi)
=
0,
\]
which implies
\[
\pi
=
\frac{\rho}{2}\|u\|_g^2
+
C,
\]
for some constant $C$.

It is also straightforward to verify that a Killing field satisfies the divergence-free condition. Indeed,
\[
\nabla_i u_j + \nabla_j u_i=0
\quad \Longrightarrow \quad
g^{ij}\left(\nabla_i u_j + \nabla_j u_i\right)=0
\quad \Longrightarrow \quad
2\nabla_i u^i=0.
\]

Therefore, any Killing field provides a steady solution of the incompressible Navier--Stokes equations.


\bibliography{mybibfile}

\end{document}